\documentclass[prd,twocolumn,floatfix,nofootinbib]{revtex4}
\usepackage{amssymb}
\usepackage{amsmath}
\usepackage{graphicx,subfigure,color,dcolumn,booktabs,bm}
\usepackage{longtable,lscape}
\usepackage{txfonts}
\usepackage{overpic}
\usepackage{indentfirst}
\usepackage{cases}
\usepackage{multirow}
\usepackage{ulem}
\usepackage[colorlinks,
            citecolor=blue,
            anchorcolor=red,
            menucolor=red,
            linkcolor=red,
            filecolor=red,
            runcolor=red,
            urlcolor=blue,
            frenchlinks=false]{hyperref}

\graphicspath{{Figures/}}
\allowdisplaybreaks
\begin{document}

\title{Masses and magnetic moments of exotic fully heavy pentaquarks}
\author{Wen-Xuan Zhang$^{1}$}
\email{zhangwx89@outlook.com}
\author{Hong-Tao An$^{2}$}
\email{anht20@lzu.edu.cn}
\author{Duojie Jia$^{1,3}$\thanks{%
Corresponding author}}
\email{jiadj@nwnu.edu.cn}
\affiliation{$^1$Institute of Theoretical Physics, College of Physics and Electronic
Engineering, Northwest Normal University, Lanzhou 730070, China \\
$^2$School of Physical Science and Technology, Lanzhou University, Lanzhou
730000, China \\
$^3$Lanzhou Center for Theoretical Physics, Lanzhou University, Lanzhou,
730000, China \\
}
\date{\today}

\begin{abstract}
Inspired by the observation of a resonant state $X(6600)$ of fully charm
tetraquark by the CMS experiment of LHCb Collaboration in double $J/\psi $
decay channel, we perform a systematical study of all configurations of
fully heavy pentaquarks $P_{Q_{1}Q_{2}Q_{3}Q_{{4}}\bar{Q_{5}}}$ ($Q_{i}=c,b$%
, $i=1,2,3,4,5$) in their ground states in unified framework of MIT bag
model. The color-spin wavefunctions of pentaquarks, classified via Young
tableau and presented in terms of the Young-Yamanouchi bases, are used to
compute masses and magnetic moments of fully heavy pentaquarks
via numerical variational method, predicting a set of masses ranging
from $8.229$ GeV for the $P_{cccc\bar{c}}$ to $24.770$ GeV for the $P_{bbbb\bar{b}}$.
Combining with computed masses of fully heavy mesons and baryons, we find
that masses of fully heavy hadrons(mesons, baryons, tetraquarks and
pentaquarks) with identical flavor rise almost linearly with the number
of valence quarks in hadrons, being consistent with the heavy quark
symmetry in the heavy-quark limit.

PACS number(s):12.39Jh, 12.40.Yx, 12.40.Nn

\end{abstract}

\maketitle
\date{\today}

\section{Introduction}

Possible existence of multiquark hadrons like tetraquarks ($q^{2}\bar{q}^{2}$) 
and pentaquarks ($q^{4}\bar{q}$) was suggested earlier at the birth of
quark model \cite{Gell-Mann:1964ewy,Zweig:1964ruk}. Later in the 1970s,
multiquark states are studied by Jaffe via the MIT bag model \cite%
{Jaffe:1976ig,Jaffe:1976ih}. Since observation of the $X(3872)$ \cite%
{Belle:2003nnu} in 2003 by the Belle, many candidates of tetraquarks have
been observed, some of which, such as the $Z_{c}(3900)$ \cite{BESIII:2013ris}
and the $T_{cc}(3875)$, are undoubtedly exotic. In 2020, a first candidate
of fully charm tetraquark, the $X(6900)/$the $X(6600)$, has been observed by
LHCb in the di-$J/\psi $ invariant mass spectrum and later confirmed by CMS
and ATLAS of the LHC at CERN \cite%
{Bouhova-Thacker:2022vnt,Zhang:2022toq,LHCb:2020bwg}. For experimental
searches of light-flavor pentaquark hadrons, it has been a long and
nontrivial history, with no undisputed candidates found in over 50 years.
Convincing evidence for pentaquark-like structures $P_{c}(4450)^{+}$ and $%
P_{c}(4380)^{+}$ with a minimal quark content of $uudc\bar{c}$ was reported
by LHCb in a study of $\Lambda _{b}^{0}\rightarrow J/\psi pK^{-}$($J/\psi
\rightarrow \mu ^{+}\mu ^{-}$) decays \cite{LHCb:2015c} in 2015, for which
the former peak(state) $P_{c}(4450)^{+}$ was resolved further into two
states $P_{c}(4440)^{+}$ and $P_{c}(4457)^{+}$ (over $5.4\sigma $) in 2019.
More recently, LHCb collaboration reported evidences of two
new charmonium pentaquarks with strangeness, $P_{cs}(4459)$ \cite%
{LHCb:2021sb}, in the $J/\psi \Lambda $ distribution(in $\Xi
_{b}^{-}\rightarrow J/\psi \Lambda K^{-}$ decays) in 2021 and $P_{\psi
s}^{\Lambda }(4338)$ \cite{LHCb:2022jad}, in flavour-untagged $%
B^{-}\rightarrow J/\psi \Lambda\bar{p}$ decays in 2022.

After observation and measurement of two $P_{c}$ states, many theoretical
groups explained the $P_{c}(4450)^{+}$ and $P_{c}(4380)^{+}$ states as
compact pentaquarks with diquarks and triquarks as building blocks 
\cite%
{Maiani:B15,Lebed:B15,Anisovich:15,LiHeHe:JH15,GhoshBC:PN17,Wang:E16,ZhuQ:B16}
, except for some few attempts via full five-body dynamics 
\cite{Richard:B17} which lead
to states below the lowest threshold for spontaneous dissociation. On the
other hand, there are molecular picture, proposed 
\cite{Yang:C12,Wu:Prl10,WuLeeZ:C12,KR:Prl15} before the first LHCb results \cite%
{LHCb:2015c}, which interprets these narrow $P_{c}$ states in terms of
deuteron-like loosely bound states of the baryon and meson, such as the $%
\Sigma _{c}\bar{D}^{(\ast )}$ states. This picture favors the narrow $P_{c}$
states, which have measured widths about $6-10$ MeV for $P_{c}(4457)^{+}$
as well as $P_{c}(4312)^{+}$ and about $21$ MeV for $P_{c}(4440)^{+}$, while the
compact picture attributes narrow width to spin-orbit interaction (via
spatial separation of $c$ and $\bar{c}$ quarks). As the observed state $%
P_{c}(4380)^{+}$ is wider (with width about $205$ MeV), the issue of width
suppression remains to be explored. Till now, there are different pictures
and approaches employed to analyze the hidden-charm pentaquarks ($P_{c}$ and $%
P_{cs}$) 
\cite{ParkLC:D22,ChenL:D22,BurnsB:E22,YangH:SC21,ShiHW:E21,LiLiuS:D21,LingXL:D21,WuPL:D21,RuangyooPC:D22,LingDW:E21,LuLS:D21,WuCJ:D21,DuGH:Jh21,XiaoWZ:D21,ZhuSH:D21,Chen:E21,YanPS:E21,YangPW:D17}, 
the hidden-strange and hidden-bottom pentaquarks 
\cite{YangC:C23,HuangZ:D21,ZhuKL:D20,YangPS:D19}. One can infer Refs.~
\cite{Burns:EA15,ChenCLZ:PR16} for the pre-2019 reviews and Refs.~
\cite{LiuCCL:PP19,Brambilla:PR20} for recent reviews.

The purpose of this work is to systematically study all possible
ground-states of fully heavy pentaquarks, $P_{Q_{1}Q_{2}Q_{3}Q_{{4}}\bar{%
Q_{5}}}$ ($Q_{i}=c,b$, $i=1,2,3,4,5$) in unified framework of MIT bag
model. For this, we construct all color-spin wavefunctions of pentaquarks
based on $SU(2)_{s}\times SU(3)_{c}$ group, classified via Young tableau and
expressed in terms of the Young-Yamanouchi bases, to compute masses and
magnetic moments of fully heavy pentaquarks in ground states and thereby
predict their masses. With the help of numerical
variation upon the bag radius, the chromomagnetic mixing among the color-spin
configurations are explored in details. Computing further spin-independent
masses of the mesons, baryons and tetraquarks all made of identical heavy
flavors, we find a linear dependence of the hadron mass upon the heavy
quark number $N$ in hadrons. This linear rise of hadron masses is notable
for the bottom sector and is demonstrated analytically via a variational
method of bag model, being consistent with heavy quark effective theory (HQET)
in heavy quark limit.

After introduction, the wavefunctions of the fully heavy pentaquarks with
various configurations are classified in Sect.~\ref{sec:WF} by the Young
tableau and expressed in terms of the Young-Yamanouchi bases. In Sect.~\ref%
{sec:bagmodel}, we briefly review basic relations of MIT bag model including
magnetic moments for each spin state, and describe how to apply the
variational method to solve the bag model for the heavy hadrons. In Sect.
\ref{sec:hadrons}, we perform numerical analysis of mass and magnetic moment
for fully heavy pentaquarks. Similar computation are given to fully heavy mesons
and baryons to show a linear relation for these fully heavy hadrons in Sect. \ref{sec:linearity}.
The paper ends with summary and conclusions in Sect.~%
\ref{sec:conclusions}.

\section{Wavefunctions of fully heavy pentaquarks}

\label{sec:WF}

Fully heavy pentaquarks $P_{Q_{1}Q_{2}Q_{3}Q_{{4}}\bar{Q_{5}}}$ may contain
two or more identical particles in the component $Q_{1}Q_{2}Q_{3}Q_{{4}}$,
for which the corresponding parts of the wavefunctions have to be
antisymmetric due to Pauli principle and colorlessness of $%
P_{Q_{1}Q_{2}Q_{3}Q_{{4}}\bar{Q_{5}}}$. Consider the pentaquark $P_{cccb\bar{%
b}}$, for instance. Since the component $ccc$ in $P_{cccb\bar{b}}$ are
identical in flavor, the antisymmetry requirement for a fermionic hadron
(color singlet) implies that the wavefunction of $P_{cccb\bar{b}}$ has to be
antisymmetric in orbital, color and spin under exchange of each $cc$
pair among the component $ccc$. To respect the overall symmetry of the
hadronic wavefunctions mentioned above, we utilize the Young tableau to
represent the irreducible representations (bases) of the permutation group
and thereby classify the pentaquark configurations with certain symmetry.
For a given classification, we use the Young-Yamanouchi basis, which
corresponds to the Young tableau, to construct explicitly the configuration
describing the pentaquark we consider.

First of all, we consider color wavefunctions of the pentaquark $%
P_{Q_{1}Q_{2}Q_{3}Q_{{4}}\bar{Q_{5}}}$, which are singlets in color space.
Utilizing direct product of the fundamental (color) representations $[3]_{c}$
and $[\bar{3}]_{c}$, one can classify, in the language of group theory, the
color wavefunctions of a pentaquark as below:
\begin{eqnarray}
&&[[3]_{c}\otimes \lbrack 3]_{c}\otimes \lbrack 3]_{c}\otimes \lbrack
3]_{c}]\otimes \lbrack \bar{3}]_{c}  \notag \\
&=&[(1_{c}\oplus 8_{c}\oplus 8_{c}\oplus 10_{c})\otimes 3_{c}]\otimes \bar{3}%
_{c}  \notag \\
&\rightarrow &(1_{c}\otimes 3_{c}\otimes \bar{3}_{c})\oplus (8_{c}\otimes
3_{c}\otimes \bar{3}_{c})\oplus (8_{c}\otimes 3_{c}\otimes \bar{3}_{c})
\notag \\
&\rightarrow &(3_{c}\otimes \bar{3}_{c})\oplus (3_{c}\otimes \bar{3}%
_{c})\oplus (3_{c}\otimes \bar{3}_{c}),
\end{eqnarray}
in which the allowed color singlet is $\{3_{c}\otimes\bar{3}_{c}\}$ only.

One is then forced to consider the color triplet in the direct product, $%
[[3]_{c}\otimes \lbrack 3]_{c}\otimes \lbrack 3]_{c}\otimes \lbrack 3]_{c}]$%
, of the component $Q_{1}Q_{2}Q_{3}Q_{{4}}$, which corresponds to Young
tableau [2,1,1] expressed by

\renewcommand{\tabcolsep}{0.1cm}
\renewcommand{\arraystretch}{1}
\begin{align}\label{eq-color1}
\left [\left(12\right)_{6}34\right ]_{3}=
\begin{tabular}{|c|c|}
\cline{1-2}
1 &  2   \\
\cline{1-2}
\multicolumn{1}{|c|}{3} \\
\cline{1-1}
\multicolumn{1}{|c|}{4}  \\
\cline{1-1}
\end{tabular},
\quad
\left [\left(12\right)_{\bar{3}}34\right ]_{3}=
\begin{tabular}{|c|c|}
\cline{1-2}
1 &  3    \\
\cline{1-2}
\multicolumn{1}{|c|}{2} \\
\cline{1-1}
\multicolumn{1}{|c|}{4}  \\
\cline{1-1}
\end{tabular},
\nonumber \\
\left [\left(123\right)_{1}4\right ]_{3}=
\begin{tabular}{|c|c|}
\cline{1-2}
1 &  4   \\
\cline{1-2}
\multicolumn{1}{|c|}{2} \\
\cline{1-1}
\multicolumn{1}{|c|}{3} \\
\cline{1-1}
\end{tabular}.
\end{align}
Here, the subscript labels the irreducible representation of the color group
$SU(3)_{c}$. This yields three color-singlet configurations of $%
P_{Q_{1}Q_{2}Q_{3}Q_{{4}}\bar{Q_{5}}}$ if one combines each of
color-triplets ($3_{c}$) listed in Eq.~(\ref{eq-color1}) with the
antitriplet ($\bar{3}_{c}$) of the remaining antiquark $\bar{Q}_{5}$. We
express the Young tableau representing the obtained colorless configurations
as \renewcommand{\tabcolsep}{0.1cm} \renewcommand{\arraystretch}{1}
\renewcommand{\tabcolsep}{0.1cm}
\renewcommand{\arraystretch}{1}
\begin{equation}\label{three}
{\begin{tabular}{|c|c|}
\cline{1-2}
1 & 2 \\
\cline{1-2}
\multicolumn{1}{|c|}{3} \\
\cline{1-1}
\multicolumn{1}{|c|}{4} \\
\cline{1-1}
\end{tabular}
\otimes\bar{5}}_{\begin{tabular}{|c|}
\multicolumn{1}{c}{$\phi_{1}^{P}$}\end{tabular}},
{\begin{tabular}{|c|c|}
\cline{1-2}
1 & 3 \\
\cline{1-2}
\multicolumn{1}{|c|}{2} \\
\cline{1-1}
\multicolumn{1}{|c|}{4}  \\
\cline{1-1}
\end{tabular}
\otimes\bar{5}}_{\begin{tabular}{|c|}
\multicolumn{1}{c}{$\phi_{2}^{P}$}\end{tabular}},
{\begin{tabular}{|c|c|}
\cline{1-2}
1 & 4 \\
\cline{1-2}
\multicolumn{1}{|c|}{2} \\
\cline{1-1}
\multicolumn{1}{|c|}{3} \\
\cline{1-1}
\end{tabular}
\otimes\bar{5}}_{\begin{tabular}{|c|}
\multicolumn{1}{c}{$\phi_{3}^{P}$}\end{tabular}},
\end{equation}
and correspond them to the following color wavefunctions explicitly:

\begin{eqnarray}
\phi _{1}^{P} &=&\frac{1}{4\sqrt{3}}\Big[(2bbgr-2bbrg+gbrb-gbbr+bgrb-bgbr
\notag \\
&&-rbgb+rbbg-brgb+brbg)\bar{b}+(2rrbg-2rrgb  \notag \\
&&+rgrb-rgbr+grrb-grbr+rbgr-rbrg+brgr  \notag \\
&&-brrg)\bar{r}+(2ggrb-2ggbr-rggb+rgbg-grgb  \notag \\
&&+grbg+gbgr-gbrg+bggr-bgrg)\bar{g}\Big],  \label{ph1p}
\end{eqnarray}

\begin{eqnarray}
\phi _{2}^{P} &=&\frac{1}{12}\Big[(3bgbr-3gbbr-3brbg+3rbbg-rbgb-2rgbb  \notag
\\
&&+2grbb+brgb+gbrb-bgrb)\bar{b}+(3grrb-3rgrb  \notag \\
&&-3brrg+3rbrg-rbgr-2gbrr+2bgrr-grbr  \notag \\
&&+rgbr+brgr)\bar{r}+(3grgb-3rggb+3bggr-3gbgr  \notag \\
&&-grbg+rgbg+2rbgg-2brgg+gbrg-bgrg)\bar{g}\Big],  \label{ph2p}
\end{eqnarray}

\begin{eqnarray}
\phi _{3}^{P} &=&\frac{1}{3\sqrt{2}}\Big[(grbb-rgbb+rbgb-brgb+bgrb-gbrb)\bar{%
b}  \notag \\
&&+(grbr-rgbr+rbgr-brgr+bgrr-gbrr)\bar{r}  \notag \\
&&+(grbg-rgbg+rbgg-brgg+bgrg-gbrg)\bar{g}\Big].  \label{ph3p}
\end{eqnarray}

In the spin space, the direct product of five fermions represented in terms
of Young tableau can be written as
\renewcommand{\tabcolsep}{0.05cm}
\renewcommand{\arraystretch}{1}
\begin{align}
&\begin{tabular}{|c|}
\cline{1-1}
\multicolumn{1}{|c|}{$\quad$}  \\
\cline{1-1}
\end{tabular}_{\begin{tabular}{|c|} \multicolumn{1}{c}{$S$}\end{tabular}}
\otimes
\begin{tabular}{|c|}
\cline{1-1}
\multicolumn{1}{|c|}{$\quad$}  \\
\cline{1-1}
\end{tabular}_{\begin{tabular}{|c|} \multicolumn{1}{c}{$S$}\end{tabular}}
\otimes
\begin{tabular}{|c|}
\cline{1-1}
\multicolumn{1}{|c|}{$\quad$}  \\
\cline{1-1}
\end{tabular}_{\begin{tabular}{|c|} \multicolumn{1}{c}{$S$}\end{tabular}}
\otimes
\begin{tabular}{|c|}
\cline{1-1}
\multicolumn{1}{|c|}{$\quad$}  \\
\cline{1-1}
\end{tabular}_{\begin{tabular}{|c|} \multicolumn{1}{c}{$S$}\end{tabular}}
\otimes
\begin{tabular}{|c|}
\cline{1-1}
\multicolumn{1}{|c|}{$\quad$}  \\
\cline{1-1}
\end{tabular}_{\begin{tabular}{|c|} \multicolumn{1}{c}{$S$}\end{tabular}} \nonumber\\
\to
&\begin{tabular}{|c|c|c|c|c|}
\cline{1-5}
$\quad$ &  $\quad$& $\quad$   &  $\quad$ &$\quad$    \\
\cline{1-5}
\end{tabular}_{\begin{tabular}{|c|} \multicolumn{1}{c}{$S$}\end{tabular}}
\oplus
\begin{tabular}{|c|c|c|c|}
\cline{1-4}
$\quad$ &  $\quad$& $\quad$& $\quad$   \\
\cline{1-4}
\multicolumn{1}{|c|}{$\quad$} \\
\cline{1-1}
\end{tabular}_{\begin{tabular}{|c|} \multicolumn{1}{c}{$S$}\end{tabular}}
\oplus
\begin{tabular}{|c|c|c|}
\cline{1-3}
$\quad$ & $\quad$  & $\quad$   \\
\cline{1-3}
$\quad$ & $\quad$   \\
\cline{1-2}
\end{tabular}_{\begin{tabular}{|c|} \multicolumn{1}{c}{$S$}\end{tabular}}.
\end{align}

For the spin $J=5/2,3/2$ and $1/2$ of the pentaquarks, the configurations
can be similarly represented in terms of Young tableau [5],[4,1], and [3,2]
of one, four, and five dimensions. For the pentaquark with $J=5/2$, we write
the spin wavefunction as,\renewcommand{\tabcolsep}{0.1cm} %
\renewcommand{\arraystretch}{1}
\begin{equation}
\begin{tabular}{|c|c|c|c|c|}
\cline{1-5}
$1$ & $2$ & $3$ & $4$ & $5$ \\
\cline{1-5}
\end{tabular}%
_{%
\begin{tabular}{c}
$\chi _{1}^{P}$%
\end{tabular}%
}\text{.}  \label{ch5}
\end{equation}%
In the case of the $J=3/2$, the spin wavefunctions become %
\renewcommand{\tabcolsep}{0.1cm} \renewcommand{\arraystretch}{1}
\begin{equation}
\begin{aligned} \begin{tabular}{|c|c|c|c|} \cline{1-4} 1 & 2 & 3 & 4 \\
\cline{1-4} \multicolumn{1}{|c|}{5} \\ \cline{1-1}
\end{tabular}_{\begin{tabular}{|c|}
\multicolumn{1}{c}{$\chi_{2}^{P}$}\end{tabular}}, \begin{tabular}{|c|c|c|c|}
\cline{1-4} 1 & 2 & 3 & 5 \\ \cline{1-4} \multicolumn{1}{|c|}{4} \\ \cline{1-1}
\end{tabular}_{\begin{tabular}{|c|}
\multicolumn{1}{c}{$\chi_{3}^{P}$}\end{tabular}}, \\
\begin{tabular}{|c|c|c|c|} \cline{1-4} 1 & 2 & 4 & 5 \\ \cline{1-4}
\multicolumn{1}{|c|}{3} \\ \cline{1-1} \end{tabular}_{\begin{tabular}{|c|}
\multicolumn{1}{c}{$\chi_{4}^{P}$}\end{tabular}}, \begin{tabular}{|c|c|c|c|}
\cline{1-4} 1 & 3 & 4 & 5 \\ \cline{1-4} \multicolumn{1}{|c|}{2} \\ \cline{1-1}
\end{tabular}_{\begin{tabular}{|c|}
\multicolumn{1}{c}{$\chi_{5}^{P}$}\end{tabular}}, \end{aligned}  \label{ch3}
\end{equation}%
while for $J=1/2$, they become, \renewcommand{\tabcolsep}{0.1cm} %
\renewcommand{\arraystretch}{1}
\begin{equation}
\begin{aligned} \begin{tabular}{|c|c|c|} \cline{1-3} 1 & 2 & 3 \\ \cline{1-3} 4 &
5 \\ \cline{1-2} \end{tabular}_{\begin{tabular}{|c|}
\multicolumn{1}{c}{$\chi_{6}^{P}$}\end{tabular}}, \begin{tabular}{|c|c|c|}
\cline{1-3} 1 & 2 & 4 \\ \cline{1-3} 3 & 5 \\ \cline{1-2}
\end{tabular}_{\begin{tabular}{|c|}
\multicolumn{1}{c}{$\chi_{7}^{P}$}\end{tabular}}, \begin{tabular}{|c|c|c|}
\cline{1-3} 1 & 3 & 4 \\ \cline{1-3} 2 & 5 \\ \cline{1-2}
\end{tabular}_{\begin{tabular}{|c|}
\multicolumn{1}{c}{$\chi_{8}^{P}$}\end{tabular}}, \\
\begin{tabular}{|c|c|c|} \cline{1-3} 1 & 2 & 5 \\ \cline{1-3} 3 & 4 \\
\cline{1-2} \end{tabular}_{\begin{tabular}{|c|}
\multicolumn{1}{c}{$\chi_{9}^{P}$}\end{tabular}}, \begin{tabular}{|c|c|c|}
\cline{1-3} 1 & 3 & 5 \\ \cline{1-3} 2 & 4 \\ \cline{1-2}
\end{tabular}_{\begin{tabular}{|c|}
\multicolumn{1}{c}{$\chi_{10}^{P}$}\end{tabular}}. \end{aligned}  \label{ch2}
\end{equation}

We see from the above analysis that to describe fully heavy pentaquarks $%
P_{Q_{1}Q_{2}Q_{3}Q_{{4}}\bar{Q_{5}}}$ with one antiquark $\bar{Q}_{5}$ having a
given flavor and color, the antiquark can be temporarily omitted from the
spin degrees of freedom. Thus, we shall identify the spin states shown in
Eqs.~(\ref{ch5}) through (\ref{ch2}) via the Young-Yamanouchi bases
associated with the Young tableau [4], [3,1], and [2,2]. Given that the
color and spin states represented in terms of the Young tableau respect
certain symmetry, one can construct the combined color-spin state of a
pentaquark which is fully antisymmetric under the exchange of any pair among
the heavy quarks $1=Q_{1}$, $2=Q_{2}$, $3=Q_{3}$, and $4=Q_{4}$(for short).

We start from the color singlets in Eq.~(\ref{three}), and combine them with
spin states by the outer product of the permutation group, $S_{4}$,
resulting in the color $\otimes $ spin states for the quarks $1$, $2$, $3$
and $4$. After isolating spin of the antiquark, one can deduce the outer
product between Young tableaus [2,1,1] of the color singlets and that of the
spin states[4], [3,1], [2,2]. For the state expression in terms of the first
four quarks ($1234$) mentioned above, one can obtain the Young tableau reps.
of color $\otimes $ spin states and write them as that in Ref.~\cite%
{An:2021vwi} (see Eq.~(17))

According to the method described in Ref.~\cite{Stancu:1999qr}, all the
possible Young-Yamanouchi bases (i.e. $\psi =\phi _{2}^{P}\chi _{1}^{P}$) for the
pentaquarks we shall address in this work can be obtained from the couplings(%
$\otimes $) of the color and spin degrees of freedom. Collectively, we
express these color-spin coupled bases, again, in terms of the Young
tableau: \renewcommand{\tabcolsep}{0.1cm} \renewcommand{\arraystretch}{1}
\begin{equation}
\begin{aligned} \begin{tabular}{|c|} \cline{1-1} 1 \\ \cline{1-1} 2 \\
\cline{1-1} 3 \\ \cline{1-1} 4 \\ \cline{1-1} \end{tabular}:\
\psi_{1}^{\prime},\psi_{1}, \quad \begin{tabular}{|c|c|} \cline{1-2} 1 & 4 \\
\cline{1-2} 2 \\ \cline{1-1} 3 \\ \cline{1-1} \end{tabular}:\
\psi_{2}^{\ast},\psi_{2}^{\prime},\psi_{5}^{\prime},\psi_{2},\psi_{5}, \\
\begin{tabular}{|c|c|} \cline{1-2} 1 & 3 \\ \cline{1-2} 2 & 4 \\ \cline{1-2}
\end{tabular}:\ \psi_{4}^{\prime},\psi_{4}, \quad \begin{tabular}{|c|c|}
\cline{1-2} 1 & 3 \\ \cline{1-2} 2 \\ \cline{1-1} 4 \\ \cline{1-1}
\end{tabular}:\
\psi_{1}^{\ast},\psi_{3}^{\prime},\psi_{6}^{\prime},\psi_{3},\psi_{6}.
\end{aligned}  \label{cswYT}
\end{equation}%
in which (for $J=5/2$)
\begin{equation}
\psi _{1}^{\ast }=\phi _{2}^{P}\chi _{1}^{P},\quad \psi _{2}^{\ast }=\phi
_{3}^{P}\chi _{1}^{P}\text{, }
\end{equation}
and (for $J=3/2$)
\begin{align}
& \psi _{1}^{\prime }=\frac{1}{\sqrt{3}}\phi _{1}^{P}\chi _{5}^{P}-\frac{1}{%
\sqrt{3}}\phi _{2}^{P}\chi _{4}^{P}+\frac{1}{\sqrt{3}}\phi _{3}^{P}\chi
_{3}^{P},  \notag \\
& \psi _{2}^{\prime }=-\frac{1}{\sqrt{6}}\phi _{1}^{P}\chi _{5}^{P}+\frac{1}{%
\sqrt{6}}\phi _{2}^{P}\chi _{4}^{P}+\sqrt{\frac{2}{3}}\phi _{3}^{P}\chi
_{3}^{P},  \notag \\
& \psi _{3}^{\prime }=\frac{1}{\sqrt{3}}\phi _{1}^{P}\chi _{5}^{P}-\frac{1}{%
\sqrt{6}}\phi _{2}^{P}\chi _{3}^{P}+\frac{1}{\sqrt{3}}\phi _{2}^{P}\chi
_{4}^{P}+\frac{1}{\sqrt{6}}\phi _{3}^{P}\chi _{4}^{P},  \notag \\
& \psi _{4}^{\prime }=-\frac{1}{\sqrt{6}}\phi _{1}^{P}\chi _{5}^{P}-\frac{1}{%
\sqrt{3}}\phi _{2}^{P}\chi _{3}^{P}-\frac{1}{\sqrt{6}}\phi _{2}^{P}\chi
_{4}^{P}+\frac{1}{\sqrt{3}}\phi _{3}^{P}\chi _{4}^{P},  \notag \\[2mm]
& \psi _{5}^{\prime }=\phi _{3}^{P}\chi _{2}^{P},  \notag \\[2mm]
& \psi _{6}^{\prime }=\phi _{2}^{P}\chi _{2}^{P}.
\end{align}%
In the case of $J=1/2$ , the color $\otimes $ spin bases in Eq.~(\ref{cswYT})
are explicitly,
\begin{align}
& \psi _{1}=\frac{1}{\sqrt{3}}\phi _{1}^{P}\chi _{8}^{P}-\frac{1}{\sqrt{3}}%
\phi _{2}^{P}\chi _{7}^{P}+\frac{1}{\sqrt{3}}\phi _{3}^{P}\chi _{6}^{P},
\notag \\
& \psi _{2}=-\frac{1}{\sqrt{6}}\phi _{1}^{P}\chi _{8}^{P}+\frac{1}{\sqrt{6}}%
\phi _{2}^{P}\chi _{7}^{P}+\sqrt{\frac{2}{3}}\phi _{3}^{P}\chi _{6}^{P},
\notag \\
& \psi _{3}=\frac{1}{\sqrt{3}}\phi _{1}^{P}\chi _{8}^{P}-\frac{1}{\sqrt{6}}%
\phi _{2}^{P}\chi _{6}^{P}+\frac{1}{\sqrt{3}}\phi _{2}^{P}\chi _{7}^{P}+%
\frac{1}{\sqrt{6}}\phi _{3}^{P}\chi _{7}^{P},  \notag \\
& \psi _{4}=-\frac{1}{\sqrt{6}}\phi _{1}^{P}\chi _{8}^{P}-\frac{1}{\sqrt{3}}%
\phi _{2}^{P}\chi _{6}^{P}-\frac{1}{\sqrt{6}}\phi _{2}^{P}\chi _{7}^{P}+%
\frac{1}{\sqrt{3}}\phi _{3}^{P}\chi _{7}^{P},  \notag \\
& \psi _{5}=\frac{1}{\sqrt{2}}\phi _{1}^{P}\chi _{10}^{P}-\frac{1}{\sqrt{2}}%
\phi _{2}^{P}\chi _{9}^{P},  \notag \\
& \psi _{6}=\frac{1}{2}\phi _{1}^{P}\chi _{10}^{P}+\frac{1}{2}\phi
_{2}^{P}\chi _{9}^{P}-\frac{1}{\sqrt{2}}\phi _{3}^{P}\chi _{9}^{P}.
\end{align}

Based on the Young tableau in Eq.~(\ref{cswYT}) and Pauli principle for each
pair of two fermions, we can construct all allowed color-spin wavefunctions and
list them in Table~\ref{tab:basis} for pentaquarks consisting of four heavy
quarks and a heavy antiquark. Based on similar principle, one can
construct the states involved in chromomagnetic interactions (CMI) (\ref{CMI})
with the help of color and spin factors (\ref{colorfc},\ref{spinfc}) via the color bases
(\ref{three}) and spin bases (\ref{ch5}-\ref{ch2}), respectively. Following Ref.~%
\cite{Zhang:2021yul}, we shall use the scaling ratios of color factors to
evaluate the matrices of the binding energy, as discussed in section \ref{sec:bagmodel}.

\renewcommand{\tabcolsep}{0.20cm} \renewcommand{\arraystretch}{1.5}
\begin{table}[!htb]
\caption{Color-spin bases of pentaquarks $q_{1}q_{2}q_{3}q_{4}\bar{q}_{5}$
with $J^{P}$ quantum number.}
\label{tab:basis}%
\begin{tabular}{lcc}
\bottomrule[1.2pt]\bottomrule[0.5pt] $q_{1}q_{2}q_{3}q_{4}$ & $J^{P}$ &
Color-spin bases \\ \hline
$bbbb$, $cccc$ & ${3/2}^{-}$ & $\psi_{1}^{\prime}$ \\
& ${1/2}^{-}$ & $\psi_{1}$ \\
$bbbc$, $cccb$ & ${5/2}^{-}$ & $\psi_{2}^{\ast}$ \\
& ${3/2}^{-}$ & $\psi_{1}^{\prime}$, $\psi_{2}^{\prime}$, $\psi_{5}^{\prime}$
\\
& ${1/2}^{-}$ & $\psi_{1}$, $\psi_{2}$, $\psi_{5}$ \\
$ccbb$ & ${5/2}^{-}$ & $\sqrt{\frac{2}{3}}\psi_{1}^{\ast}-\sqrt{\frac{1}{3}}%
\psi_{2}^{\ast}$ \\
& ${3/2}^{-}$ & $\psi_{1}^{\prime}$, $\sqrt{\frac{2}{3}}\psi_{3}^{\prime}-%
\sqrt{\frac{1}{3}}\psi_{2}^{\prime}$, $\psi_{4}^{\prime}$, $\sqrt{\frac{2}{3}%
}\psi_{6}^{\prime}-\sqrt{\frac{1}{3}}\psi_{5}^{\prime}$ \\
& ${1/2}^{-}$ & $\psi_{1}$, $\sqrt{\frac{2}{3}}\psi_{3}-\sqrt{\frac{1}{3}}%
\psi_{2}$, $\psi_{4}$, $\sqrt{\frac{2}{3}}\psi_{6}-\sqrt{\frac{1}{3}}%
\psi_{5} $ \\
\bottomrule[0.5pt]\bottomrule[1.2pt]
\end{tabular}%
\end{table}

\section{MIT bag model and CMI}

\label{sec:bagmodel}

MIT bag model describes hadron as a sphere bag containing confined valence
quarks. The model also includes perturbative inter-quark interactions via
considering the lowest-order gluon exchange among quarks, which is referred
as chromomagnetic interaction \cite{DeGrand:1975cf,Johnson:1975zp}.
Mass of hadrons for a bag of radius $R$ is \cite%
{DeGrand:1975cf,Johnson:1975zp},
\begin{equation}
M\left( R\right) =\sum_{i}\omega _{i}+\frac{4}{3}\pi R^{3}B-\frac{Z_{0}}{R}%
+M_{BD}+M_{CMI},  \label{MBm}
\end{equation}%
\begin{equation}
\omega _{i}=\left( m_{i}^{2}+\frac{x_{i}^{2}}{R^{2}}\right) ^{1/2},
\label{omega}
\end{equation}%
where the first term is total sum of the kinematic energy of relativistic
quark $i$ with mass $m_{i}$, the second is bag volume energy with $B$ the
bag constant, the third is zero point energy with coefficient $Z_{0}$, the
forth is binding energy ($M_{BD}$) between heavy quarks or between heavy and
strange quarks \cite{Karliner:2014gca,Karliner:2017elp}, and the fifth is
chromomagnetic interaction. In Eq.~(\ref{MBm}), the bag radius $R$ is bag
radius to be determined variationally, and the quark momentum $x_{i}$ in
unit of $R^{-1}$ satisfies a boundary condition on bag surface,
\begin{equation}
\tan x_{i}=\frac{x_{i}}{1-m_{i}R-\left( m_{i}^{2}R^{2}+x_{i}^{2}\right)
^{1/2}}\text{,}  \label{transc}
\end{equation}%
which is to be solved iteratively in this work.

The interaction in this work enters among quarks via two parts, $\mathrm{\Delta}%
M=M_{BD}+M_{CMI}$. The first part, the binding energy $M_{BD}=%
\sum_{M<N}B_{MN}$ ($M,N=s$, $c$ and $b$), proposed in Refs.~\cite%
{Karliner:2014gca,Karliner:2017elp}, rises mainly from short-range
chromoelectric interactions between heavy quarks or within heavy-strange
pairs as they are (relatively) massive and move nonrelativistically. This
part was supported by our previous work \cite{Zhang:2021yul}, which reconciles
bag dynamics of the light and heavy hadrons. Five binding energies $%
B_{cs}$, $B_{cc}$, $B_{bs}$, $B_{bc}$ and $B_{bb}$ in the color $\bar{3}_{c}$
rep., extracted from the bag mass corrections to the baryons when heavy
pairs ($cs,cc,bs,bc$ and $bb$) involved, can be scaled to the binding
energies $B_{MN}$ in $6_{c}$ reps. via scaling factor ratios between two
reps. \cite{Karliner:2014gca,Zhang:2021yul}. As there are two reps. ($\bar{3}%
_{c}$ and $6_{c}$) for color wavefunctions of quark pairs, as discussed in section \ref%
{sec:WF}, the binding energies could be matrices in color-spin space given
in Table~\ref{tab:basis}, where spin bases are orthogonal.

The second part of interaction in Eq.~(\ref{MBm}) is chromomagnetic
interaction $M_{CMI}$, which can be due to the short-range gluon exchange %
\cite{DeRujula:1975qlm}. Similar to the interaction of magnetic moments of
the quark spins, this part has the form of
\begin{equation}
M_{CMI}=-\sum_{i<j}\left( \mathbf{\lambda }_{i}\cdot \mathbf{\lambda }%
_{j}\right) \left( \mathbf{\sigma }_{i}\cdot \mathbf{\sigma }_{j}\right)
C_{ij}\text{,}  \label{CMI}
\end{equation}%
where $i$ and $j$ denote the quark (anti-quark) indices, $\lambda $ the
Gell-Mann matrices, $\sigma $ the Pauli matrices, and $C_{ij}$ the coupling
parameters of the CMI.

To compute the color and spin factor in Eq.~(\ref{CMI}), we employ the
following formula of the matrix element:
\begin{equation}
{\left\langle \mathbf{\lambda }_{i}\cdot \mathbf{\lambda }_{j}\right\rangle }%
_{nm}=\sum_{\alpha =1}^{8}\mathrm{Tr}\left( c_{in}^{\dagger }\lambda
^{\alpha }c_{im}\right) \mathrm{Tr}\left( c_{jn}^{\dagger }\lambda ^{\alpha
}c_{jm}\right) \text{,}  \label{colorfc}
\end{equation}%
\begin{equation}
{\left\langle \mathbf{\sigma }_{i}\cdot \mathbf{\sigma }_{j}\right\rangle }%
_{xy}=\sum_{\alpha =1}^{3}\mathrm{Tr}\left( \chi _{ix}^{\dagger }\sigma
^{\alpha }\chi _{iy}\right) \mathrm{Tr}\left( \chi _{jx}^{\dagger }\sigma
^{\alpha }\chi _{jy}\right) \text{,}  \label{spinfc}
\end{equation}%
where $n$, $m$ and $x$, $y$ indicate components of the basis vectors of
color and spin wavefunctions of a hadron, respectively. The symbols $c$ and $%
\chi $ denote bases of color and spin vectors associated with a quark,
respectively. The matrices of two factors can be calculated with the help of
Eqs.~(\ref{colorfc})-(\ref{spinfc}) once color-spin wavefunctions $\psi
=\sum\phi ^{P}\chi ^{P}$ were determined.

In the MIT bag model, the CMI parameter $C_{ij}$ is known analytically,
given by \cite{DeGrand:1975cf}
\begin{equation}
C_{ij}=3\frac{\alpha _{s}\left( R\right) }{R^{3}}\bar{\mu}_{i}\bar{\mu}%
_{j}I_{ij}\text{,}  \label{Cij}
\end{equation}%
where the reduced magnetic moment $\bar{\mu}_{i}$ and the (running) strong
coupling $\alpha _{s}$ are given by
\begin{equation}
\bar{\mu}_{i}=\frac{R}{6}\frac{4\alpha _{i}+2\lambda _{i}-3}{2\alpha
_{i}\left( \alpha _{i}-1\right) +\lambda _{i}}\text{,}  \label{muBari}
\end{equation}%
\begin{equation}
\alpha _{s}(R)=\frac{0.296}{ln\left[ 1+{\left( 0.281R\right) }^{-1}\right] }%
\text{,}  \label{alphaS-mine}
\end{equation}%
and
\begin{equation}
I_{ij}=1+2\int_{0}^{R}\frac{dr}{r^{4}}\bar{\mu}_{i}\bar{\mu}%
_{j}=1+F(x_{i},x_{j})\text{.}  \label{Iij}
\end{equation}%
Here, $\alpha _{i}=\omega _{i}R$ and $\lambda _{i}=m_{i}R$ \cite%
{DeGrand:1975cf}.
The function of $\alpha _{s}(R)$ in Eq.~(\ref%
{alphaS-mine}) is discussed in Ref.~\cite{Zhang:2021yul}. The
function $F(x_{i},x_{j})$ in Eq.~(\ref{Iij}) is rational in terms of
parameters $x_{i}$ and $x_{j}$,
\begin{equation}
\begin{aligned} F(x_{i}, x_{j})={\left(x_{i}
\mathrm{sin}^{2}x_{i}-\frac{3}{2}y_{i}\right)}^{-1} {\left(x_{j}
\mathrm{sin}^{2}x_{j}-\frac{3}{2}y_{j}\right)}^{-1} \\
\left\{-\frac{3}{2}y_{i}y_{j}-2x_{i}x_{j} \mathrm{sin}^{2}x_{i}
\mathrm{sin}^{2}x_{j} +\frac{1}{2}x_{i}x_{j}\left[2x_{i}
\mathrm{Si}(2x_{i})\right.\right.\\ \left.\left. +2x_{j} \mathrm{Si}(2x_{j})
-(x_{i}+x_{j}) \mathrm{Si}(2(x_{i}+x_{j})) \right.\right.\\ \left.\left.
-(x_{i}-x_{j}) \mathrm{Si}(2(x_{i}-x_{j})) \right] \right\} \text{,}
\end{aligned}  \label{F}
\end{equation}%
where $y_{i}=x_{i}-\mathrm{sin}(x_{i})\mathrm{cos}(x_{i})$ and
\begin{equation}
\mathrm{Si}(x)=\int_{0}^{x}\ \frac{\mathrm{sin}(t)}{t}\mathrm{d}t\text{.}
\end{equation}

From Eq.~(\ref{muBari}), one can write the magnetic moment $\mu _{i}$ of the
quark (anti-quark) $i$ as,  in bag model,
\begin{equation}
\mu _{i}=Q_{i}\bar{\mu}_{i}=Q_{i}\frac{R}{6}\frac{4\alpha _{i}+2\lambda
_{i}-3}{2\alpha _{i}\left( \alpha _{i}-1\right) +\lambda _{i}}\text{,}
\label{mui}
\end{equation}%
with electric charge $Q_{i}$. As a result, the magnetic moment of a hadron
with color-spin wavefunction $\psi $ is,
\begin{equation}
\mu =\left\langle \psi \left\vert \sum\nolimits_{i}g_{i}\mu
_{i}S_{iz}\right\vert \psi \right\rangle \text{,}  \label{musum}
\end{equation}%
where $g_{i}=2$, and $S_{iz}$ is the third component of spin of the
individual quark $i$ \cite{Wang:2016dzu}.

We shall calculate the magnetic moments of the pentaquarks in unit of
the magnetic moment of the proton $\mu _{p}$ and further transform the obtained
results into that in unit of $\mu _{N}$ with the help of measured proton magnetic moment $\mu
_{p}=2.79285\mu _{N}$ \cite{ParticleDataGroup:2022pth,Tiesinga:2021myr}.
Note that Eq.~(\ref{musum}) also holds true in the case of chromomagnetic
mixing, for which one can expand $\psi $ in terms of the color-spin bases to
find magnetic moment by Eq.~(\ref{musum}). For the spin wavefunctions in
section \ref{sec:WF}, one can derive each diagonal element of magnetic moment. The
results are listed in Table~\ref{tab:mu} in detail.

\renewcommand{\tabcolsep}{0.8cm} \renewcommand{\arraystretch}{1.5}
\begin{table}[!htb]
\caption{Sum rule of magnetic moments for pentaquarks $q_{1}q_{2}q_{3}q_{4}{%
\bar{q}}_{5}$.}
\label{tab:mu}%
\begin{tabular}{cc}
\bottomrule[1.2pt]\bottomrule[0.5pt] Spin basis & $\mu$ \\ \hline
$\chi_{1}^{P}$ & $\mu_{1}+\mu_{2}+\mu_{3}+\mu_{4}+\mu_{5}$ \\
$\chi_{2}^{P}$ & $\frac{9}{10}\left(\mu_{1}+\mu_{2}+\mu_{3}+\mu_{4}\right)-%
\frac{3}{5}\mu_{5}$ \\
$\chi_{3}^{P}$ & $\frac{5}{6}\left(\mu_{1}+\mu_{2}+\mu_{3}\right)-\frac{1}{2}%
\mu_{4}+\mu_{5}$ \\
$\chi_{4}^{P}$ & $\frac{2}{3}\left(\mu_{1}+\mu_{2}\right)-\frac{1}{3}\mu_{3}+%
\frac{5}{6}\left(\mu_{4}+\mu_{5}\right)$ \\
$\chi_{5}^{P}$ & $\mu_{3}+\mu_{4}+\mu_{5}$ \\
$\chi_{6}^{P}$ & $\frac{5}{9}\left(\mu_{1}+\mu_{2}+\mu_{3}\right)-\frac{1}{3}%
\left(\mu_{4}+\mu_{5}\right)$ \\
$\chi_{7}^{P}$ & $\frac{4}{9}\left(\mu_{1}+\mu_{2}\right)-\frac{2}{9}\mu_{3}+%
\frac{2}{3}\mu_{4}-\frac{1}{3}\mu_{5}$ \\
$\chi_{8}^{P}$ & $\frac{2}{3}\left(\mu_{3}+\mu_{4}\right)-\frac{1}{3}\mu_{5}$
\\
$\chi_{9}^{P}$ & $\mu_{5}$ \\
$\chi_{10}^{P}$ & $\mu_{5}$ \\
\bottomrule[0.5pt]\bottomrule[1.2pt]
\end{tabular}%
\end{table}

For the model parameters, we use the values in our
previous work \cite{Zhang:2021yul} which reconcile dynamics of the light and heavy
hadrons. These values are $Z_{0}$, the constant $B$ and
masses of the quarks (the nonstrange quarks $n=u,d$, the strange quark $s$, the charm quark $c$
and the bottom quark $b$) in this work,
\begin{equation}
\begin{Bmatrix}
Z_{0}=1.83, & B^{1/4}=0.145\,\text{GeV,} \\
m_{n}=0\,\text{GeV,} & m_{s}=0.279\,\text{GeV,} \\
m_{c}=1.641\,\text{GeV,} & m_{b}=5.093\,\text{GeV.}%
\end{Bmatrix}
\label{oripar}
\end{equation}

Meanwhile, using the binding energies $B_{QQ^{\prime }}$ for the quark pair $%
QQ^{\prime }$ in color antitriplet rep. \cite{Zhang:2021yul},
\begin{equation}
\bar{3}_{c}:%
\begin{Bmatrix}
B_{cs}=-0.025\,\text{GeV,} & B_{cc}=-0.077\,\text{GeV,} \\
B_{bs}=-0.032\,\text{GeV,} & B_{bb}=-0.128\,\text{GeV,} \\
B_{bc}=-0.101\,\text{GeV,} &
\end{Bmatrix}
\label{Binds}
\end{equation}%
one can obtain the corresponding values of the binding energies for color-$%
\boldsymbol{6}_{c}$ rep., via multiplying the color-factor ratios (scaling factors).
The results of the binding energies between quark pairs (denoted in subscripts) are:
\begin{equation}
6_{c}:%
\begin{Bmatrix}
B_{cs}=0.013\,\text{GeV,} & B_{cc}=0.039\,\text{GeV,} \\
B_{bs}=0.016\,\text{GeV,} & B_{bb}=0.064\,\text{GeV,} \\
B_{bc}=0.051\,\text{GeV. } &
\end{Bmatrix}
\label{bind2}
\end{equation}

Given the parameter inputs in Eqs.~(\ref{oripar}), (\ref{Binds}) and (\ref{bind2}%
), one can apply variational method to Eq.~(\ref{MBm}) to determine the bag
radius $R$ and the respective momentum $x_{i}$ via Eq.~(\ref{transc}) for a
given hadron with color-spin wavefunction $\psi $, given in
Table~\ref{tab:basis}. It is then straightforward to calculate masses and magnetic moments of
fully heavy pentaquarks we address in this work, as detailed in section \ref{sec:hadrons}.

\section{Masses and magnetic moments of fully heavy hadrons}

\label{sec:hadrons}

\renewcommand{\tabcolsep}{0.52cm} \renewcommand{\arraystretch}{1.2}
\begin{table*}[tbp]
\caption{Computed masses (MeV) of triply heavy baryons, compared to other
works cited. Bag radius $R_{0}$ is in GeV$^{-1}$. }%
\begin{tabular}{ccccccc}
\bottomrule[1.2pt]\bottomrule[0.5pt] \textrm{Ref} & $\Omega_{ccc}$ & $%
\Omega_{ccb}$ & $\Omega_{ccb}^{\ast}$ & $\Omega_{cbb}$ & $%
\Omega_{cbb}^{\ast} $ & $\Omega_{bbb}$ \\ \hline
$R_{0}$ & 4.22 & 3.75 & 3.83 & 3.18 & 3.31 & 2.59 \\
$M$ & 4841 & 8112 & 8133 & 11373 & 11402 & 14626 \\
\cite{Zhang:2009re} & 4670$\pm$150 & 7410$\pm$130 & 7450$\pm$160 & 10300$\pm$%
100 & 10540$\pm$110 & 13280$\pm$100 \\
\cite{Aliev:2014lxa} & 4720$\pm$120 & - & 8070$\pm$100 & - & 11350$\pm$150 &
14300$\pm$200 \\
\cite{Qin:2019hgk} & 4760 & 7867 & 7963 & 11077 & 11167 & 14370 \\
\cite{Faessler:2006ft} & 4760$\pm$60 & 7980$\pm$70 & 7980$\pm$70 & 11190$\pm$%
80 & 11190$\pm$80 & 14370$\pm$80 \\
\cite{Bernotas:2008bu} & 4777 & 7984 & 8005 & 11139 & 11163 & 14276 \\
\cite{Brown:2014ena} & 4796$\pm$8$\pm$18 & 8007$\pm$9$\pm$20 & 8037$\pm$9$%
\pm $20 & 11195$\pm$8$\pm$20 & 11229$\pm$8$\pm$20 & 14366$\pm$9$\pm$20 \\
\cite{Yang:2019lsg} & 4798 & 8004 & 8023 & 11200 & 11221 & 14396 \\
\cite{Flynn:2011gf} & 4799 & 8018 & 8046 & 11214 & 11245 & 14398 \\
\cite{Martynenko:2007je} & 4803 & 8018 & 8025 & 11280 & 11287 & 14569 \\
\cite{Wang:2020avt} & 4810$\pm$100 & 8020$\pm$80 & 8030$\pm$80 & 11220$\pm$80
& 11230$\pm$80 & 14430$\pm$90 \\
\cite{Wei:2015gsa} & 4834 & - & - & - & - & - \\
\cite{Wei:2016jyk} & - & - & - & - & - & 14788 \\
\cite{Patel:2008mv} & 4897 & 8262 & 8273 & 11546 & 11589 & 14688 \\
\cite{Llanes-Estrada:2013rwa} & 4900$\pm$250 & 8150$\pm$300 & - & 11400$\pm$%
300 & - & 14700$\pm$300 \\
\cite{Gutierrez-Guerrero:2019uwa} & 4930 & 8010 & 8030 & 11090 & 11120 &
14230 \\
\cite{Roberts:2007ni} & 4965 & 8245 & 8265 & 11535 & 11554 & 14834 \\
\cite{Wang:2011ae} & 4990$\pm$140 & 8230$\pm$130 & 8230$\pm$130 & 11500$\pm$%
110 & 11490$\pm$110 & 14830$\pm$100 \\
\cite{Yin:2019bxe} & 5000 & 8190 & - & - & - & 14570 \\
\bottomrule[0.5pt]\bottomrule[1.2pt]
\end{tabular}%
\label{tab:3heavybaryon}
\end{table*}

Before discussing fully heavy pentaquarks, we first consider triply (fully) heavy
baryons, the fully heavy binding systems of three quarks, which include
the five baryons $\Omega _{ccc}$, $\Omega _{ccb}$, $\Omega _{ccb}^{\ast }$, $%
\Omega _{cbb}$, $\Omega _{cbb}^{\ast }$ and $\Omega _{bbb}$,
as listed in Table~\ref{tab:3heavybaryon}. For these baryons, we apply the
same framework of MIT bag model (\ref{MBm}) with the inputs in Eqs.~(\ref%
{oripar}), (\ref{Binds}) and (\ref{bind2}) to compute the mass $M$ and the
bag radii $R_{0}$ of the fully heavy baryons, with other works cited for
comparison.

Given the color-spin bases \cite{Zhang:2021yul} (see Eq.~(15)), defined
as $\phi ^{B}\chi _{1}^{B}$ and $\phi ^{B}\chi _{2}^{B}$ for $J^{P}=3/2^{+}$
and $1/2^{+}$, respectively, the CMI matrices and binding energies are well
determined. When pentaquarks included, the computed results could help us
further understand the general features of fully heavy systems, as we shall
discuss in Sect.~\ref{sec:linearity}.

In Table~\ref{tab:3heavybaryon}, the second and third rows present bag radii
$R_{0}$ and masses of triplet heavy baryons, and the data followed rows present the
masses (in unit of MeV) cited from various works. One sees that the predicted masses of the $\Omega_{ccc}$,
for instance,  ranges from $4520\,$MeV to $5130\,$MeV for which various methods including QCD sum rules \cite%
{Zhang:2009re,Aliev:2014lxa,Wang:2020avt,Wang:2011ae} are employed. Up to $%
610\,$MeV, our prediction is close to the prediction $4810\pm 100\,$MeV in
Ref.~\cite{Wang:2020avt}. The lattice QCD was also applied to study the $%
\Omega _{ccc}$ in Ref.~\cite{Brown:2014ena}, with the prediction of $4796\pm
8\pm 18\,$MeV. Comparing with the bag model computation of the doubly heavy
baryons and tetraquarks \cite{Zhang:2021yul} and the fully charm tetraquarks $%
cc\bar{c}\bar{c}$ in Ref.~\cite{Yan:2023lvm}, our results for fully heavy
baryons and pentaquarks are reasonable, considering that the inputs of the
parameters are fitted only from light baryons and heavy mesons.

\renewcommand{\tabcolsep}{0.31cm} \renewcommand{\arraystretch}{1.2}
\begin{table*}[!htb]
\caption{Predicted spectra of pentaquarks $P_{bbbb\bar{b}}$ and $P_{cccc\bar{%
c}}$ with comparisons from various works. Bag radius $R_{0}$ is in GeV$^{-1}$%
. Masses are in GeV. Magnetic moments $\protect\mu$ are in unit of $\protect%
\mu_N$.}
\label{tab:QQQQQ}%
\begin{tabular}{ccccccccccc}
\bottomrule[1.2pt]\bottomrule[0.5pt] \textrm{State} & $J^{P}$ & $R_{0}$ & $M$
& $M$\cite{An:2020jix} & $M$\cite{An:2022fvs} & $M$\cite{Yang:2022bfu} & $M$%
\cite{Yan:2021glh} & $M$\cite{Wang:2021xao} & $M$\cite{Zhang:2020vpz} & $\mu$
\\ \hline
$P_{bbbb\bar{b}}$ & ${3/2}^{-}$ & 3.48 & 24.761 & 23.775 & 24.211 & 24.035 &
23.748-23.752 & - & $21.60^{+0.73}_{-0.22}$ & -0.08 \\
& ${1/2}^{-}$ & 3.53 & 24.770 & 23.821 & 24.248 & 24.035 & 23.810-23.814 &
23.91$\pm$0.15 & - & -0.14 \\
$P_{cccc\bar{c}}$ & ${3/2}^{-}$ & 4.99 & 8.229 & 7.864 & 8.145 & 8.095 & - &
- & $7.41^{+0.27}_{-0.31}$ & 0.50 \\
& ${1/2}^{-}$ & 5.08 & 8.262 & 7.949 & 8.193 & 8.045 & 7.892-7.893 & 7.93$%
\pm $0.15 & - & 0.83 \\
\bottomrule[0.5pt]\bottomrule[1.2pt]
\end{tabular}%
\end{table*}

Using the method described in sections \ref{sec:WF} and \ref{sec:bagmodel},
one can construct a total mass formula (\ref{MBm}) for the fully heavy hadrons
containing the CMI and binding energies and calculate masses, magnetic
moments and eigenvectors for all fully heavy systems containing the
fully heavy mesons, baryons and pentaquarks, especially of the
ten pentaquarks $P_{bbbb\bar{b}}$, $P_{bbbb\bar{c}}$, $P_{cccc\bar{b}}$, $%
P_{cccc\bar{c}}$, $P_{bbbc\bar{b}}$, $P_{bbbc\bar{c}}$, $P_{cccb\bar{b}}$, $%
P_{cccb\bar{c}}$, $P_{ccbb\bar{b}}$ and $P_{ccbb\bar{c}}$, whose color-spin
bases have been listed in Table~\ref{tab:basis}. Among these heavy hadrons,
more attention is given to fully bottom and charm systems, which attracts
extensive attention \cite%
{An:2020jix,An:2022fvs,Yang:2022bfu,Yan:2021glh,Wang:2021xao,Zhang:2020vpz}.

In Table~\ref{tab:QQQQQ}, we present our computed results for the masses and magnetic
moments of the fully bottom system $P_{bbbb\bar{b}}$ and the fully charm
system $P_{cccc\bar{c}}$, compared to other works via the CMI model \cite%
{An:2020jix} and constituent quark model \cite{An:2022fvs}, which adopt the
same color-spin wavefunctions, and via the lattice-QCD inspired model \cite%
{Yang:2022bfu} and the chiral quark model \cite{Yan:2021glh}. The comparison is
also given with the predictions via the QCD sum rules \cite%
{Wang:2021xao,Zhang:2020vpz}. Agreement of our predictions with that by the
constituent quark model \cite{An:2022fvs} is achieved. Given the masses of
triply heavy baryons computed in Table~\ref{tab:3heavybaryon} and that of
the measured heavy mesons \cite{ParticleDataGroup:2022pth}, one finds
that our predicted masses of fully heavy pentaquarks are all above the
thresholds of the heavy baryons and mesons, as shown in FIG.~\ref%
{fig:threshold1}.

\begin{figure}[t]
\centering
\subfigure[\
$bbbb\bar{b}$]{\includegraphics[width=0.45\textwidth]{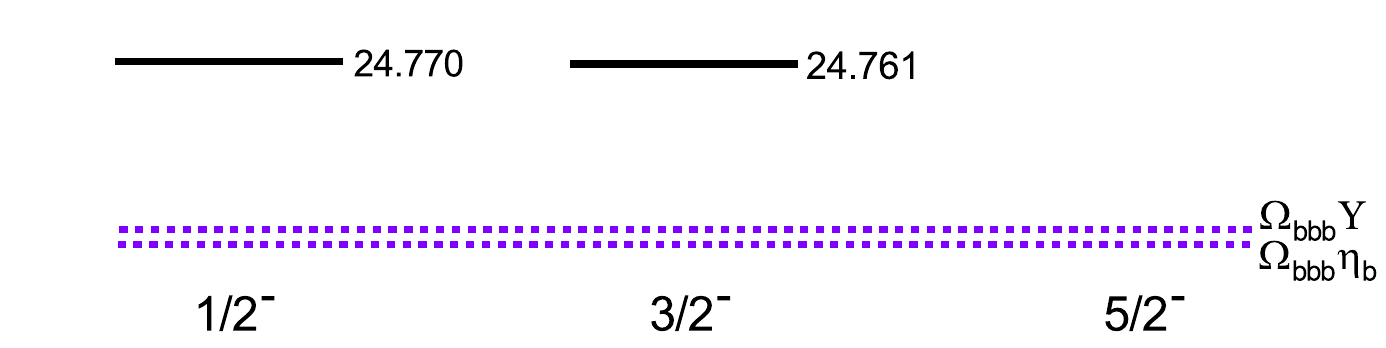}}
\par
\subfigure[\
$cccc\bar{c}$]{\includegraphics[width=0.45\textwidth]{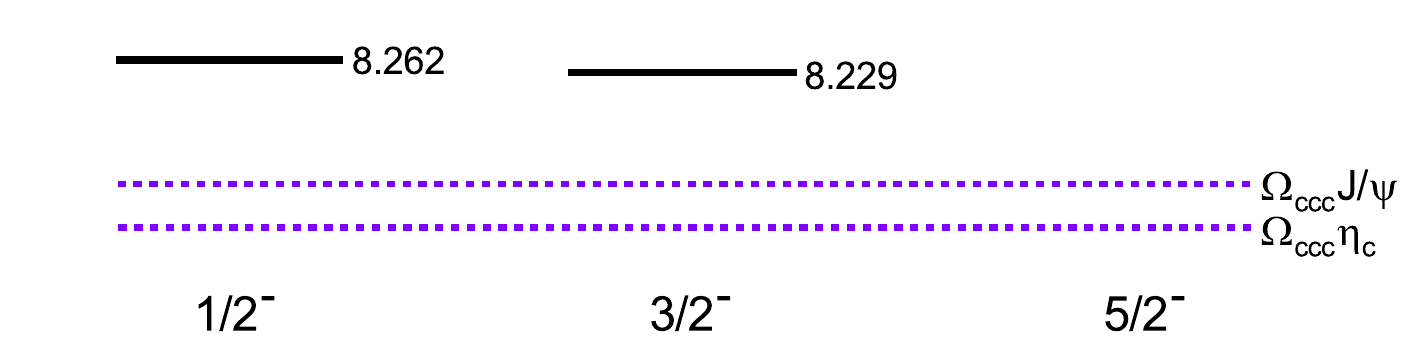}}
\caption{Computed masses(in GeV, short solid lines) of pentaquarks $P_{bbbb%
\bar{b}}$ and $P_{cccc\bar{c}}$ for $J^{P}=5/2^-,3/2^-,1/2^-$, compared to the
respective thresholds plotted as long dotted lines.}
\label{fig:threshold1}
\end{figure}

\renewcommand{\tabcolsep}{0.23cm} \renewcommand{\arraystretch}{1.2}
\begin{table}[!htb]
\caption{Calculated masses $M$ (in GeV) of fully heavy mesons \protect\cite%
{Zhang:2021yul}, tetraquarks \cite{Yan:2023lvm} and pentaquarks, with their mass
splittings $\mathrm{\Delta} M$ (in GeV) due to the lowest two $J^P$ quantum numbers.}
\label{tab:splittings}%
\begin{tabular}{c|ccc|ccc}
\bottomrule[1.2pt]\bottomrule[0.5pt] $J^{P(C)}$ & State & $M$ & $\mathrm{\Delta} M$ &
State & $M$ & $\mathrm{\Delta} M$ \\ \hline
$0^{-+}$ & $\eta_c$ & 3.002 & \multirow{2}{*}{0.095} & $\eta_b$ & 9.396 & %
\multirow{2}{*}{0.064} \\
$1^{--}$ & $J/\psi$ & 3.097 &  & $\Upsilon$ & 9.460 &  \\ \hline
$0^{++}$ & \multirow{3}{*}{$cc\bar{c}\bar{c}$} & 6.469 & %
\multirow{2}{*}{0.050} & \multirow{3}{*}{$bb\bar{b}\bar{b}$} & 19.685 & %
\multirow{2}{*}{0.015} \\
$1^{+-}$ &  & 6.519 & \multirow{2}{*}{0.053} &  & 19.700 & %
\multirow{2}{*}{0.017} \\
$0^{++}$ &  & 6.572 &  &  & 19.717 &  \\ \hline
${1/2}^{-}$ & \multirow{2}{*}{$cccc\bar{c}$} & 8.262 & \multirow{2}{*}{0.033}
& \multirow{2}{*}{$bbbb\bar{b}$} & 24.770 & \multirow{2}{*}{0.009} \\
${3/2}^{-}$ &  & 8.229 &  &  & 24.761 &  \\
\bottomrule[0.5pt]\bottomrule[1.2pt]
\end{tabular}%
\end{table}

\renewcommand{\tabcolsep}{0.22cm} \renewcommand{\arraystretch}{1.2}
\begin{table}[!htb]
\caption{Predicted spectra of pentaquarks $P_{bbbb\bar{c}}$ and $P_{cccc\bar{%
b}}$ with comparisons from various works. Bag radius $R_{0}$ is in GeV$^{-1}$%
. Masses are in GeV. Magnetic moments $\protect\mu$ are in unit of $\protect%
\mu_N$.}
\label{tab:QQQQq}%
\begin{tabular}{ccccccc}
\bottomrule[1.2pt]\bottomrule[0.5pt] \textrm{State} & $J^{P}$ & $R_{0}$ & $M$
& $M$\cite{An:2020jix} & $M$\cite{An:2022fvs} & $\mu$ \\ \hline
$P_{bbbb\bar{c}}$ & ${3/2}^{-}$ & 3.89 & 21.472 & 20.652 & 20.975 & -0.65 \\
& ${1/2}^{-}$ & 3.96 & 21.491 & 20.699 & 21.026 & 0.05 \\
$P_{cccc\bar{b}}$ & ${3/2}^{-}$ & 4.83 & 11.569 & 11.130 & 11.478 & 1.07 \\
& ${1/2}^{-}$ & 4.86 & 11.582 & 11.177 & 11.502 & 0.63 \\
\bottomrule[0.5pt]\bottomrule[1.2pt]
\end{tabular}%
\end{table}

\renewcommand{\tabcolsep}{0.17cm} \renewcommand{\arraystretch}{1.2}
\begin{table*}[!htb]
\caption{Predicted spectra of pentaquarks $P_{bbbc\bar{b}}$ and $P_{bbbc\bar{%
c}}$ with comparisons from various works. Bag radius $R_{0}$ is in GeV$^{-1}$%
. Masses are in GeV. Magnetic moments $\protect\mu$ are in unit of $\protect%
\mu_N$. The states denoted by asterisks couples strongly to scattering states.}
\label{tab:bbbcQ}%
\begin{tabular}{c|cccccc|cccccc}
\bottomrule[1.2pt]\bottomrule[0.5pt] \multirow{2}{*}{$J^{P}$} &
\multicolumn{6}{l|}{$P_{bbbc\bar{b}}$} & \multicolumn{6}{l}{$P_{bbbc\bar{c}}$%
} \\
& $R_{0}$ & EigenVector & $M$ & $M$\cite{An:2020jix} & $M$\cite{An:2022fvs}
& $\mu$ & $R_{0}$ & EigenVector & $M$ & $M$\cite{An:2020jix} & $M$\cite%
{An:2022fvs} & $\mu$ \\ \hline
${5/2}^{-}$ & 3.92 & 1.00 & 21.480$\ast$ & 20.648 & - & 0.31 & 4.28 & 1.00 & 18.183$\ast$
& 17.407 & - & -0.26 \\
${3/2}^{-}$ & 3.93 & (0.92,0.21,0.34) & 21.484 & 20.654 & 21.092 & 0.13 &
4.26 & (0.98,-0.04,0.17) & 18.180 & 17.535 & 17.891 & -0.27 \\
& 3.90 & (-0.39,0.70,0.60) & 21.475 & 20.644 & - & 0.35 & 4.23 &
(-0.08,0.75,0.65) & 18.171 & 17.406 & - & 0.05 \\
& 3.80 & (-0.11,-0.68,0.72) & 21.451$\ast$ & 20.578 & - & -0.21 & 4.16 &
(-0.15,-0.66,0.74) & 18.149$\ast$ & 17.291 & - & -0.30 \\
${1/2}^{-}$ & 3.97 & (-0.99,-0.10,0.03) & 21.493 & 20.691 & 21.079 & -0.03 &
4.32 & (-1.00,-0.08,0.07) & 18.198 & 17.578 & 17.884 & 0.16 \\
& 3.89 & (0.11,-0.94,0.31) & 21.472 & 20.653 & - & 0.06 & 4.25 &
(0.10,-0.88,0.46) & 18.177 & 17.523 & - & 0.16 \\
& 3.84 & (0.00,0.31,0.95) & 21.461 & 20.607 & - & -0.05 & 4.18 &
(0.02,0.46,0.89) & 18.159 & 17.399 & - & -0.52 \\
\bottomrule[0.5pt]\bottomrule[1.2pt]
\end{tabular}%
\end{table*}

\renewcommand{\tabcolsep}{0.18cm} \renewcommand{\arraystretch}{1.2}
\begin{table*}[!htb]
\caption{Predicted spectra of pentaquarks $P_{cccb\bar{b}}$ and $P_{cccb\bar{%
c}}$ with comparisons from various works. Bag radius $R_{0}$ is in GeV$^{-1}$%
. Masses are in GeV. Magnetic moments $\protect\mu$ are in unit of $\protect%
\mu_N$. The states denoted by asterisks couples strongly to scattering states.}
\label{tab:cccbQ}%
\begin{tabular}{c|cccccc|cccccc}
\bottomrule[1.2pt]\bottomrule[0.5pt] \multirow{2}{*}{$J^{P}$} &
\multicolumn{6}{l|}{$P_{cccb\bar{b}}$} & \multicolumn{6}{l}{$P_{cccb\bar{c}}$%
} \\
& $R_{0}$ & EigenVector & $M$ & $M$\cite{An:2020jix} & $M$\cite{An:2022fvs}
& $\mu$ & $R_{0}$ & EigenVector & $M$ & $M$\cite{An:2020jix} & $M$\cite%
{An:2022fvs} & $\mu$ \\ \hline
${5/2}^{-}$ & 4.56 & 1.00 & 14.873$\ast$ & 14.246 & - & 1.47 & 4.78 & 1.00 & 11.554$\ast$
& 11.124 & - & 0.90 \\
${3/2}^{-}$ & 4.58 & (-0.89,0.33,0.32) & 14.885 & 14.373 & 14.687 & 0.65 &
4.81 & (-0.42,0.62,0.67) & 11.564 & 11.137 & 11.444 & 0.62 \\
& 4.56 & (0.46,0.65,0.61) & 14.873 & 14.246 & - & 1.08 & 4.77 &
(0.87,0.49,0.09) & 11.549 & 11.101 & - & 0.59 \\
& 4.52 & (0.01,-0.69,0.72) & 14.862$\ast$ & 14.182 & - & 1.31 & 4.70 &
(0.27,-0.62,0.74) & 11.525$\ast$ & 11.038 & - & 1.03 \\
${1/2}^{-}$ & 4.61 & (0.97,-0.22,0.09) & 14.895 & 14.411 & 14.676 & 0.27 &
4.86 & (0.97,-0.19,0.14) & 11.581 & 11.175 & 11.438 & 0.43 \\
& 4.56 & (-0.20,-0.58,0.79) & 14.878 & 14.357 & - & 0.26 & 4.81 &
(-0.23,-0.68,0.69) & 11.564 & 11.137 & - & 0.17 \\
& 4.51 & (0.11,0.79,0.61) & 14.862 & 14.238 & - & 0.62 & 4.71 &
(0.04,0.70,0.71) & 11.526 & 11.048 & - & 0.35 \\
\bottomrule[0.5pt]\bottomrule[1.2pt]
\end{tabular}%
\end{table*}

\renewcommand{\tabcolsep}{0.11cm} \renewcommand{\arraystretch}{1.2}
\begin{table*}[!htb]
\caption{Predicted spectra of pentaquarks $P_{ccbb\bar{b}}$ and $P_{ccbb\bar{%
c}}$ with comparisons from various works. Bag radius $R_{0}$ is in GeV$^{-1}$%
. Masses are in GeV. Magnetic moments $\protect\mu$ are in unit of $\protect%
\mu_N$.}
\label{tab:ccbbQ}%
\begin{tabular}{c|cccccc|cccccc}
\bottomrule[1.2pt]\bottomrule[0.5pt] \multirow{2}{*}{$J^{P}$} &
\multicolumn{6}{l|}{$P_{ccbb\bar{b}}$} & \multicolumn{6}{l}{$P_{ccbb\bar{c}}$%
} \\
& $R_{0}$ & EigenVector & $M$ & $M$\cite{An:2020jix} & $M$\cite{An:2022fvs}
& $\mu$ & $R_{0}$ & EigenVector & $M$ & $M$\cite{An:2020jix} & $M$\cite%
{An:2022fvs} & $\mu$ \\ \hline
${5/2}^{-}$ & 4.27 & 1.00 & 18.182 & 17.477 & - & 0.88 & 4.55 & 1.00 & 14.872
& 14.295 & - & 0.32 \\
${3/2}^{-}$ & 4.29 & (0.83,0.36,-0.13,0.40) & 18.191 & 17.554 & 17.785 & 0.28
& 4.57 & (0.40,0.63,-0.26,0.61) & 14.880 & 14.375 & 14.579 & -0.01 \\
& 4.26 & (-0.54,0.66,-0.21,0.47) & 18.181 & 17.479 & - & 0.65 & 4.52 &
(-0.86,0.42,-0.28,0.01) & 14.866 & 14.298 & - & 0.17 \\
& 4.22 & (-0.04,-0.02,0.92,0.40) & 18.170 & 17.457 & - & 0.65 & 4.51 &
(-0.16,0.28,0.93,0.20) & 14.862 & 14.274 & - & 0.24 \\
& 4.20 & (-0.09,-0.66,-0.31,0.68) & 18.164 & 17.416 & - & 0.54 & 4.41 &
(-0.26,-0.59,-0.04,0.76) & 14.828 & 14.197 & - & 0.36 \\
${1/2}^{-}$ & 4.32 & (-0.98,-0.20,0.03,0.07) & 18.200 & 17.576 & 17.785 &
0.08 & 4.61 & (-0.98,-0.16,0.01,0.10) & 14.893 & 14.406 & 14.566 & 0.26 \\
& 4.26 & (0.20,-0.85,0.01,0.49) & 18.180 & 17.496 & - & 0.37 & 4.55 &
(0.19,-0.81,-0.06,0.55) & 14.875 & 14.318 & - & 0.31 \\
& 4.22 & (-0.03,0.48,0.30,0.82) & 18.168 & 17.437 & - & 0.41 & 4.48 &
(0.01,0.48,0.46,0.75) & 14.852 & 14.253 & - & 0.31 \\
& 4.16 & (-0.03,0.14,-0.95,0.27) & 18.154 & 17.405 & - & -0.06 & 4.39 &
(-0.02,0.30,-0.89,0.35) & 14.821 & 14.185 & - & -0.07 \\
\bottomrule[0.5pt]\bottomrule[1.2pt]
\end{tabular}%
\end{table*}

\begin{figure*}[!t]
\centering
\subfigure[\
    $bbbb\bar{c}$]{\includegraphics[width=0.45\textwidth]{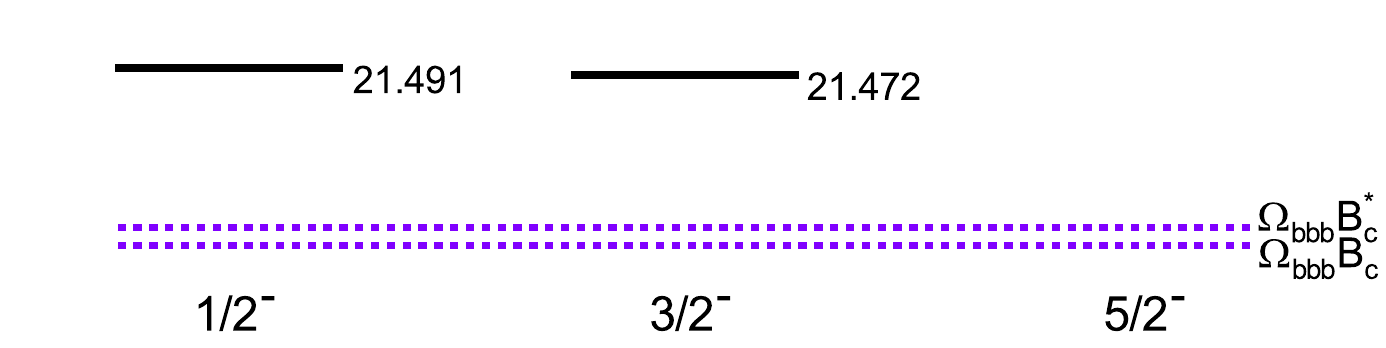}}
\subfigure[\
    $cccc\bar{b}$]{\includegraphics[width=0.49\textwidth]{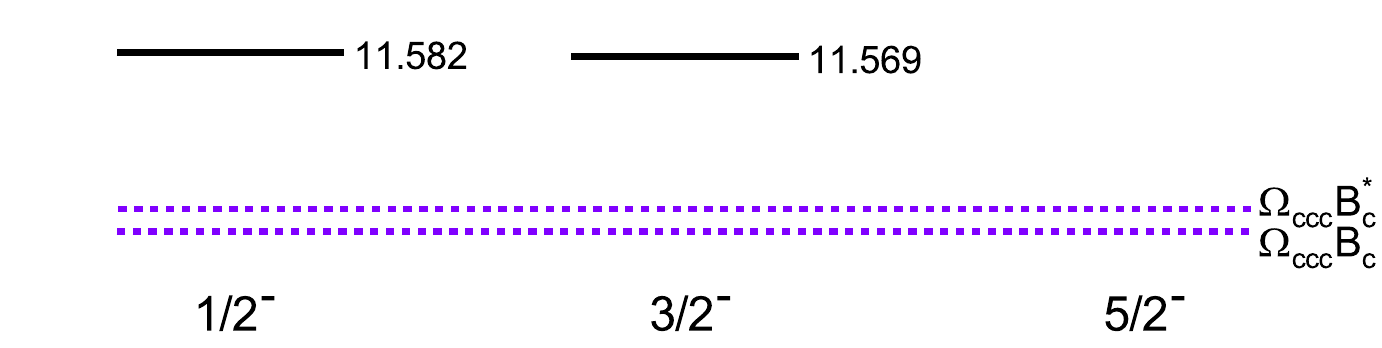}}
\par
\subfigure[\
    $bbbc\bar{b}$]{\includegraphics[width=0.49\textwidth]{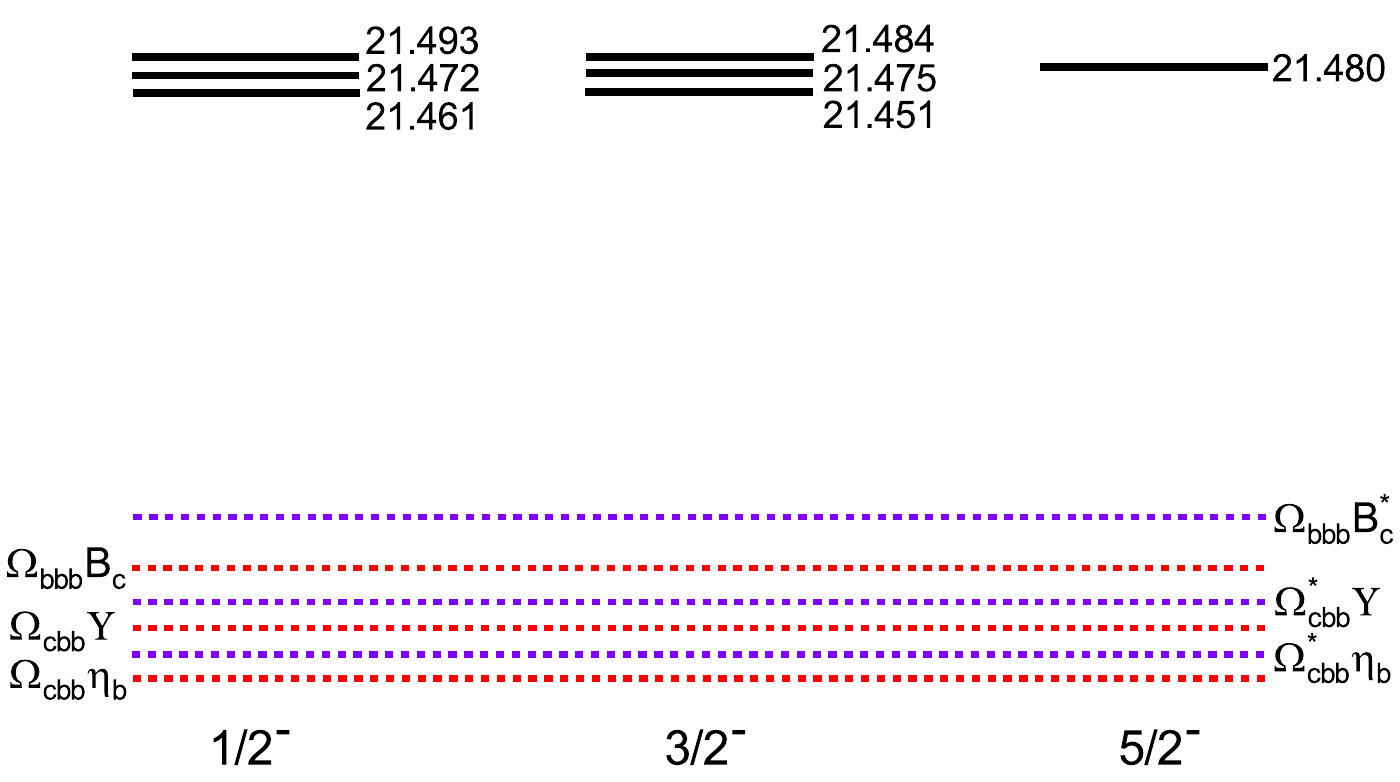}}
\subfigure[\
    $bbbc\bar{c}$]{\includegraphics[width=0.49\textwidth]{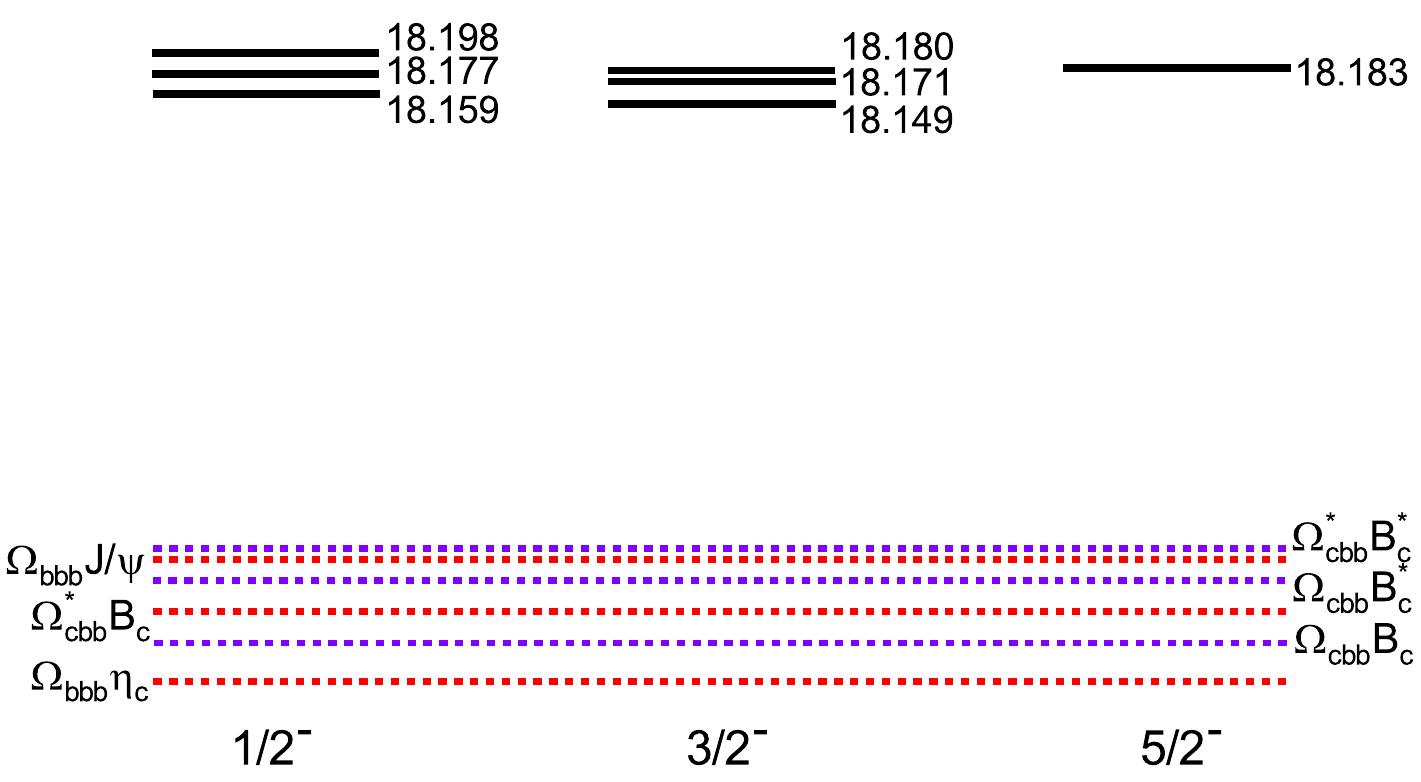}}
\par
\subfigure[\
    $cccb\bar{b}$]{\includegraphics[width=0.49\textwidth]{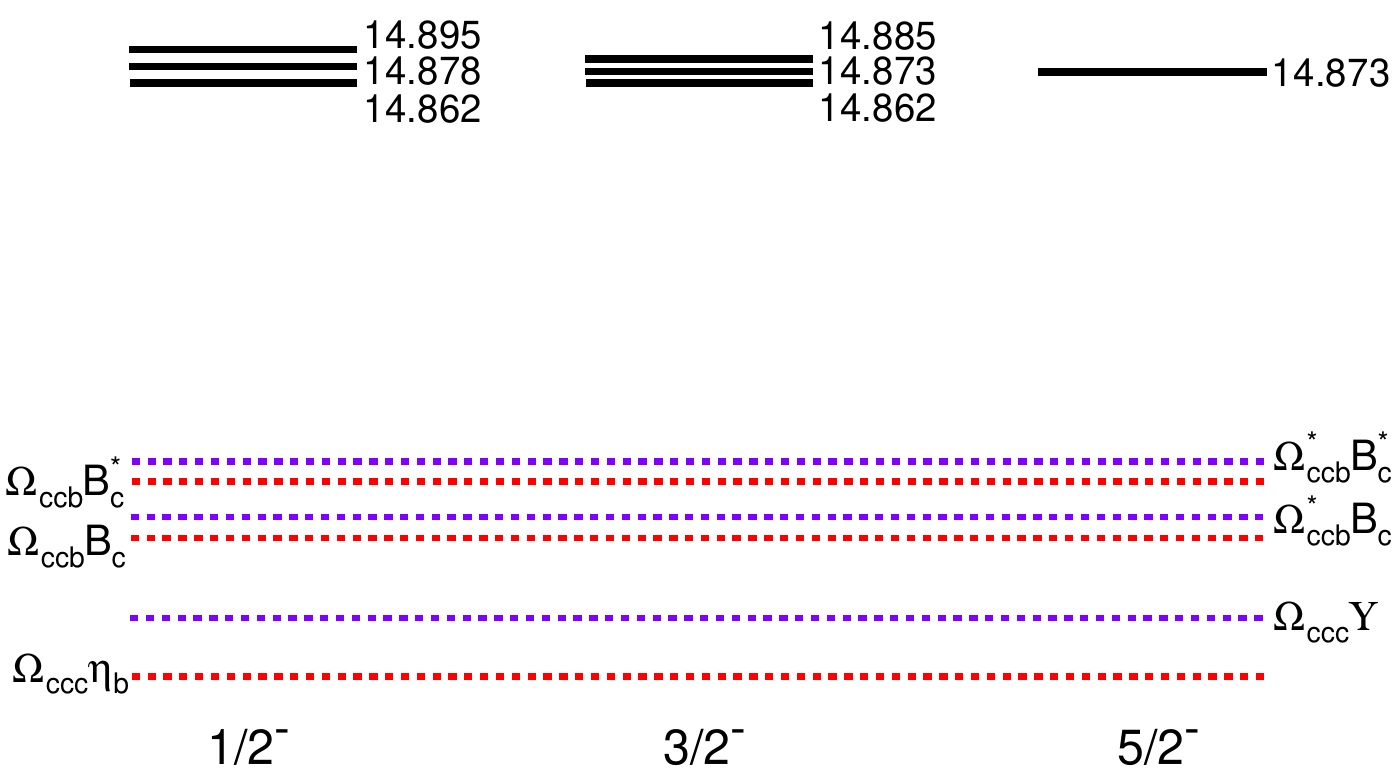}}
\subfigure[\
    $cccb\bar{c}$]{\includegraphics[width=0.49\textwidth]{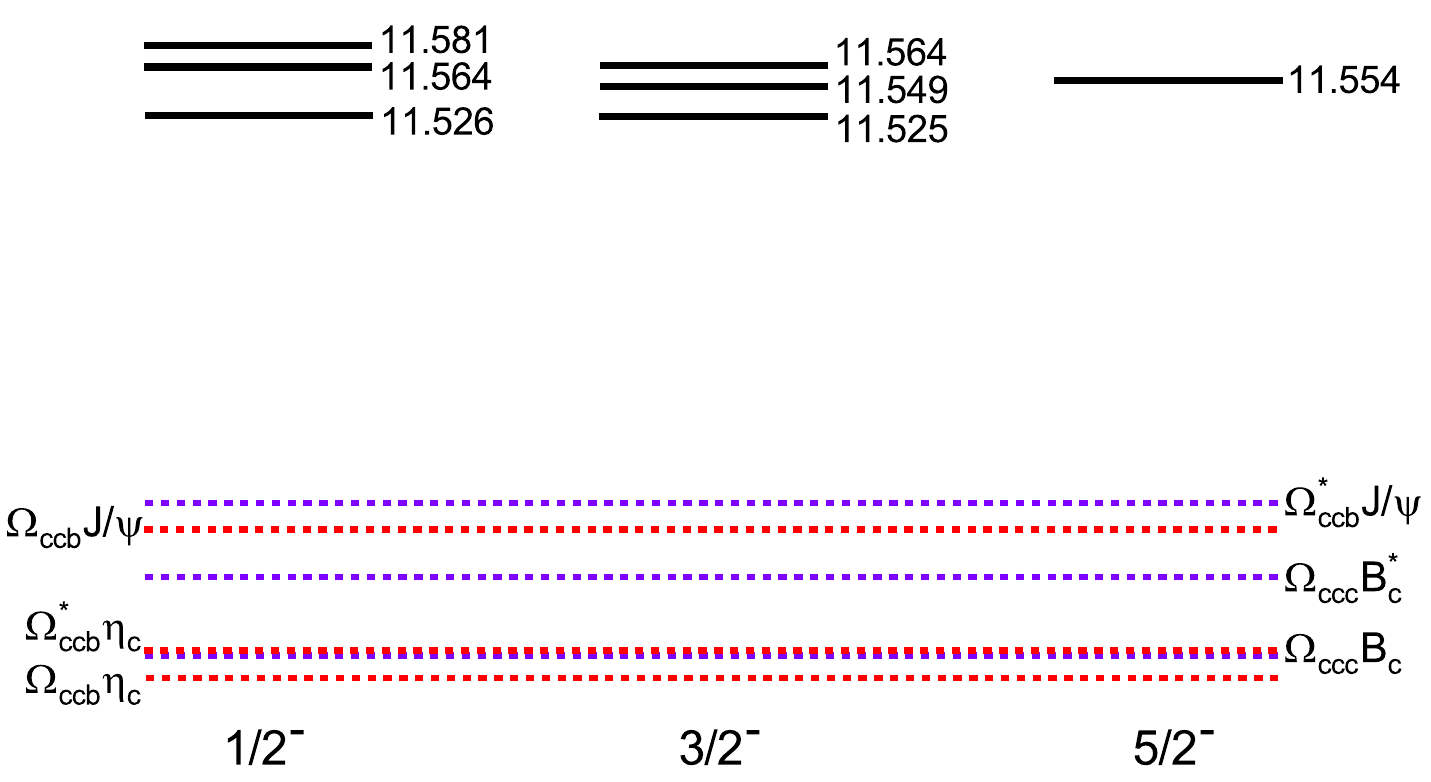}}
\par
\subfigure[\
    $ccbb\bar{b}$]{\includegraphics[width=0.49\textwidth]{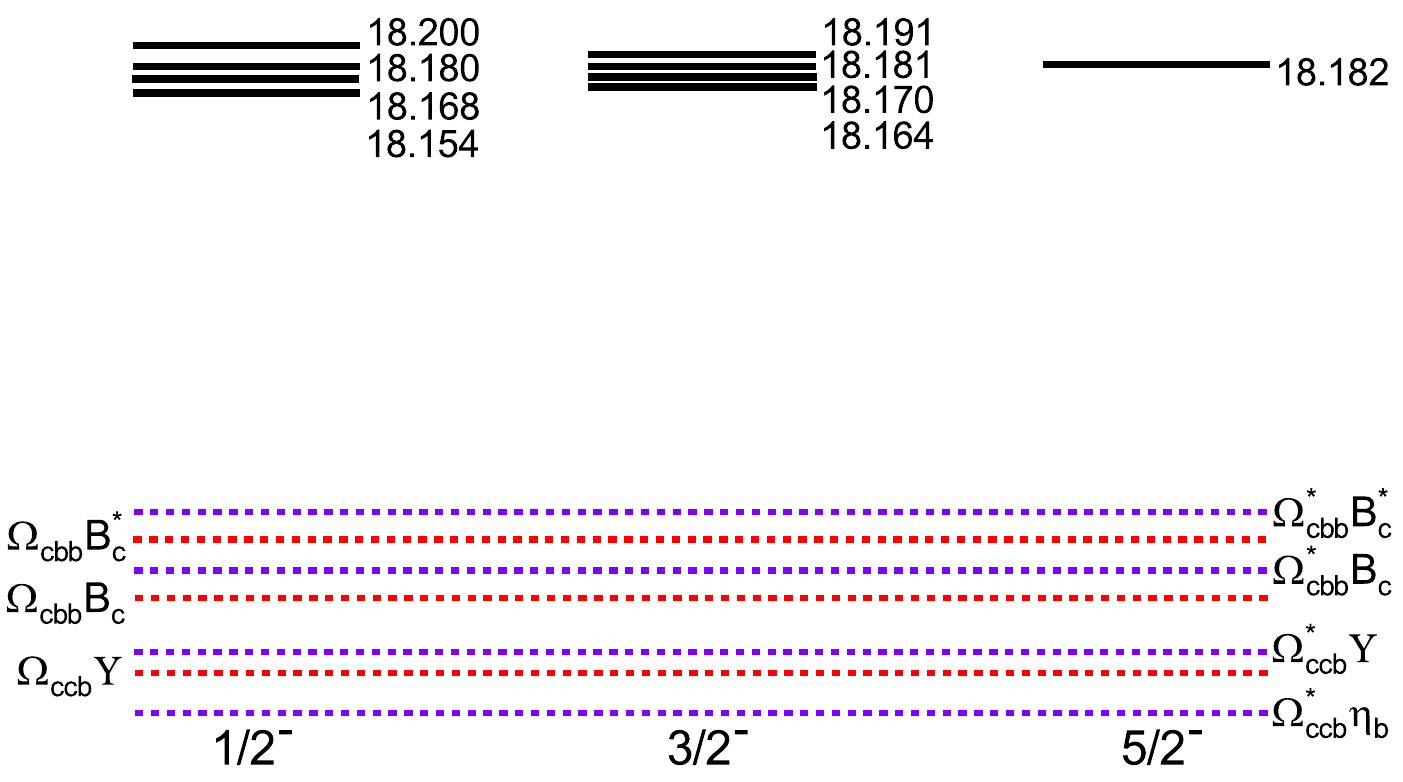}}
\subfigure[\
    $ccbb\bar{c}$]{\includegraphics[width=0.49\textwidth]{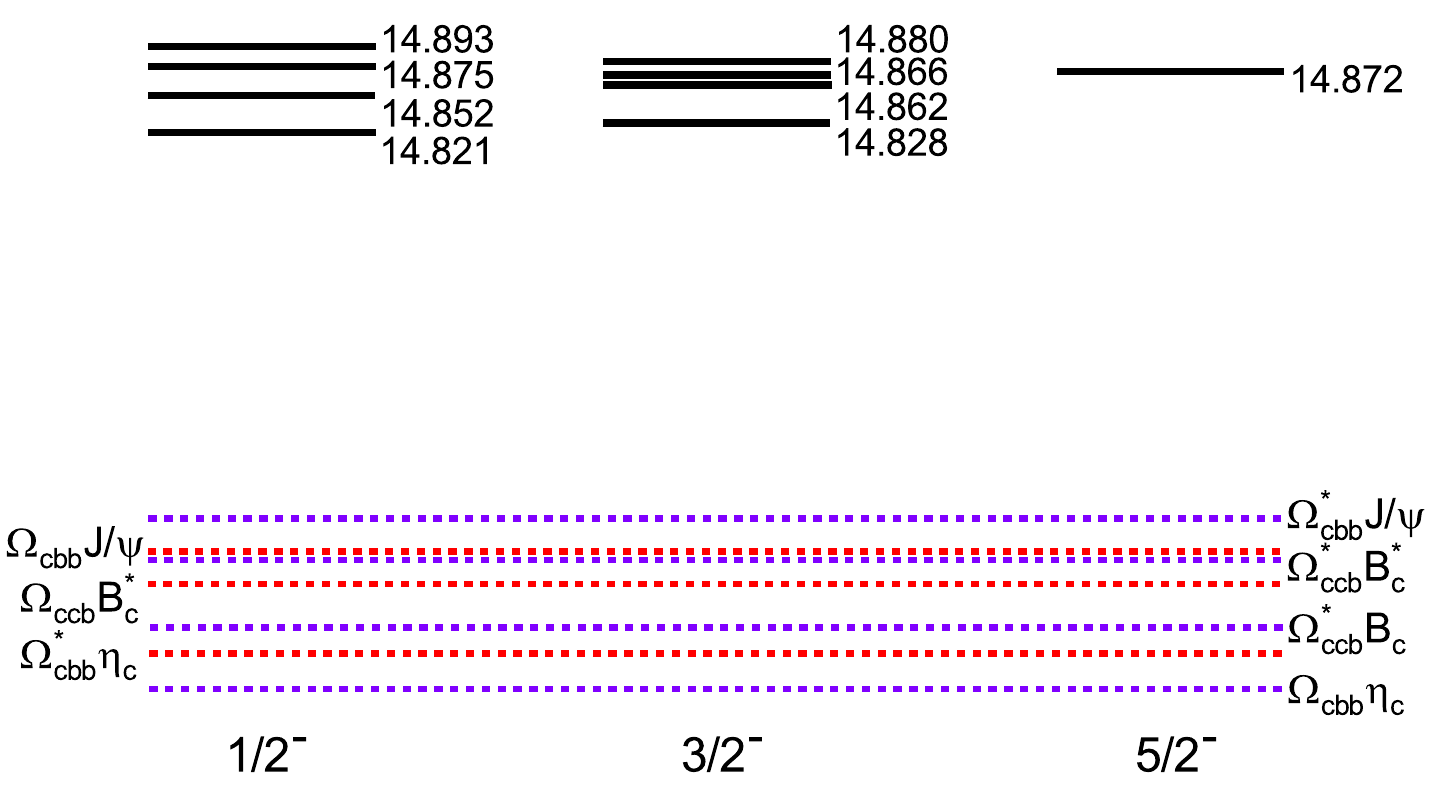}}
\caption{Computed mass spectra (in GeV) of pentaquarks $P_{bbbb\bar{c}}$, $%
P_{cccc\bar{b}}$, $P_{bbbc\bar{b}}$, $P_{bbbc\bar{c}}$, $P_{cccb\bar{b}}$, $%
P_{cccb\bar{c}}$, $P_{ccbb\bar{b}}$, and $P_{ccbb\bar{c}}$ for $%
J^{P}=5/2^-,3/2^-,1/2^-$, plotted by short solid lines, with respective threshold
energy as blue (labeled right) and red (labeled left) dotted lines.}
\label{fig:threshold2}
\end{figure*}

Further more, we examine the respective mass splittings $\mathrm{\Delta} M$ between
spin multiplets due to the lowest two quantum numbers $J$ for the fully heavy
mesons, tetraquarks and pentaquarks and show them collectively in Table~\ref%
{tab:splittings}. Here, $J=\{0,1\}$ or $J=\{1/2,3/2\}$. For the fully heavy
systems with identical quarks, it is found that $\mathrm{\Delta} M$ are
suppressed significantly with the heavy quark number $N$ increasing and
hadron becoming heavier. We see that the mass splittings are narrower
for the bottom sector compared to the charm sector.

For the fully heavy pentaquarks with both bottom and charm flavors,
we list the corresponding results computed via the MIT bag model
in Tables \ref{tab:QQQQq}, \ref{tab:bbbcQ}, \ref{tab:cccbQ}, and \ref%
{tab:ccbbQ}, in comparison with other similar works cited \cite%
{An:2020jix,An:2022fvs}. The obtained masses are also compared to the respective
thresholds and plotted in FIG.~\ref{fig:threshold2}. All hadrons computed are
above the thresholds associated with the baryons and mesons,
meaning that these fully heavy pentaquarks are unstable against strong
decays to two-hadron final states.

The instability of pentaquarks can be investigated by analyzing the eigenvectors obtained in this study.
In order to examine the color-spin wavefunction $\psi$ of pentaquark state, 
we utilize the corresponding eigenvectors and bases defined in Table \ref{tab:basis},
and express it in the form as shown below:
\begin{equation}
    \psi = c_{1} \left(q_{1}q_{2}q_{3}\right)_{\bold{1_c}} \otimes \left(q_{4}\bar{q_{3}}\right)_{\bold{1_c}} + \dots.
\end{equation}
Here, $c_{1}$ represents the overlap between the wavefunctions of the pentaquark 
and the specific baryon $\otimes$ meson component, which corresponds to a scattering state. 
If the pentaquark state couples strongly to a scattering state, indicating that the 
probability $|c_1|^2$ of the collapse of $\psi$ into a color-singlet (\ref{ph3p}) approaches 1, 
we can infer that it is unstable against strong decay with a very broad width. 

For the pentaquark states presented in Table \ref{tab:QQQQQ} and \ref{tab:QQQQq}, 
the values of $|c_1|^2$ are determined to be 1/3. Similarly, in Table \ref{tab:ccbbQ}, 
all the calculated values are approximately 1/3. Notably, the pentaquark states denoted by 
asterisks in Table \ref{tab:bbbcQ} and \ref{tab:cccbQ} have dominant components of 
scattering states with $|c_1|^2$ values greater than 0.83, while the remaining states 
range from 0.34 to 0.80. Based on the above discussions, we conclude that the pentaquark states 
$P_{bbbb\bar{Q}}$, $P_{cccc\bar{Q}}$, and $P_{ccbb\bar{Q}}$ ($Q$ = $c$, $b$) are likely to 
possess a compact structure with relatively narrow widths.

On the other hand, the bag radius $R_{0}$ ranging from $3.48\,$GeV$^{-1}=0.7$
fm to $5.08\,$GeV$^{-1}=1.0$ fm, which are in the order of the typical
hadrons in size (about $1\,$fm), implies that these compact hadrons of
pentaquarks may exist during hadronization in experiments of LHCb.

\section{Mass pattern and linearity upon quark number}

\label{sec:linearity}

To explain the mass pattern obtained in this work, we consider question as to
if the bag radius $R$ leads to the larger volume energy $%
M_{V}=4\pi R^{3}B/3$ for the larger heavy quark number $N$, so that
the hadron masses tend to be of suprathreshold. For this, we ignore the mass
splittings due to the CMI and rewrite Eq.~(\ref{MBm}) in the form that hadrons
consist of the identical flavor $Q$($=c,b$). The result is,
\begin{equation}
\bar{M}\left( R,N\right) =N\left( m_{Q}^{2}+\frac{x_{Q}^{2}}{R^{2}}\right)
^{1/2}+\frac{4}{3}\pi R^{3}B-\frac{Z_{0}}{R}+NB_{QQ},  \label{MBmN}
\end{equation}%
where $\bar{M}$ stands for the spin-independent mass, $N$ the total number of
valence heavy quarks, and $B_{QQ}$ the binding energy given in Eq.~(\ref{Binds}).

Owing to the scaling of the color factor, as mentioned in Sect.~\ref%
{sec:bagmodel}, the fully heavy hadrons with $N$ identical quarks happen to
have total binding energy of $NB_{QQ}$. Given Eq.~(\ref{MBmN}), one can use
it to examine each part of the energy for the fully heavy
systems with identical quark number $N=2\sim 5$. The numerical results are
listed in Table~\ref{tab:contri}. One sees that the volume energy $M_{V}$
grows quickly while the suppression of zero point energy $M_{Z}=-Z_{0}/R$
becomes weaker significantly with more valence quarks involved.
This implies that fully heavy hadrons tend to be heavier
when $N$ becomes larger and they are all above thresholds of the
decaying final states.

\renewcommand{\tabcolsep}{0.2cm} \renewcommand{\arraystretch}{1.2}
\begin{table}[tbh]
\caption{Numerical results for the mass components (all in GeV),
the kinematic energy $\omega _{Q}$ (in GeV), the volume
energy $M_{V}$, zero point energy $M_{Z}$ and spin-independent mass $\bar{M}$
of fully heavy systems (mesons, baryons and pentaquarks) in detail. The
parameters of the bag radius $R_{0}$(GeV$^{-1}$) and $x_{Q}$ are also shown for the heavy quark $Q$.}%
\begin{tabular}{lcccccc}
\bottomrule[1.2pt]\bottomrule[0.5pt] $\mathrm{System}$ & $R_{0}$ & $x_{Q}$ & $%
\omega _{Q}$ & $M_{V}$ & $M_{Z}$ & $\bar{M}$ \\ \hline
$c\bar{c}$ & $3.45$ & $2.886$ & $1.842$ & $0.076$ & $-0.530$ & $3.075$ \\
$ccc$ & $4.14$ & $2.924$ & $1.786$ & $0.132$ & $-0.442$ & $4.818$ \\
$cc\bar{c}\bar{c}$ & $4.59$ & $2.943$ & $1.762$ & $0.179$ & $-0.399$ & $6.520
$ \\
$cccc\bar{c}$ & $4.92$ & $2.955$ & $1.748$ & $0.220$ & $-0.372$ & $8.201$ \\
\hline
$b\bar{b}$ & $1.75$ & $2.971$ & $5.369$ & $0.010$ & $-1.047$ & $9.445$ \\
$bbb$ & $2.54$ & $3.022$ & $5.230$ & $0.030$ & $-0.720$ & $14.616$ \\
$bb\bar{b}\bar{b}$ & $3.07$ & $3.042$ & $5.188$ & $0.054$ & $-0.596$ & $%
19.700$ \\
$bbbb\bar{b}$ & $3.45$ & $3.053$ & $5.169$ & $0.076$ & $-0.531$ & $24.752$ \\
\bottomrule[0.5pt]\bottomrule[1.2pt]
\end{tabular}%
\label{tab:contri}
\end{table}

\begin{figure}[!ht]
\centering
\subfigure[\ fully bottom
system]{\includegraphics[width=0.45\textwidth]{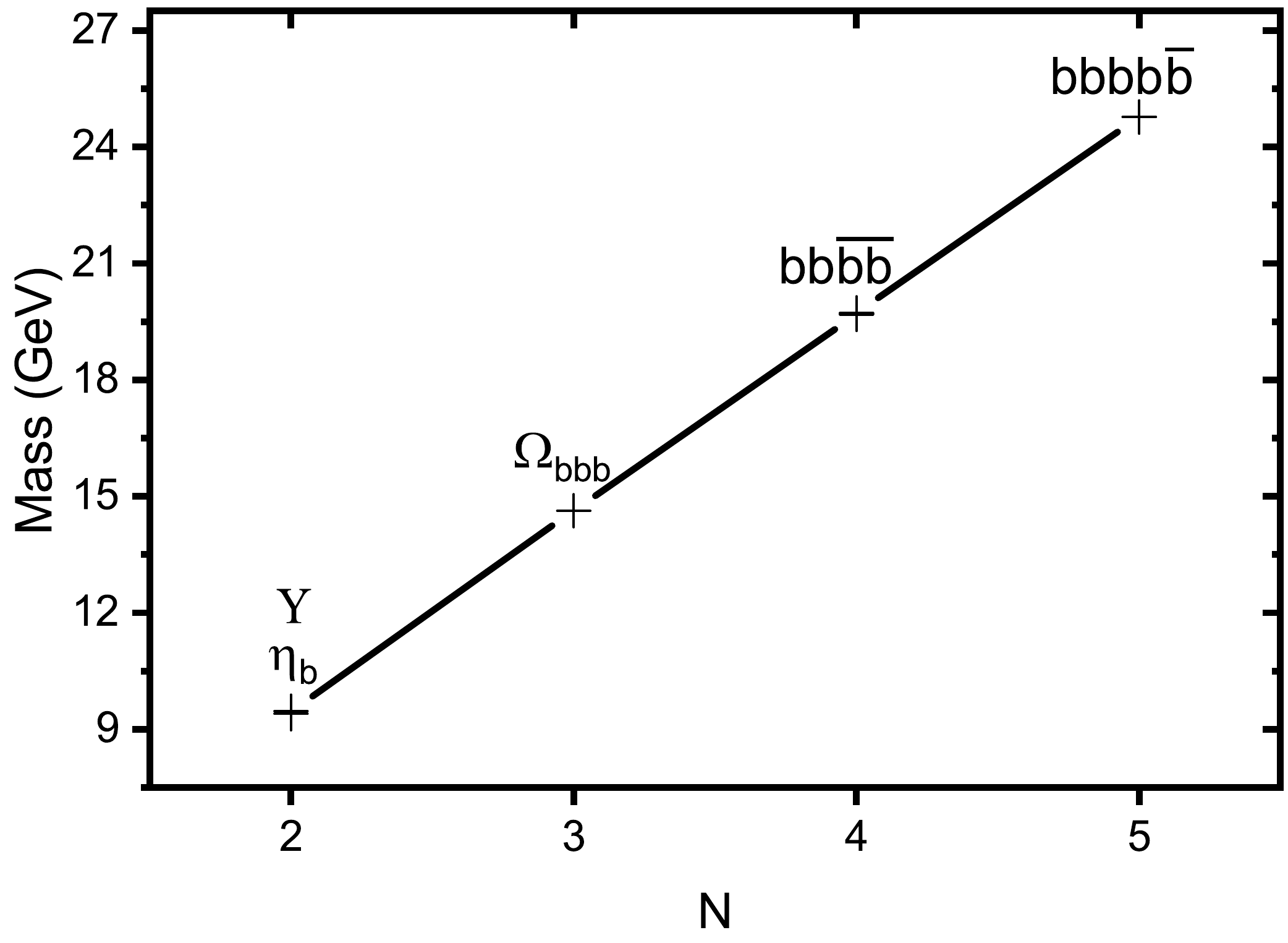}}
\subfigure[\
fully charm system]{\includegraphics[width=0.45\textwidth]{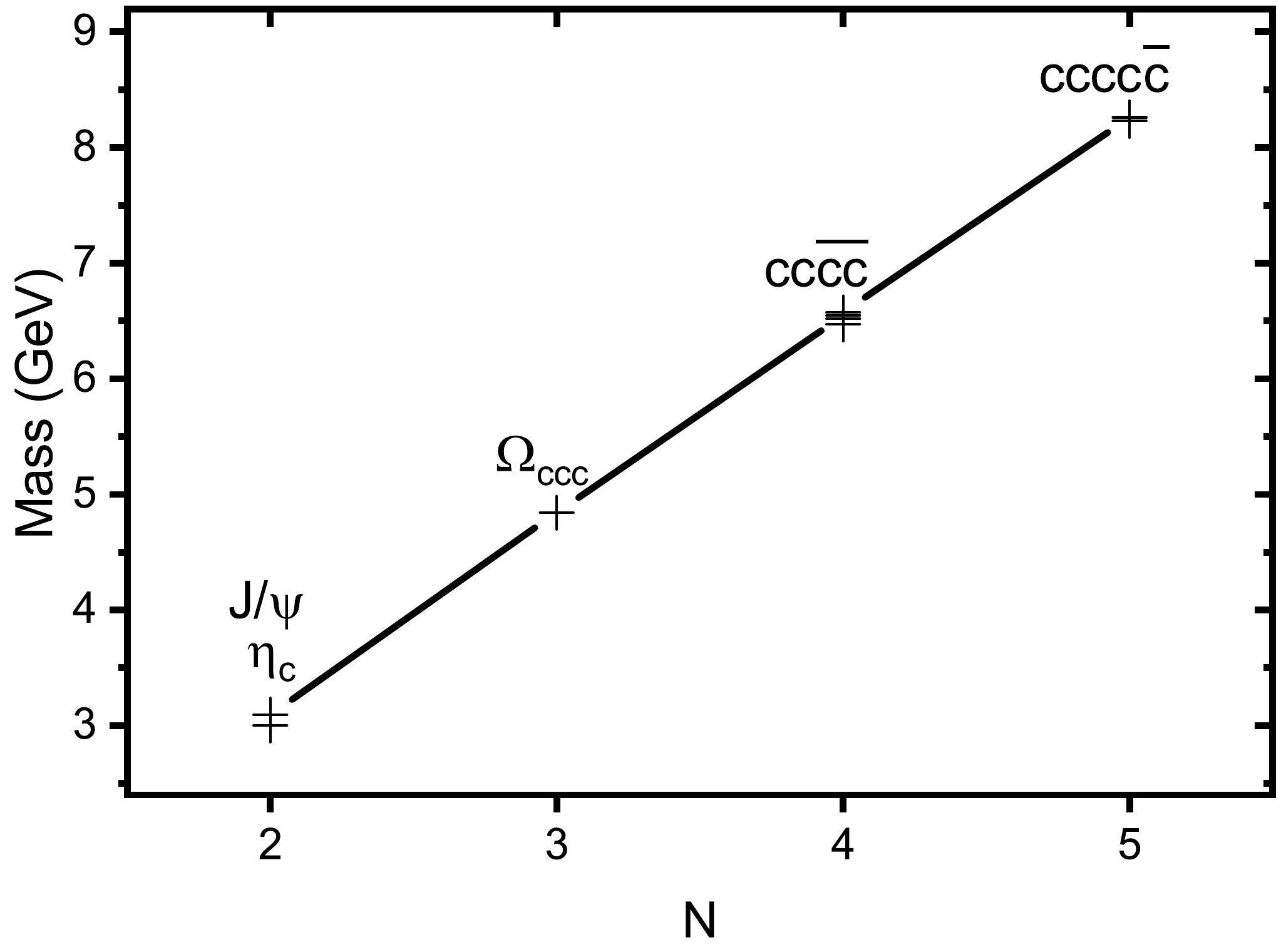}}
\caption{Masses of fully heavy systems ranging from mesons ($N=2$) to
pentaquarks ($N=5$) with $N$ the number of quarks and antiquarks, calculated
in the unified framework of MIT bag model. The results of mesons are quoted
from Ref.~\protect\cite{Zhang:2021yul}, and that of tetraquarks are from Ref.~%
\protect\cite{Yan:2023lvm}.}
\label{fig:FullyH}
\end{figure}

We further examine this mass pattern for fully heavy systems of the mesons,
baryons, tetraquarks and pentaquarks within the same framework of MIT bag model.
Beside the masses of baryons and pentaquarks which are computed via the MIT bag model,
we cite the results for the mesons and tetraquarks in Refs.~\cite{Zhang:2021yul}
and \cite{Yan:2023lvm}, respectively. The obtained results are plotted in
FIG.~\ref{fig:FullyH}. For the fully bottom and charm hadrons one sees that
hadron masses rise almost linearly with $N$
ranging from $2$ to $5$. It is notable that our mass predictions of the
mesons and tetraquark $cc\bar{c}\bar{c}$ are in good agreements with
experimental data, especially, the computed mass $6572\,$MeV of the fully
charm tetraquark $cc\bar{c}\bar{c}$ agrees well with the measured mass $%
6552\pm 10\pm 12\,$MeV of the newly-discovered resonance $X(6600)$ \cite%
{Zhang:2022toq}.

\begin{figure}[!ht]
\centering
\subfigure[\ fully heavy
system\label{fig:MNh}]{\includegraphics[width=0.45\textwidth]{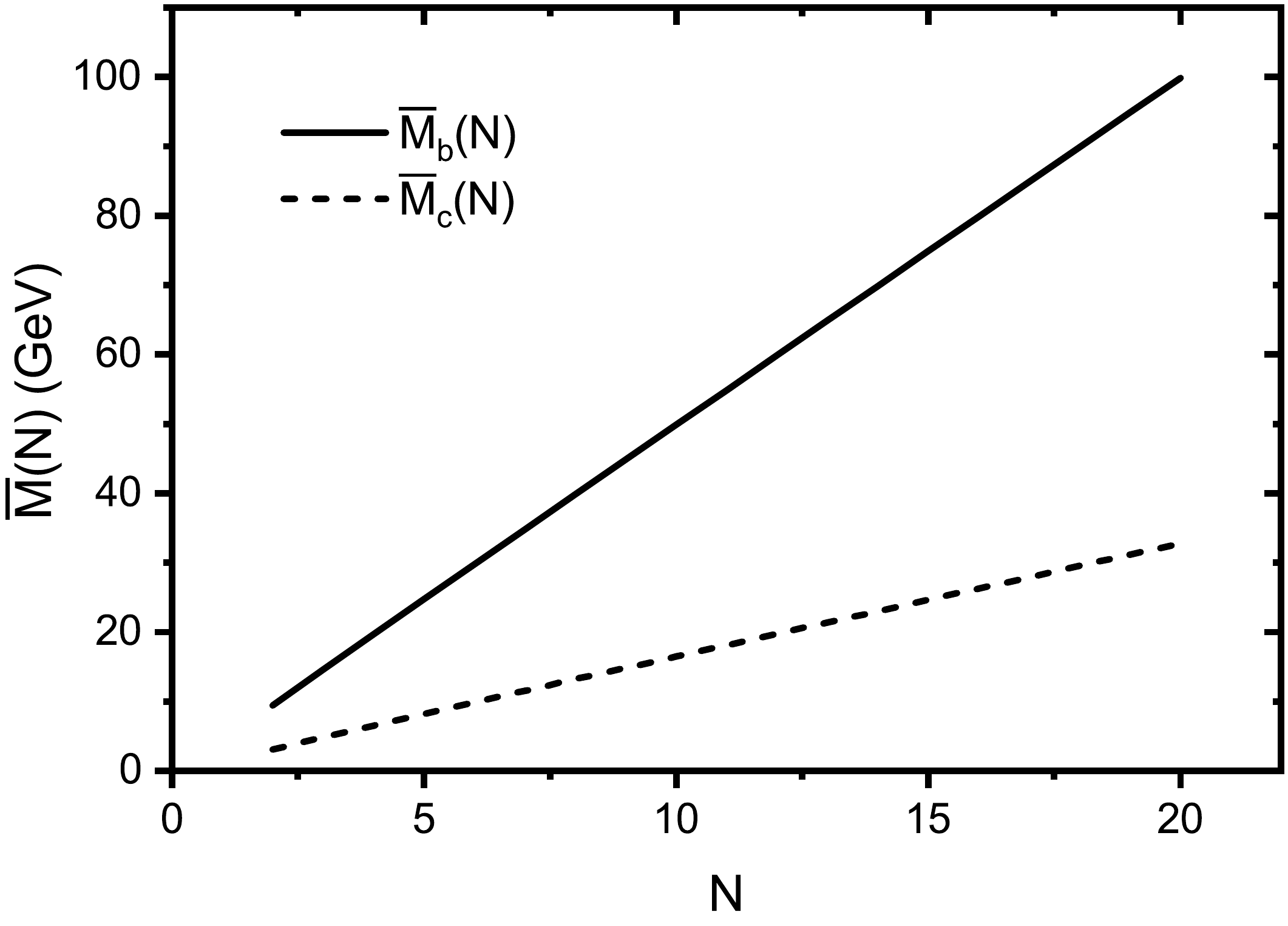}}
\subfigure[\ fully light
system\label{fig:MNl}]{\includegraphics[width=0.45\textwidth]{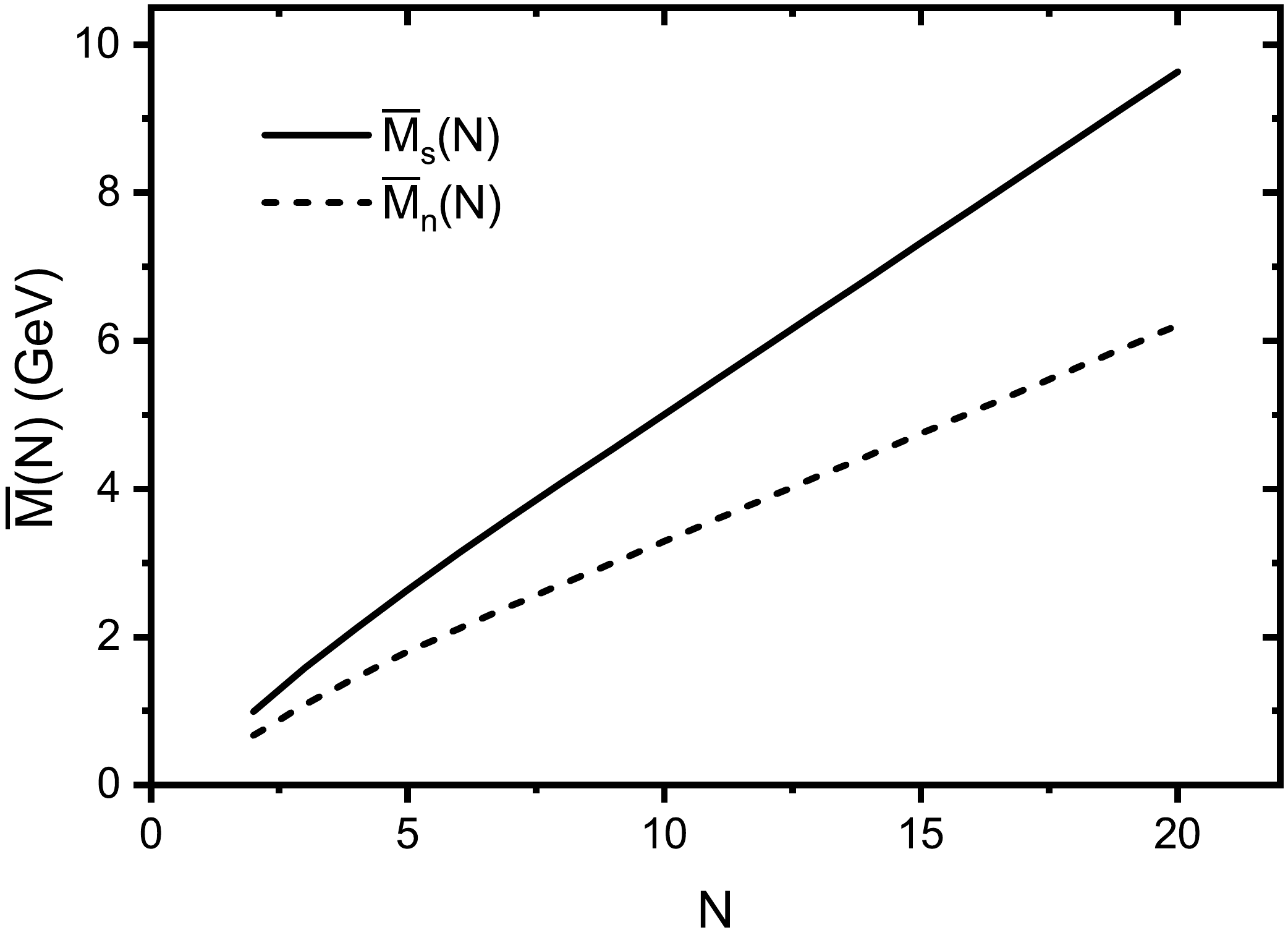}}
\caption{The calculated spin-independent masses of fully heavy and light
systems by Eq.~(\protect\ref{MBmN}). Quark number $N$ is from 2 up to 20 for
mathematical extrapolation.}
\end{figure}

Inspired by Eq.~(\ref{MBmN}) and the numerical results above, we promote
the heavy quark number $N$ to be as large as $20$ mathematically, despite
that a hadron with $20$ valence quarks, if exist, may result in a very
large and thereby unphysical bag radius $R$. We apply this procedure to $%
N=2,3,4$ and $5$ to test if the linear dependence of the hadron mass holds
true as far as this work is involved. Similar to FIG.~\ref{fig:FullyH}, we
plot the numerical results in FIG.~\ref{fig:MNh} for the spin-independent
masses v.s. quark number $N$ for fully bottom and charm systems
respectively, with the data from $N=2\sim 20$. From these results, we find
the following numerical fits for the fully bottom (denoted by subscript b)
and charm (denoted by subscript c) systems (in GeV,
$2\leq N\leq 5$):
\begin{equation}
\begin{aligned} \bar{M}_{b}\left(N\right) &=5.1004 N-0.7229, \\
\bar{M}_{c}\left(N\right) &=1.7079 N-0.3241. \end{aligned}  \label{linearM}
\end{equation}

In FIG.~\ref{fig:MNh}, one observes the linear lines numerically, corresponding to Eq.~(\ref{linearM}),
but the marching of it with the data of the spin-independent masses
in Table~\ref{tab:contri} is approximated. To see this,
we apply Eq.~(\ref{MBmN}) to the fully light hadron systems with
binding energy ignored, $B_{QQ}=0$, and obtain the results
plotted in FIG.~\ref{fig:MNl}. The linear dependence breaks slightly when $N$ is small,
as it should be for the light hadrons. Notice that the (solid) line for the bottom sector
is nearly linear in contrast with that for the charm sector, this implies
that linearity of mass dependence holds true at heavy quark limit.

In the following, we show such a mass linearity upon $N$ in heavy quark limit analytically.
We expand the mass formula (\ref{MBmN}) in term of the inverse heavy-quark
mass $m_{Q}^{-1}$,
\begin{equation}
\bar{M}\left(R, N\right) \approx
Nm_{Q}\left(1+\frac{x_{Q}^{2}}{2m^{2}_{Q}R^{2}}\right)+\frac{4}{3}\pi
R^{3}B-\frac{Z_{0}}{R}+NB_{QQ}, \label{expand}
\end{equation}%
and minimize the mean mass Eq.~(\ref{expand}) to find the bag radius,
\begin{equation}
\frac{\partial \bar{M}\left( R,N\right) }{\partial R}=0\Rightarrow R=R_{0}.
\label{variation}
\end{equation}%
The solution of Eq.~(\ref{variation}) in the large-$N$ limit is $R_{0}\sim N^{1/5}$.
This gives rise to the following mass
formula for fully heavy hadrons
\begin{equation}
\bar{M}\left( N\right) =Nm_{Q}+\bar{\Lambda}_{N}+\frac{x_{Q}^{2}N^{\frac{3}{5%
}}}{2m_{Q}},  \label{NMQ}
\end{equation}%
with $\bar{\Lambda}_{N}=NB_{QQ}+\left( 4\pi B/3\right) N^{3/5}-Z_{0}N^{-1/5}$,
by analogy with the mass relation $m_{H_{Q}}=m_{Q}+\bar{\Lambda}+\frac{\mathrm{\Delta} m^{2}}{2m_{Q}}$
in heavy quark effective theory (HQET).
This explains the linear mass dependence of fully heavy hadrons with heavy
quark flavor $N$, as demonstrated in FIG.~\ref{fig:FullyH}.

\section{Conclusions and remarks}

\label{sec:conclusions} In this work, we employ a unified framework of MIT
bag model to perform a systematical study of all ground-state configurations
of fully heavy pentaquarks and other fully heavy hadrons. With the help of
the color-spin wavefunctions of pentaquarks, which are classified via Young
tableaux which respect symmetry of the $SU(2)_{s}\otimes SU(3)_{c}$ group and
present in terms of the Young-Yamanouchi bases, we have computed the masses
and magnetic moments of the fully charm, bottom and bottom-charm pentaquarks in
ground states via numerical variational method that applies to the MIT bag
model. Our computation predicts a set of masses of fully heavy pentaquarks,
ranging from $8.229$ GeV for the $P_{cccc\bar{c}}$ to $24.770$ GeV for the $%
P_{bbbb\bar{b}}$. Combining with the similar bag-model computation of masses of
fully heavy systems, we find that the masses of fully heavy
hadrons (mesons, baryons, tetraquarks and pentaquarks) with identical flavors
depend almost linearly upon the number $N$ of the valence quarks in hadrons,
and we demonstrate this linear behavior of the hadron masses using the MIT bag
model, being consistent with the heavy quark symmetry at the heavy-quark
limit. Our results also indicate that the heavier the fully heavy hadrons,
the narrower the mass splittings of their lowest two multiplets (in $J$).

We remark that with the increase of the heavy quark number $N$, the bag radius $%
R_{0}$ of the fully heavy hadrons can rise up as high as $1\,$fm, and it
results in the larger volume energy $M_{V}$ and the less suppression of zero
point energy $M_{Z}$, which makes them suprathreshold and unstable against
strong decay. On the other hand, the shorter bag radius of the compact
pentaquarks implies some larger possibilities for their formation via
hadronization in experiments, compared to the formation of the hadrons
containing light degrees of freedom. We hope that our predictions can help
the future LHCb experiments to search the fully heavy pentaquarks or baryons
addressed in this work.

\textbf{Acknowledgments}

D. J. is supported by the National Natural Science Foundation of China under
Grant No. 12165017.


\begin{thebibliography}{86}%
\makeatletter
\providecommand \@ifxundefined [1]{%
 \@ifx{#1\undefined}
}%
\providecommand \@ifnum [1]{%
 \ifnum #1\expandafter \@firstoftwo
 \else \expandafter \@secondoftwo
 \fi
}%
\providecommand \@ifx [1]{%
 \ifx #1\expandafter \@firstoftwo
 \else \expandafter \@secondoftwo
 \fi
}%
\providecommand \natexlab [1]{#1}%
\providecommand \enquote  [1]{``#1''}%
\providecommand \bibnamefont  [1]{#1}%
\providecommand \bibfnamefont [1]{#1}%
\providecommand \citenamefont [1]{#1}%
\providecommand \href@noop [0]{\@secondoftwo}%
\providecommand \href [0]{\begingroup \@sanitize@url \@href}%
\providecommand \@href[1]{\@@startlink{#1}\@@href}%
\providecommand \@@href[1]{\endgroup#1\@@endlink}%
\providecommand \@sanitize@url [0]{\catcode `\\12\catcode `\$12\catcode
  `\&12\catcode `\#12\catcode `\^12\catcode `\_12\catcode `\%12\relax}%
\providecommand \@@startlink[1]{}%
\providecommand \@@endlink[0]{}%
\providecommand \url  [0]{\begingroup\@sanitize@url \@url }%
\providecommand \@url [1]{\endgroup\@href {#1}{\urlprefix }}%
\providecommand \urlprefix  [0]{URL }%
\providecommand \Eprint [0]{\href }%
\providecommand \doibase [0]{http://dx.doi.org/}%
\providecommand \selectlanguage [0]{\@gobble}%
\providecommand \bibinfo  [0]{\@secondoftwo}%
\providecommand \bibfield  [0]{\@secondoftwo}%
\providecommand \translation [1]{[#1]}%
\providecommand \BibitemOpen [0]{}%
\providecommand \bibitemStop [0]{}%
\providecommand \bibitemNoStop [0]{.\EOS\space}%
\providecommand \EOS [0]{\spacefactor3000\relax}%
\providecommand \BibitemShut  [1]{\csname bibitem#1\endcsname}%
\let\auto@bib@innerbib\@empty
\bibitem [{\citenamefont {Gell-Mann}(1964)}]{Gell-Mann:1964ewy}%
  \BibitemOpen
  \bibfield  {author} {\bibinfo {author} {\bibfnamefont {M.}~\bibnamefont
  {Gell-Mann}},\ }\href {\doibase 10.1016/S0031-9163(64)92001-3} {\bibfield
  {journal} {\bibinfo  {journal} {Phys. Lett.}\ }\textbf {\bibinfo {volume}
  {8}},\ \bibinfo {pages} {214} (\bibinfo {year} {1964})}\BibitemShut {NoStop}%
\bibitem [{\citenamefont {Zweig}(1964)}]{Zweig:1964ruk}%
  \BibitemOpen
  \bibfield  {author} {\bibinfo {author} {\bibfnamefont {G.}~\bibnamefont
  {Zweig}},\ }\href@noop {} {\bibfield  {journal} {\bibinfo  {journal}
  {CERN-TH-401}\ } (\bibinfo {year} {1964})}\BibitemShut {NoStop}%
\bibitem [{\citenamefont {Jaffe}(1977{\natexlab{a}})}]{Jaffe:1976ig}%
  \BibitemOpen
  \bibfield  {author} {\bibinfo {author} {\bibfnamefont {R.~L.}\ \bibnamefont
  {Jaffe}},\ }\href {\doibase 10.1103/PhysRevD.15.267} {\bibfield  {journal}
  {\bibinfo  {journal} {Phys. Rev. D}\ }\textbf {\bibinfo {volume} {15}},\
  \bibinfo {pages} {267} (\bibinfo {year} {1977}{\natexlab{a}})}\BibitemShut
  {NoStop}%
\bibitem [{\citenamefont {Jaffe}(1977{\natexlab{b}})}]{Jaffe:1976ih}%
  \BibitemOpen
  \bibfield  {author} {\bibinfo {author} {\bibfnamefont {R.~L.}\ \bibnamefont
  {Jaffe}},\ }\href {\doibase 10.1103/PhysRevD.15.281} {\bibfield  {journal}
  {\bibinfo  {journal} {Phys. Rev. D}\ }\textbf {\bibinfo {volume} {15}},\
  \bibinfo {pages} {281} (\bibinfo {year} {1977}{\natexlab{b}})}\BibitemShut
  {NoStop}%
\bibitem [{\citenamefont {Choi}\ \emph {et~al.}(2003)\citenamefont {Choi} \emph
  {et~al.}}]{Belle:2003nnu}%
  \BibitemOpen
  \bibfield  {author} {\bibinfo {author} {\bibfnamefont {S.~K.}\ \bibnamefont
  {Choi}} \emph {et~al.} (\bibinfo {collaboration} {Belle}),\ }\href {\doibase
  10.1103/PhysRevLett.91.262001} {\bibfield  {journal} {\bibinfo  {journal}
  {Phys. Rev. Lett.}\ }\textbf {\bibinfo {volume} {91}},\ \bibinfo {pages}
  {262001} (\bibinfo {year} {2003})},\ \Eprint
  {http://arxiv.org/abs/hep-ex/0309032} {arXiv:hep-ex/0309032} \BibitemShut
  {NoStop}%
\bibitem [{\citenamefont {Ablikim}\ \emph {et~al.}(2013)\citenamefont {Ablikim}
  \emph {et~al.}}]{BESIII:2013ris}%
  \BibitemOpen
  \bibfield  {author} {\bibinfo {author} {\bibfnamefont {M.}~\bibnamefont
  {Ablikim}} \emph {et~al.} (\bibinfo {collaboration} {BESIII}),\ }\href
  {\doibase 10.1103/PhysRevLett.110.252001} {\bibfield  {journal} {\bibinfo
  {journal} {Phys. Rev. Lett.}\ }\textbf {\bibinfo {volume} {110}},\ \bibinfo
  {pages} {252001} (\bibinfo {year} {2013})},\ \Eprint
  {http://arxiv.org/abs/1303.5949} {arXiv:1303.5949 [hep-ex]} \BibitemShut
  {NoStop}%
\bibitem [{\citenamefont {Bouhova-Thacker}(2022)}]{Bouhova-Thacker:2022vnt}%
  \BibitemOpen
  \bibfield  {author} {\bibinfo {author} {\bibfnamefont {E.}~\bibnamefont
  {Bouhova-Thacker}} (\bibinfo {collaboration} {ATLAS}),\ }\href {\doibase
  10.22323/1.414.0806} {\bibfield  {journal} {\bibinfo  {journal} {PoS}\
  }\textbf {\bibinfo {volume} {ICHEP2022}},\ \bibinfo {pages} {806} (\bibinfo
  {year} {2022})}\BibitemShut {NoStop}%
\bibitem [{\citenamefont {Zhang}\ and\ \citenamefont
  {Yi}(2022)}]{Zhang:2022toq}%
  \BibitemOpen
  \bibfield  {author} {\bibinfo {author} {\bibfnamefont {J.}~\bibnamefont
  {Zhang}}\ and\ \bibinfo {author} {\bibfnamefont {K.}~\bibnamefont {Yi}}
  (\bibinfo {collaboration} {CMS}),\ }\href {\doibase 10.22323/1.414.0775}
  {\bibfield  {journal} {\bibinfo  {journal} {PoS}\ }\textbf {\bibinfo {volume}
  {ICHEP2022}},\ \bibinfo {pages} {775} (\bibinfo {year} {2022})},\ \Eprint
  {http://arxiv.org/abs/2212.00504} {arXiv:2212.00504 [hep-ex]} \BibitemShut
  {NoStop}%
\bibitem [{\citenamefont {Aaij}\ \emph {et~al.}(2020)\citenamefont {Aaij} \emph
  {et~al.}}]{LHCb:2020bwg}%
  \BibitemOpen
  \bibfield  {author} {\bibinfo {author} {\bibfnamefont {R.}~\bibnamefont
  {Aaij}} \emph {et~al.} (\bibinfo {collaboration} {LHCb}),\ }\href {\doibase
  10.1016/j.scib.2020.08.032} {\bibfield  {journal} {\bibinfo  {journal} {Sci.
  Bull.}\ }\textbf {\bibinfo {volume} {65}},\ \bibinfo {pages} {1983} (\bibinfo
  {year} {2020})},\ \Eprint {http://arxiv.org/abs/2006.16957} {arXiv:2006.16957
  [hep-ex]} \BibitemShut {NoStop}%
\bibitem [{\citenamefont {Aaij}\ \emph {et~al.}(2015)\citenamefont {Aaij} \emph
  {et~al.}}]{LHCb:2015c}%
  \BibitemOpen
  \bibfield  {author} {\bibinfo {author} {\bibfnamefont {R.}~\bibnamefont
  {Aaij}} \emph {et~al.} (\bibinfo {collaboration} {LHCb}),\ }\href {\doibase
  10.1103/PhysRevLett.115.072001} {\bibfield  {journal} {\bibinfo  {journal}
  {Phys. Rev. Lett.}\ }\textbf {\bibinfo {volume} {115}},\ \bibinfo {pages}
  {072001} (\bibinfo {year} {2015})},\ \Eprint
  {http://arxiv.org/abs/1507.03414} {arXiv:1507.03414 [hep-ex]} \BibitemShut
  {NoStop}%
\bibitem [{\citenamefont {Aaij}\ \emph {et~al.}(2021)\citenamefont {Aaij} \emph
  {et~al.}}]{LHCb:2021sb}%
  \BibitemOpen
  \bibfield  {author} {\bibinfo {author} {\bibfnamefont {R.}~\bibnamefont
  {Aaij}} \emph {et~al.} (\bibinfo {collaboration} {LHCb}),\ }\href {\doibase
  10.1016/j.scib.2021.02.030} {\bibfield  {journal} {\bibinfo  {journal} {Sci.
  Bull.}\ }\textbf {\bibinfo {volume} {66}},\ \bibinfo {pages} {1278} (\bibinfo
  {year} {2021})},\ \Eprint {http://arxiv.org/abs/2012.10380} {arXiv:2012.10380
  [hep-ex]} \BibitemShut {NoStop}%
\bibitem [{\citenamefont {[LHCb]}(2022)}]{LHCb:2022jad}%
  \BibitemOpen
  \bibfield  {author} {\bibinfo {author} {\bibnamefont {[LHCb]}} (\bibinfo
  {collaboration} {LHCb}),\ }\href@noop {} {\bibfield  {journal} {\bibinfo
  {journal} {CERN-EP-2022-198, LHCb-PAPER-2022-031}\ } (\bibinfo {year}
  {2022})},\ \Eprint {http://arxiv.org/abs/2210.10346} {arXiv:2210.10346
  [hep-ex]} \BibitemShut {NoStop}%
\bibitem [{\citenamefont {Maiani}\ \emph {et~al.}(2015)\citenamefont {Maiani},
  \citenamefont {Polosa},\ and\ \citenamefont {Riquer}}]{Maiani:B15}%
  \BibitemOpen
  \bibfield  {author} {\bibinfo {author} {\bibfnamefont {L.}~\bibnamefont
  {Maiani}}, \bibinfo {author} {\bibfnamefont {A.~D.}\ \bibnamefont {Polosa}},
  \ and\ \bibinfo {author} {\bibfnamefont {V.}~\bibnamefont {Riquer}},\ }\href
  {\doibase 10.1016/j.physletb.2015.08.008} {\bibfield  {journal} {\bibinfo
  {journal} {Phys. Lett. B}\ }\textbf {\bibinfo {volume} {749}},\ \bibinfo
  {pages} {289} (\bibinfo {year} {2015})},\ \Eprint
  {http://arxiv.org/abs/1507.04980} {arXiv:1507.04980 [hep-ph]} \BibitemShut
  {NoStop}%
\bibitem [{\citenamefont {Lebed}(2015)}]{Lebed:B15}%
  \BibitemOpen
  \bibfield  {author} {\bibinfo {author} {\bibfnamefont {R.~F.}\ \bibnamefont
  {Lebed}},\ }\href {\doibase 10.1016/j.physletb.2015.08.032} {\bibfield
  {journal} {\bibinfo  {journal} {Phys. Lett. B}\ }\textbf {\bibinfo {volume}
  {749}},\ \bibinfo {pages} {454} (\bibinfo {year} {2015})},\ \Eprint
  {http://arxiv.org/abs/1507.05867} {arXiv:1507.05867 [hep-ph]} \BibitemShut
  {NoStop}%
\bibitem [{\citenamefont {Anisovich}\ \emph {et~al.}(2015)\citenamefont
  {Anisovich}, \citenamefont {Matveev}, \citenamefont {Nyiri}, \citenamefont
  {Sarantsev},\ and\ \citenamefont {Semenova}}]{Anisovich:15}%
  \BibitemOpen
  \bibfield  {author} {\bibinfo {author} {\bibfnamefont {V.~V.}\ \bibnamefont
  {Anisovich}}, \bibinfo {author} {\bibfnamefont {M.~A.}\ \bibnamefont
  {Matveev}}, \bibinfo {author} {\bibfnamefont {J.}~\bibnamefont {Nyiri}},
  \bibinfo {author} {\bibfnamefont {A.~V.}\ \bibnamefont {Sarantsev}}, \ and\
  \bibinfo {author} {\bibfnamefont {A.~N.}\ \bibnamefont {Semenova}},\
  }\href@noop {} {\  (\bibinfo {year} {2015})},\ \Eprint
  {http://arxiv.org/abs/1507.07652} {arXiv:1507.07652 [hep-ph]} \BibitemShut
  {NoStop}%
\bibitem [{\citenamefont {Li}\ \emph {et~al.}(2015)\citenamefont {Li},
  \citenamefont {He},\ and\ \citenamefont {He}}]{LiHeHe:JH15}%
  \BibitemOpen
  \bibfield  {author} {\bibinfo {author} {\bibfnamefont {G.-N.}\ \bibnamefont
  {Li}}, \bibinfo {author} {\bibfnamefont {X.-G.}\ \bibnamefont {He}}, \ and\
  \bibinfo {author} {\bibfnamefont {M.}~\bibnamefont {He}},\ }\href {\doibase
  10.1007/JHEP12(2015)128} {\bibfield  {journal} {\bibinfo  {journal} {JHEP}\
  }\textbf {\bibinfo {volume} {12}},\ \bibinfo {pages} {128} (\bibinfo {year}
  {2015})},\ \Eprint {http://arxiv.org/abs/1507.08252} {arXiv:1507.08252
  [hep-ph]} \BibitemShut {NoStop}%
\bibitem [{\citenamefont {Ghosh}\ \emph {et~al.}(2017)\citenamefont {Ghosh},
  \citenamefont {Bhattacharya},\ and\ \citenamefont
  {Chakrabarti}}]{GhoshBC:PN17}%
  \BibitemOpen
  \bibfield  {author} {\bibinfo {author} {\bibfnamefont {R.}~\bibnamefont
  {Ghosh}}, \bibinfo {author} {\bibfnamefont {A.}~\bibnamefont {Bhattacharya}},
  \ and\ \bibinfo {author} {\bibfnamefont {B.}~\bibnamefont {Chakrabarti}},\
  }\href {\doibase 10.1134/S1547477117040100} {\bibfield  {journal} {\bibinfo
  {journal} {Phys. Part. Nucl. Lett.}\ }\textbf {\bibinfo {volume} {14}},\
  \bibinfo {pages} {550} (\bibinfo {year} {2017})},\ \Eprint
  {http://arxiv.org/abs/1508.00356} {arXiv:1508.00356 [hep-ph]} \BibitemShut
  {NoStop}%
\bibitem [{\citenamefont {Wang}(2016)}]{Wang:E16}%
  \BibitemOpen
  \bibfield  {author} {\bibinfo {author} {\bibfnamefont {Z.-G.}\ \bibnamefont
  {Wang}},\ }\href {\doibase 10.1140/epjc/s10052-016-3920-4} {\bibfield
  {journal} {\bibinfo  {journal} {Eur. Phys. J. C}\ }\textbf {\bibinfo {volume}
  {76}},\ \bibinfo {pages} {70} (\bibinfo {year} {2016})},\ \Eprint
  {http://arxiv.org/abs/1508.01468} {arXiv:1508.01468 [hep-ph]} \BibitemShut
  {NoStop}%
\bibitem [{\citenamefont {Zhu}\ and\ \citenamefont {Qiao}(2016)}]{ZhuQ:B16}%
  \BibitemOpen
  \bibfield  {author} {\bibinfo {author} {\bibfnamefont {R.}~\bibnamefont
  {Zhu}}\ and\ \bibinfo {author} {\bibfnamefont {C.-F.}\ \bibnamefont {Qiao}},\
  }\href {\doibase 10.1016/j.physletb.2016.03.022} {\bibfield  {journal}
  {\bibinfo  {journal} {Phys. Lett. B}\ }\textbf {\bibinfo {volume} {756}},\
  \bibinfo {pages} {259} (\bibinfo {year} {2016})},\ \Eprint
  {http://arxiv.org/abs/1510.08693} {arXiv:1510.08693 [hep-ph]} \BibitemShut
  {NoStop}%
\bibitem [{\citenamefont {Richard}\ \emph {et~al.}(2017)\citenamefont
  {Richard}, \citenamefont {Valcarce},\ and\ \citenamefont
  {Vijande}}]{Richard:B17}%
  \BibitemOpen
  \bibfield  {author} {\bibinfo {author} {\bibfnamefont {J.~M.}\ \bibnamefont
  {Richard}}, \bibinfo {author} {\bibfnamefont {A.}~\bibnamefont {Valcarce}}, \
  and\ \bibinfo {author} {\bibfnamefont {J.}~\bibnamefont {Vijande}},\ }\href
  {\doibase 10.1016/j.physletb.2017.10.036} {\bibfield  {journal} {\bibinfo
  {journal} {Phys. Lett. B}\ }\textbf {\bibinfo {volume} {774}},\ \bibinfo
  {pages} {710} (\bibinfo {year} {2017})},\ \Eprint
  {http://arxiv.org/abs/1710.08239} {arXiv:1710.08239 [hep-ph]} \BibitemShut
  {NoStop}%
\bibitem [{\citenamefont {Yang}\ \emph {et~al.}(2012)\citenamefont {Yang},
  \citenamefont {Sun}, \citenamefont {He}, \citenamefont {Liu},\ and\
  \citenamefont {Zhu}}]{Yang:C12}%
  \BibitemOpen
  \bibfield  {author} {\bibinfo {author} {\bibfnamefont {Z.-C.}\ \bibnamefont
  {Yang}}, \bibinfo {author} {\bibfnamefont {Z.-F.}\ \bibnamefont {Sun}},
  \bibinfo {author} {\bibfnamefont {J.}~\bibnamefont {He}}, \bibinfo {author}
  {\bibfnamefont {X.}~\bibnamefont {Liu}}, \ and\ \bibinfo {author}
  {\bibfnamefont {S.-L.}\ \bibnamefont {Zhu}},\ }\href {\doibase
  10.1088/1674-1137/36/1/002} {\bibfield  {journal} {\bibinfo  {journal} {Chin.
  Phys. C}\ }\textbf {\bibinfo {volume} {36}},\ \bibinfo {pages} {6} (\bibinfo
  {year} {2012})},\ \Eprint {http://arxiv.org/abs/1105.2901} {arXiv:1105.2901
  [hep-ph]} \BibitemShut {NoStop}%
\bibitem [{\citenamefont {Wu}\ \emph {et~al.}(2010)\citenamefont {Wu},
  \citenamefont {Molina}, \citenamefont {Oset},\ and\ \citenamefont
  {Zou}}]{Wu:Prl10}%
  \BibitemOpen
  \bibfield  {author} {\bibinfo {author} {\bibfnamefont {J.-J.}\ \bibnamefont
  {Wu}}, \bibinfo {author} {\bibfnamefont {R.}~\bibnamefont {Molina}}, \bibinfo
  {author} {\bibfnamefont {E.}~\bibnamefont {Oset}}, \ and\ \bibinfo {author}
  {\bibfnamefont {B.~S.}\ \bibnamefont {Zou}},\ }\href {\doibase
  10.1103/PhysRevLett.105.232001} {\bibfield  {journal} {\bibinfo  {journal}
  {Phys. Rev. Lett.}\ }\textbf {\bibinfo {volume} {105}},\ \bibinfo {pages}
  {232001} (\bibinfo {year} {2010})},\ \Eprint {http://arxiv.org/abs/1007.0573}
  {arXiv:1007.0573 [nucl-th]} \BibitemShut {NoStop}%
\bibitem [{\citenamefont {Wu}\ \emph {et~al.}(2012)\citenamefont {Wu},
  \citenamefont {Lee},\ and\ \citenamefont {Zou}}]{WuLeeZ:C12}%
  \BibitemOpen
  \bibfield  {author} {\bibinfo {author} {\bibfnamefont {J.-J.}\ \bibnamefont
  {Wu}}, \bibinfo {author} {\bibfnamefont {T.~S.~H.}\ \bibnamefont {Lee}}, \
  and\ \bibinfo {author} {\bibfnamefont {B.~S.}\ \bibnamefont {Zou}},\ }\href
  {\doibase 10.1103/PhysRevC.85.044002} {\bibfield  {journal} {\bibinfo
  {journal} {Phys. Rev. C}\ }\textbf {\bibinfo {volume} {85}},\ \bibinfo
  {pages} {044002} (\bibinfo {year} {2012})},\ \Eprint
  {http://arxiv.org/abs/1202.1036} {arXiv:1202.1036 [nucl-th]} \BibitemShut
  {NoStop}%
\bibitem [{\citenamefont {Karliner}\ and\ \citenamefont
  {Rosner}(2015)}]{KR:Prl15}%
  \BibitemOpen
  \bibfield  {author} {\bibinfo {author} {\bibfnamefont {M.}~\bibnamefont
  {Karliner}}\ and\ \bibinfo {author} {\bibfnamefont {J.~L.}\ \bibnamefont
  {Rosner}},\ }\href {\doibase 10.1103/PhysRevLett.115.122001} {\bibfield
  {journal} {\bibinfo  {journal} {Phys. Rev. Lett.}\ }\textbf {\bibinfo
  {volume} {115}},\ \bibinfo {pages} {122001} (\bibinfo {year} {2015})},\
  \Eprint {http://arxiv.org/abs/1506.06386} {arXiv:1506.06386 [hep-ph]}
  \BibitemShut {NoStop}%
\bibitem [{\citenamefont {Park}\ \emph {et~al.}(2022)\citenamefont {Park},
  \citenamefont {Cho}, \citenamefont {Kim},\ and\ \citenamefont
  {Lee}}]{ParkLC:D22}%
  \BibitemOpen
  \bibfield  {author} {\bibinfo {author} {\bibfnamefont {I.~W.}\ \bibnamefont
  {Park}}, \bibinfo {author} {\bibfnamefont {S.}~\bibnamefont {Cho}}, \bibinfo
  {author} {\bibfnamefont {Y.}~\bibnamefont {Kim}}, \ and\ \bibinfo {author}
  {\bibfnamefont {S.~H.}\ \bibnamefont {Lee}},\ }\href {\doibase
  10.1103/PhysRevD.105.114023} {\bibfield  {journal} {\bibinfo  {journal}
  {Phys. Rev. D}\ }\textbf {\bibinfo {volume} {105}},\ \bibinfo {pages}
  {114023} (\bibinfo {year} {2022})},\ \Eprint
  {http://arxiv.org/abs/2202.11631} {arXiv:2202.11631 [hep-ph]} \BibitemShut
  {NoStop}%
\bibitem [{\citenamefont {Chen}\ and\ \citenamefont {Liu}(2022)}]{ChenL:D22}%
  \BibitemOpen
  \bibfield  {author} {\bibinfo {author} {\bibfnamefont {R.}~\bibnamefont
  {Chen}}\ and\ \bibinfo {author} {\bibfnamefont {X.}~\bibnamefont {Liu}},\
  }\href {\doibase 10.1103/PhysRevD.105.014029} {\bibfield  {journal} {\bibinfo
   {journal} {Phys. Rev. D}\ }\textbf {\bibinfo {volume} {105}},\ \bibinfo
  {pages} {014029} (\bibinfo {year} {2022})},\ \Eprint
  {http://arxiv.org/abs/2201.07603} {arXiv:2201.07603 [hep-ph]} \BibitemShut
  {NoStop}%
\bibitem [{\citenamefont {Burns}\ and\ \citenamefont
  {Swanson}(2022)}]{BurnsB:E22}%
  \BibitemOpen
  \bibfield  {author} {\bibinfo {author} {\bibfnamefont {T.~J.}\ \bibnamefont
  {Burns}}\ and\ \bibinfo {author} {\bibfnamefont {E.~S.}\ \bibnamefont
  {Swanson}},\ }\href {\doibase 10.1140/epja/s10050-022-00723-9} {\bibfield
  {journal} {\bibinfo  {journal} {Eur. Phys. J. A}\ }\textbf {\bibinfo {volume}
  {58}},\ \bibinfo {pages} {68} (\bibinfo {year} {2022})},\ \Eprint
  {http://arxiv.org/abs/2112.11527} {arXiv:2112.11527 [hep-ph]} \BibitemShut
  {NoStop}%
\bibitem [{\citenamefont {Yang}\ \emph {et~al.}(2021)\citenamefont {Yang},
  \citenamefont {Huang},\ and\ \citenamefont {Zhu}}]{YangH:SC21}%
  \BibitemOpen
  \bibfield  {author} {\bibinfo {author} {\bibfnamefont {F.}~\bibnamefont
  {Yang}}, \bibinfo {author} {\bibfnamefont {Y.}~\bibnamefont {Huang}}, \ and\
  \bibinfo {author} {\bibfnamefont {H.~Q.}\ \bibnamefont {Zhu}},\ }\href
  {\doibase 10.1007/s11433-021-1796-0} {\bibfield  {journal} {\bibinfo
  {journal} {Sci. China Phys. Mech. Astron.}\ }\textbf {\bibinfo {volume}
  {64}},\ \bibinfo {pages} {121011} (\bibinfo {year} {2021})},\ \Eprint
  {http://arxiv.org/abs/2107.13267} {arXiv:2107.13267 [hep-ph]} \BibitemShut
  {NoStop}%
\bibitem [{\citenamefont {Shi}\ \emph {et~al.}(2021)\citenamefont {Shi},
  \citenamefont {Huang},\ and\ \citenamefont {Wang}}]{ShiHW:E21}%
  \BibitemOpen
  \bibfield  {author} {\bibinfo {author} {\bibfnamefont {P.-P.}\ \bibnamefont
  {Shi}}, \bibinfo {author} {\bibfnamefont {F.}~\bibnamefont {Huang}}, \ and\
  \bibinfo {author} {\bibfnamefont {W.-L.}\ \bibnamefont {Wang}},\ }\href
  {\doibase 10.1140/epja/s10050-021-00542-4} {\bibfield  {journal} {\bibinfo
  {journal} {Eur. Phys. J. A}\ }\textbf {\bibinfo {volume} {57}},\ \bibinfo
  {pages} {237} (\bibinfo {year} {2021})},\ \Eprint
  {http://arxiv.org/abs/2107.08680} {arXiv:2107.08680 [hep-ph]} \BibitemShut
  {NoStop}%
\bibitem [{\citenamefont {Li}\ \emph {et~al.}(2021)\citenamefont {Li},
  \citenamefont {Liu}, \citenamefont {Sun},\ and\ \citenamefont
  {Chen}}]{LiLiuS:D21}%
  \BibitemOpen
  \bibfield  {author} {\bibinfo {author} {\bibfnamefont {M.-W.}\ \bibnamefont
  {Li}}, \bibinfo {author} {\bibfnamefont {Z.-W.}\ \bibnamefont {Liu}},
  \bibinfo {author} {\bibfnamefont {Z.-F.}\ \bibnamefont {Sun}}, \ and\
  \bibinfo {author} {\bibfnamefont {R.}~\bibnamefont {Chen}},\ }\href {\doibase
  10.1103/PhysRevD.104.054016} {\bibfield  {journal} {\bibinfo  {journal}
  {Phys. Rev. D}\ }\textbf {\bibinfo {volume} {104}},\ \bibinfo {pages}
  {054016} (\bibinfo {year} {2021})},\ \Eprint
  {http://arxiv.org/abs/2106.15053} {arXiv:2106.15053 [hep-ph]} \BibitemShut
  {NoStop}%
\bibitem [{\citenamefont {Ling}\ \emph
  {et~al.}(2021{\natexlab{a}})\citenamefont {Ling}, \citenamefont {Lu},
  \citenamefont {Liu},\ and\ \citenamefont {Geng}}]{LingXL:D21}%
  \BibitemOpen
  \bibfield  {author} {\bibinfo {author} {\bibfnamefont {X.-Z.}\ \bibnamefont
  {Ling}}, \bibinfo {author} {\bibfnamefont {J.-X.}\ \bibnamefont {Lu}},
  \bibinfo {author} {\bibfnamefont {M.-Z.}\ \bibnamefont {Liu}}, \ and\
  \bibinfo {author} {\bibfnamefont {L.-S.}\ \bibnamefont {Geng}},\ }\href
  {\doibase 10.1103/PhysRevD.104.074022} {\bibfield  {journal} {\bibinfo
  {journal} {Phys. Rev. D}\ }\textbf {\bibinfo {volume} {104}},\ \bibinfo
  {pages} {074022} (\bibinfo {year} {2021}{\natexlab{a}})},\ \Eprint
  {http://arxiv.org/abs/2106.12250} {arXiv:2106.12250 [hep-ph]} \BibitemShut
  {NoStop}%
\bibitem [{\citenamefont {Wu}\ \emph {et~al.}(2021{\natexlab{a}})\citenamefont
  {Wu}, \citenamefont {Pan}, \citenamefont {Liu}, \citenamefont {Lu},
  \citenamefont {Geng},\ and\ \citenamefont {Liu}}]{WuPL:D21}%
  \BibitemOpen
  \bibfield  {author} {\bibinfo {author} {\bibfnamefont {T.-W.}\ \bibnamefont
  {Wu}}, \bibinfo {author} {\bibfnamefont {Y.-W.}\ \bibnamefont {Pan}},
  \bibinfo {author} {\bibfnamefont {M.-Z.}\ \bibnamefont {Liu}}, \bibinfo
  {author} {\bibfnamefont {J.-X.}\ \bibnamefont {Lu}}, \bibinfo {author}
  {\bibfnamefont {L.-S.}\ \bibnamefont {Geng}}, \ and\ \bibinfo {author}
  {\bibfnamefont {X.-H.}\ \bibnamefont {Liu}},\ }\href {\doibase
  10.1103/PhysRevD.104.094032} {\bibfield  {journal} {\bibinfo  {journal}
  {Phys. Rev. D}\ }\textbf {\bibinfo {volume} {104}},\ \bibinfo {pages}
  {094032} (\bibinfo {year} {2021}{\natexlab{a}})},\ \Eprint
  {http://arxiv.org/abs/2106.11450} {arXiv:2106.11450 [hep-ph]} \BibitemShut
  {NoStop}%
\bibitem [{\citenamefont {Ruangyoo}\ \emph {et~al.}(2022)\citenamefont
  {Ruangyoo}, \citenamefont {Phumphan}, \citenamefont {Chen}, \citenamefont
  {Limphirat},\ and\ \citenamefont {Yan}}]{RuangyooPC:D22}%
  \BibitemOpen
  \bibfield  {author} {\bibinfo {author} {\bibfnamefont {W.}~\bibnamefont
  {Ruangyoo}}, \bibinfo {author} {\bibfnamefont {K.}~\bibnamefont {Phumphan}},
  \bibinfo {author} {\bibfnamefont {C.-C.}\ \bibnamefont {Chen}}, \bibinfo
  {author} {\bibfnamefont {A.}~\bibnamefont {Limphirat}}, \ and\ \bibinfo
  {author} {\bibfnamefont {Y.}~\bibnamefont {Yan}},\ }\href {\doibase
  10.1088/1361-6471/ac58af} {\bibfield  {journal} {\bibinfo  {journal} {J.
  Phys. G}\ }\textbf {\bibinfo {volume} {49}},\ \bibinfo {pages} {075001}
  (\bibinfo {year} {2022})},\ \Eprint {http://arxiv.org/abs/2105.14249}
  {arXiv:2105.14249 [hep-ph]} \BibitemShut {NoStop}%
\bibitem [{\citenamefont {Ling}\ \emph
  {et~al.}(2021{\natexlab{b}})\citenamefont {Ling}, \citenamefont {Dai},
  \citenamefont {Du},\ and\ \citenamefont {Wang}}]{LingDW:E21}%
  \BibitemOpen
  \bibfield  {author} {\bibinfo {author} {\bibfnamefont {P.}~\bibnamefont
  {Ling}}, \bibinfo {author} {\bibfnamefont {X.-H.}\ \bibnamefont {Dai}},
  \bibinfo {author} {\bibfnamefont {M.-L.}\ \bibnamefont {Du}}, \ and\ \bibinfo
  {author} {\bibfnamefont {Q.}~\bibnamefont {Wang}},\ }\href {\doibase
  10.1140/epjc/s10052-021-09613-8} {\bibfield  {journal} {\bibinfo  {journal}
  {Eur. Phys. J. C}\ }\textbf {\bibinfo {volume} {81}},\ \bibinfo {pages} {819}
  (\bibinfo {year} {2021}{\natexlab{b}})},\ \Eprint
  {http://arxiv.org/abs/2104.11133} {arXiv:2104.11133 [hep-ph]} \BibitemShut
  {NoStop}%
\bibitem [{\citenamefont {Lu}\ \emph {et~al.}(2021)\citenamefont {Lu},
  \citenamefont {Liu}, \citenamefont {Shi},\ and\ \citenamefont
  {Geng}}]{LuLS:D21}%
  \BibitemOpen
  \bibfield  {author} {\bibinfo {author} {\bibfnamefont {J.-X.}\ \bibnamefont
  {Lu}}, \bibinfo {author} {\bibfnamefont {M.-Z.}\ \bibnamefont {Liu}},
  \bibinfo {author} {\bibfnamefont {R.-X.}\ \bibnamefont {Shi}}, \ and\
  \bibinfo {author} {\bibfnamefont {L.-S.}\ \bibnamefont {Geng}},\ }\href
  {\doibase 10.1103/PhysRevD.104.034022} {\bibfield  {journal} {\bibinfo
  {journal} {Phys. Rev. D}\ }\textbf {\bibinfo {volume} {104}},\ \bibinfo
  {pages} {034022} (\bibinfo {year} {2021})},\ \Eprint
  {http://arxiv.org/abs/2104.10303} {arXiv:2104.10303 [hep-ph]} \BibitemShut
  {NoStop}%
\bibitem [{\citenamefont {Wu}\ \emph {et~al.}(2021{\natexlab{b}})\citenamefont
  {Wu}, \citenamefont {Chen},\ and\ \citenamefont {Ji}}]{WuCJ:D21}%
  \BibitemOpen
  \bibfield  {author} {\bibinfo {author} {\bibfnamefont {Q.}~\bibnamefont
  {Wu}}, \bibinfo {author} {\bibfnamefont {D.-Y.}\ \bibnamefont {Chen}}, \ and\
  \bibinfo {author} {\bibfnamefont {R.}~\bibnamefont {Ji}},\ }\href {\doibase
  10.1088/0256-307X/38/7/071301} {\bibfield  {journal} {\bibinfo  {journal}
  {Chin. Phys. Lett.}\ }\textbf {\bibinfo {volume} {38}},\ \bibinfo {pages}
  {071301} (\bibinfo {year} {2021}{\natexlab{b}})},\ \Eprint
  {http://arxiv.org/abs/2103.05257} {arXiv:2103.05257 [hep-ph]} \BibitemShut
  {NoStop}%
\bibitem [{\citenamefont {Du}\ \emph {et~al.}(2021)\citenamefont {Du},
  \citenamefont {Baru}, \citenamefont {Guo}, \citenamefont {Hanhart},
  \citenamefont {Mei\ss{}ner}, \citenamefont {Oller},\ and\ \citenamefont
  {Wang}}]{DuGH:Jh21}%
  \BibitemOpen
  \bibfield  {author} {\bibinfo {author} {\bibfnamefont {M.-L.}\ \bibnamefont
  {Du}}, \bibinfo {author} {\bibfnamefont {V.}~\bibnamefont {Baru}}, \bibinfo
  {author} {\bibfnamefont {F.-K.}\ \bibnamefont {Guo}}, \bibinfo {author}
  {\bibfnamefont {C.}~\bibnamefont {Hanhart}}, \bibinfo {author} {\bibfnamefont
  {U.-G.}\ \bibnamefont {Mei\ss{}ner}}, \bibinfo {author} {\bibfnamefont
  {J.~A.}\ \bibnamefont {Oller}}, \ and\ \bibinfo {author} {\bibfnamefont
  {Q.}~\bibnamefont {Wang}},\ }\href {\doibase 10.1007/JHEP08(2021)157}
  {\bibfield  {journal} {\bibinfo  {journal} {JHEP}\ }\textbf {\bibinfo
  {volume} {08}},\ \bibinfo {pages} {157} (\bibinfo {year} {2021})},\ \Eprint
  {http://arxiv.org/abs/2102.07159} {arXiv:2102.07159 [hep-ph]} \BibitemShut
  {NoStop}%
\bibitem [{\citenamefont {Xiao}\ \emph {et~al.}(2021)\citenamefont {Xiao},
  \citenamefont {Wu},\ and\ \citenamefont {Zou}}]{XiaoWZ:D21}%
  \BibitemOpen
  \bibfield  {author} {\bibinfo {author} {\bibfnamefont {C.~W.}\ \bibnamefont
  {Xiao}}, \bibinfo {author} {\bibfnamefont {J.~J.}\ \bibnamefont {Wu}}, \ and\
  \bibinfo {author} {\bibfnamefont {B.~S.}\ \bibnamefont {Zou}},\ }\href
  {\doibase 10.1103/PhysRevD.103.054016} {\bibfield  {journal} {\bibinfo
  {journal} {Phys. Rev. D}\ }\textbf {\bibinfo {volume} {103}},\ \bibinfo
  {pages} {054016} (\bibinfo {year} {2021})},\ \Eprint
  {http://arxiv.org/abs/2102.02607} {arXiv:2102.02607 [hep-ph]} \BibitemShut
  {NoStop}%
\bibitem [{\citenamefont {Zhu}\ \emph {et~al.}(2021)\citenamefont {Zhu},
  \citenamefont {Song},\ and\ \citenamefont {He}}]{ZhuSH:D21}%
  \BibitemOpen
  \bibfield  {author} {\bibinfo {author} {\bibfnamefont {J.-T.}\ \bibnamefont
  {Zhu}}, \bibinfo {author} {\bibfnamefont {L.-Q.}\ \bibnamefont {Song}}, \
  and\ \bibinfo {author} {\bibfnamefont {J.}~\bibnamefont {He}},\ }\href
  {\doibase 10.1103/PhysRevD.103.074007} {\bibfield  {journal} {\bibinfo
  {journal} {Phys. Rev. D}\ }\textbf {\bibinfo {volume} {103}},\ \bibinfo
  {pages} {074007} (\bibinfo {year} {2021})},\ \Eprint
  {http://arxiv.org/abs/2101.12441} {arXiv:2101.12441 [hep-ph]} \BibitemShut
  {NoStop}%
\bibitem [{\citenamefont {Chen}(2021)}]{Chen:E21}%
  \BibitemOpen
  \bibfield  {author} {\bibinfo {author} {\bibfnamefont {R.}~\bibnamefont
  {Chen}},\ }\href {\doibase 10.1140/epjc/s10052-021-08904-4} {\bibfield
  {journal} {\bibinfo  {journal} {Eur. Phys. J. C}\ }\textbf {\bibinfo {volume}
  {81}},\ \bibinfo {pages} {122} (\bibinfo {year} {2021})},\ \Eprint
  {http://arxiv.org/abs/2101.10614} {arXiv:2101.10614 [hep-ph]} \BibitemShut
  {NoStop}%
\bibitem [{\citenamefont {Yan}\ \emph {et~al.}(2022{\natexlab{a}})\citenamefont
  {Yan}, \citenamefont {Peng}, \citenamefont {S\'anchez~S\'anchez},\ and\
  \citenamefont {Pavon~Valderrama}}]{YanPS:E21}%
  \BibitemOpen
  \bibfield  {author} {\bibinfo {author} {\bibfnamefont {M.-J.}\ \bibnamefont
  {Yan}}, \bibinfo {author} {\bibfnamefont {F.-Z.}\ \bibnamefont {Peng}},
  \bibinfo {author} {\bibfnamefont {M.}~\bibnamefont {S\'anchez~S\'anchez}}, \
  and\ \bibinfo {author} {\bibfnamefont {M.}~\bibnamefont {Pavon~Valderrama}},\
  }\href {\doibase 10.1140/epjc/s10052-022-10522-7} {\bibfield  {journal}
  {\bibinfo  {journal} {Eur. Phys. J. C}\ }\textbf {\bibinfo {volume} {82}},\
  \bibinfo {pages} {574} (\bibinfo {year} {2022}{\natexlab{a}})},\ \Eprint
  {http://arxiv.org/abs/2108.05306} {arXiv:2108.05306 [hep-ph]} \BibitemShut
  {NoStop}%
\bibitem [{\citenamefont {Yang}\ and\ \citenamefont {Ping}(2017)}]{YangPW:D17}%
  \BibitemOpen
  \bibfield  {author} {\bibinfo {author} {\bibfnamefont {G.}~\bibnamefont
  {Yang}}\ and\ \bibinfo {author} {\bibfnamefont {J.}~\bibnamefont {Ping}},\
  }\href {\doibase 10.1103/PhysRevD.95.014010} {\bibfield  {journal} {\bibinfo
  {journal} {Phys. Rev. D}\ }\textbf {\bibinfo {volume} {95}},\ \bibinfo
  {pages} {014010} (\bibinfo {year} {2017})},\ \Eprint
  {http://arxiv.org/abs/1511.09053} {arXiv:1511.09053 [hep-ph]} \BibitemShut
  {NoStop}%
\bibitem [{\citenamefont {Yang}\ and\ \citenamefont {Chen}(2023)}]{YangC:C23}%
  \BibitemOpen
  \bibfield  {author} {\bibinfo {author} {\bibfnamefont {P.}~\bibnamefont
  {Yang}}\ and\ \bibinfo {author} {\bibfnamefont {W.}~\bibnamefont {Chen}},\
  }\href {\doibase 10.1088/1674-1137/ac9889} {\bibfield  {journal} {\bibinfo
  {journal} {Chin. Phys. C}\ }\textbf {\bibinfo {volume} {47}},\ \bibinfo
  {pages} {013105} (\bibinfo {year} {2023})},\ \Eprint
  {http://arxiv.org/abs/2203.15616} {arXiv:2203.15616 [hep-ph]} \BibitemShut
  {NoStop}%
\bibitem [{\citenamefont {Huang}\ and\ \citenamefont {Zhu}(2021)}]{HuangZ:D21}%
  \BibitemOpen
  \bibfield  {author} {\bibinfo {author} {\bibfnamefont {Y.}~\bibnamefont
  {Huang}}\ and\ \bibinfo {author} {\bibfnamefont {H.~Q.}\ \bibnamefont
  {Zhu}},\ }\href {\doibase 10.1103/PhysRevD.104.056027} {\bibfield  {journal}
  {\bibinfo  {journal} {Phys. Rev. D}\ }\textbf {\bibinfo {volume} {104}},\
  \bibinfo {pages} {056027} (\bibinfo {year} {2021})},\ \Eprint
  {http://arxiv.org/abs/2107.03773} {arXiv:2107.03773 [hep-ph]} \BibitemShut
  {NoStop}%
\bibitem [{\citenamefont {Zhu}\ \emph {et~al.}(2020)\citenamefont {Zhu},
  \citenamefont {Kong}, \citenamefont {Liu},\ and\ \citenamefont
  {He}}]{ZhuKL:D20}%
  \BibitemOpen
  \bibfield  {author} {\bibinfo {author} {\bibfnamefont {J.-T.}\ \bibnamefont
  {Zhu}}, \bibinfo {author} {\bibfnamefont {S.-Y.}\ \bibnamefont {Kong}},
  \bibinfo {author} {\bibfnamefont {Y.}~\bibnamefont {Liu}}, \ and\ \bibinfo
  {author} {\bibfnamefont {J.}~\bibnamefont {He}},\ }\href {\doibase
  10.1140/epjc/s10052-020-8410-z} {\bibfield  {journal} {\bibinfo  {journal}
  {Eur. Phys. J. C}\ }\textbf {\bibinfo {volume} {80}},\ \bibinfo {pages}
  {1016} (\bibinfo {year} {2020})},\ \Eprint {http://arxiv.org/abs/2007.07596}
  {arXiv:2007.07596 [hep-ph]} \BibitemShut {NoStop}%
\bibitem [{\citenamefont {Yang}\ \emph {et~al.}(2019)\citenamefont {Yang},
  \citenamefont {Ping},\ and\ \citenamefont {Segovia}}]{YangPS:D19}%
  \BibitemOpen
  \bibfield  {author} {\bibinfo {author} {\bibfnamefont {G.}~\bibnamefont
  {Yang}}, \bibinfo {author} {\bibfnamefont {J.}~\bibnamefont {Ping}}, \ and\
  \bibinfo {author} {\bibfnamefont {J.}~\bibnamefont {Segovia}},\ }\href
  {\doibase 10.1103/PhysRevD.99.014035} {\bibfield  {journal} {\bibinfo
  {journal} {Phys. Rev. D}\ }\textbf {\bibinfo {volume} {99}},\ \bibinfo
  {pages} {014035} (\bibinfo {year} {2019})},\ \Eprint
  {http://arxiv.org/abs/1809.06193} {arXiv:1809.06193 [hep-ph]} \BibitemShut
  {NoStop}%
\bibitem [{\citenamefont {Burns}(2015)}]{Burns:EA15}%
  \BibitemOpen
  \bibfield  {author} {\bibinfo {author} {\bibfnamefont {T.~J.}\ \bibnamefont
  {Burns}},\ }\href {\doibase 10.1140/epja/i2015-15152-6} {\bibfield  {journal}
  {\bibinfo  {journal} {Eur. Phys. J. A}\ }\textbf {\bibinfo {volume} {51}},\
  \bibinfo {pages} {152} (\bibinfo {year} {2015})},\ \Eprint
  {http://arxiv.org/abs/1509.02460} {arXiv:1509.02460 [hep-ph]} \BibitemShut
  {NoStop}%
\bibitem [{\citenamefont {Chen}\ \emph {et~al.}(2016)\citenamefont {Chen},
  \citenamefont {Chen}, \citenamefont {Liu},\ and\ \citenamefont
  {Zhu}}]{ChenCLZ:PR16}%
  \BibitemOpen
  \bibfield  {author} {\bibinfo {author} {\bibfnamefont {H.-X.}\ \bibnamefont
  {Chen}}, \bibinfo {author} {\bibfnamefont {W.}~\bibnamefont {Chen}}, \bibinfo
  {author} {\bibfnamefont {X.}~\bibnamefont {Liu}}, \ and\ \bibinfo {author}
  {\bibfnamefont {S.-L.}\ \bibnamefont {Zhu}},\ }\href {\doibase
  10.1016/j.physrep.2016.05.004} {\bibfield  {journal} {\bibinfo  {journal}
  {Phys. Rept.}\ }\textbf {\bibinfo {volume} {639}},\ \bibinfo {pages} {1}
  (\bibinfo {year} {2016})},\ \Eprint {http://arxiv.org/abs/1601.02092}
  {arXiv:1601.02092 [hep-ph]} \BibitemShut {NoStop}%
\bibitem [{\citenamefont {Liu}\ \emph {et~al.}(2019)\citenamefont {Liu},
  \citenamefont {Chen}, \citenamefont {Chen}, \citenamefont {Liu},\ and\
  \citenamefont {Zhu}}]{LiuCCL:PP19}%
  \BibitemOpen
  \bibfield  {author} {\bibinfo {author} {\bibfnamefont {Y.-R.}\ \bibnamefont
  {Liu}}, \bibinfo {author} {\bibfnamefont {H.-X.}\ \bibnamefont {Chen}},
  \bibinfo {author} {\bibfnamefont {W.}~\bibnamefont {Chen}}, \bibinfo {author}
  {\bibfnamefont {X.}~\bibnamefont {Liu}}, \ and\ \bibinfo {author}
  {\bibfnamefont {S.-L.}\ \bibnamefont {Zhu}},\ }\href {\doibase
  10.1016/j.ppnp.2019.04.003} {\bibfield  {journal} {\bibinfo  {journal} {Prog.
  Part. Nucl. Phys.}\ }\textbf {\bibinfo {volume} {107}},\ \bibinfo {pages}
  {237} (\bibinfo {year} {2019})},\ \Eprint {http://arxiv.org/abs/1903.11976}
  {arXiv:1903.11976 [hep-ph]} \BibitemShut {NoStop}%
\bibitem [{\citenamefont {Brambilla}\ \emph {et~al.}(2020)\citenamefont
  {Brambilla}, \citenamefont {Eidelman}, \citenamefont {Hanhart}, \citenamefont
  {Nefediev}, \citenamefont {Shen}, \citenamefont {Thomas}, \citenamefont
  {Vairo},\ and\ \citenamefont {Yuan}}]{Brambilla:PR20}%
  \BibitemOpen
  \bibfield  {author} {\bibinfo {author} {\bibfnamefont {N.}~\bibnamefont
  {Brambilla}}, \bibinfo {author} {\bibfnamefont {S.}~\bibnamefont {Eidelman}},
  \bibinfo {author} {\bibfnamefont {C.}~\bibnamefont {Hanhart}}, \bibinfo
  {author} {\bibfnamefont {A.}~\bibnamefont {Nefediev}}, \bibinfo {author}
  {\bibfnamefont {C.-P.}\ \bibnamefont {Shen}}, \bibinfo {author}
  {\bibfnamefont {C.~E.}\ \bibnamefont {Thomas}}, \bibinfo {author}
  {\bibfnamefont {A.}~\bibnamefont {Vairo}}, \ and\ \bibinfo {author}
  {\bibfnamefont {C.-Z.}\ \bibnamefont {Yuan}},\ }\href {\doibase
  10.1016/j.physrep.2020.05.001} {\bibfield  {journal} {\bibinfo  {journal}
  {Phys. Rept.}\ }\textbf {\bibinfo {volume} {873}},\ \bibinfo {pages} {1}
  (\bibinfo {year} {2020})},\ \Eprint {http://arxiv.org/abs/1907.07583}
  {arXiv:1907.07583 [hep-ex]} \BibitemShut {NoStop}%
\bibitem [{\citenamefont {An}\ \emph {et~al.}(2021{\natexlab{a}})\citenamefont
  {An}, \citenamefont {Chen}, \citenamefont {Liu},\ and\ \citenamefont
  {Liu}}]{An:2021vwi}%
  \BibitemOpen
  \bibfield  {author} {\bibinfo {author} {\bibfnamefont {H.-T.}\ \bibnamefont
  {An}}, \bibinfo {author} {\bibfnamefont {K.}~\bibnamefont {Chen}}, \bibinfo
  {author} {\bibfnamefont {Z.-W.}\ \bibnamefont {Liu}}, \ and\ \bibinfo
  {author} {\bibfnamefont {X.}~\bibnamefont {Liu}},\ }\href {\doibase
  10.1103/PhysRevD.103.114027} {\bibfield  {journal} {\bibinfo  {journal}
  {Phys. Rev. D}\ }\textbf {\bibinfo {volume} {103}},\ \bibinfo {pages}
  {114027} (\bibinfo {year} {2021}{\natexlab{a}})},\ \Eprint
  {http://arxiv.org/abs/2106.02837} {arXiv:2106.02837 [hep-ph]} \BibitemShut
  {NoStop}%
\bibitem [{\citenamefont {Stancu}\ and\ \citenamefont
  {Pepin}(1999)}]{Stancu:1999qr}%
  \BibitemOpen
  \bibfield  {author} {\bibinfo {author} {\bibfnamefont {F.}~\bibnamefont
  {Stancu}}\ and\ \bibinfo {author} {\bibfnamefont {S.}~\bibnamefont {Pepin}},\
  }\href {\doibase 10.1007/s006010050109} {\bibfield  {journal} {\bibinfo
  {journal} {Few Body Syst.}\ }\textbf {\bibinfo {volume} {26}},\ \bibinfo
  {pages} {113} (\bibinfo {year} {1999})}\BibitemShut {NoStop}%
\bibitem [{\citenamefont {Zhang}\ \emph {et~al.}(2021)\citenamefont {Zhang},
  \citenamefont {Xu},\ and\ \citenamefont {Jia}}]{Zhang:2021yul}%
  \BibitemOpen
  \bibfield  {author} {\bibinfo {author} {\bibfnamefont {W.-X.}\ \bibnamefont
  {Zhang}}, \bibinfo {author} {\bibfnamefont {H.}~\bibnamefont {Xu}}, \ and\
  \bibinfo {author} {\bibfnamefont {D.}~\bibnamefont {Jia}},\ }\href {\doibase
  10.1103/PhysRevD.104.114011} {\bibfield  {journal} {\bibinfo  {journal}
  {Phys. Rev. D}\ }\textbf {\bibinfo {volume} {104}},\ \bibinfo {pages}
  {114011} (\bibinfo {year} {2021})},\ \Eprint
  {http://arxiv.org/abs/2109.07040} {arXiv:2109.07040 [hep-ph]} \BibitemShut
  {NoStop}%
\bibitem [{\citenamefont {DeGrand}\ \emph {et~al.}(1975)\citenamefont
  {DeGrand}, \citenamefont {Jaffe}, \citenamefont {Johnson},\ and\
  \citenamefont {Kiskis}}]{DeGrand:1975cf}%
  \BibitemOpen
  \bibfield  {author} {\bibinfo {author} {\bibfnamefont {T.~A.}\ \bibnamefont
  {DeGrand}}, \bibinfo {author} {\bibfnamefont {R.~L.}\ \bibnamefont {Jaffe}},
  \bibinfo {author} {\bibfnamefont {K.}~\bibnamefont {Johnson}}, \ and\
  \bibinfo {author} {\bibfnamefont {J.~E.}\ \bibnamefont {Kiskis}},\ }\href
  {\doibase 10.1103/PhysRevD.12.2060} {\bibfield  {journal} {\bibinfo
  {journal} {Phys. Rev. D}\ }\textbf {\bibinfo {volume} {12}},\ \bibinfo
  {pages} {2060} (\bibinfo {year} {1975})}\BibitemShut {NoStop}%
\bibitem [{\citenamefont {Johnson}(1975)}]{Johnson:1975zp}%
  \BibitemOpen
  \bibfield  {author} {\bibinfo {author} {\bibfnamefont {K.}~\bibnamefont
  {Johnson}},\ }\href@noop {} {\bibfield  {journal} {\bibinfo  {journal} {Acta
  Phys. Polon. B}\ }\textbf {\bibinfo {volume} {6}},\ \bibinfo {pages} {865}
  (\bibinfo {year} {1975})}\BibitemShut {NoStop}%
\bibitem [{\citenamefont {Karliner}\ and\ \citenamefont
  {Rosner}(2014)}]{Karliner:2014gca}%
  \BibitemOpen
  \bibfield  {author} {\bibinfo {author} {\bibfnamefont {M.}~\bibnamefont
  {Karliner}}\ and\ \bibinfo {author} {\bibfnamefont {J.~L.}\ \bibnamefont
  {Rosner}},\ }\href {\doibase 10.1103/PhysRevD.90.094007} {\bibfield
  {journal} {\bibinfo  {journal} {Phys. Rev. D}\ }\textbf {\bibinfo {volume}
  {90}},\ \bibinfo {pages} {094007} (\bibinfo {year} {2014})},\ \Eprint
  {http://arxiv.org/abs/1408.5877} {arXiv:1408.5877 [hep-ph]} \BibitemShut
  {NoStop}%
\bibitem [{\citenamefont {Karliner}\ and\ \citenamefont
  {Rosner}(2017)}]{Karliner:2017elp}%
  \BibitemOpen
  \bibfield  {author} {\bibinfo {author} {\bibfnamefont {M.}~\bibnamefont
  {Karliner}}\ and\ \bibinfo {author} {\bibfnamefont {J.~L.}\ \bibnamefont
  {Rosner}},\ }\href {\doibase 10.1038/nature24289} {\bibfield  {journal}
  {\bibinfo  {journal} {Nature}\ }\textbf {\bibinfo {volume} {551}},\ \bibinfo
  {pages} {89} (\bibinfo {year} {2017})},\ \Eprint
  {http://arxiv.org/abs/1708.02547} {arXiv:1708.02547 [hep-ph]} \BibitemShut
  {NoStop}%
\bibitem [{\citenamefont {De~Rujula}\ \emph {et~al.}(1975)\citenamefont
  {De~Rujula}, \citenamefont {Georgi},\ and\ \citenamefont
  {Glashow}}]{DeRujula:1975qlm}%
  \BibitemOpen
  \bibfield  {author} {\bibinfo {author} {\bibfnamefont {A.}~\bibnamefont
  {De~Rujula}}, \bibinfo {author} {\bibfnamefont {H.}~\bibnamefont {Georgi}}, \
  and\ \bibinfo {author} {\bibfnamefont {S.~L.}\ \bibnamefont {Glashow}},\
  }\href {\doibase 10.1103/PhysRevD.12.147} {\bibfield  {journal} {\bibinfo
  {journal} {Phys. Rev. D}\ }\textbf {\bibinfo {volume} {12}},\ \bibinfo
  {pages} {147} (\bibinfo {year} {1975})}\BibitemShut {NoStop}%
\bibitem [{\citenamefont {Wang}\ \emph {et~al.}(2016)\citenamefont {Wang},
  \citenamefont {Chen}, \citenamefont {Ma}, \citenamefont {Liu},\ and\
  \citenamefont {Zhu}}]{Wang:2016dzu}%
  \BibitemOpen
  \bibfield  {author} {\bibinfo {author} {\bibfnamefont {G.-J.}\ \bibnamefont
  {Wang}}, \bibinfo {author} {\bibfnamefont {R.}~\bibnamefont {Chen}}, \bibinfo
  {author} {\bibfnamefont {L.}~\bibnamefont {Ma}}, \bibinfo {author}
  {\bibfnamefont {X.}~\bibnamefont {Liu}}, \ and\ \bibinfo {author}
  {\bibfnamefont {S.-L.}\ \bibnamefont {Zhu}},\ }\href {\doibase
  10.1103/PhysRevD.94.094018} {\bibfield  {journal} {\bibinfo  {journal} {Phys.
  Rev. D}\ }\textbf {\bibinfo {volume} {94}},\ \bibinfo {pages} {094018}
  (\bibinfo {year} {2016})},\ \Eprint {http://arxiv.org/abs/1605.01337}
  {arXiv:1605.01337 [hep-ph]} \BibitemShut {NoStop}%
\bibitem [{\citenamefont {Workman}\ \emph {et~al.}(2022)\citenamefont {Workman}
  \emph {et~al.}}]{ParticleDataGroup:2022pth}%
  \BibitemOpen
  \bibfield  {author} {\bibinfo {author} {\bibfnamefont {R.~L.}\ \bibnamefont
  {Workman}} \emph {et~al.} (\bibinfo {collaboration} {Particle Data Group}),\
  }\href {\doibase 10.1093/ptep/ptac097} {\bibfield  {journal} {\bibinfo
  {journal} {PTEP}\ }\textbf {\bibinfo {volume} {2022}},\ \bibinfo {pages}
  {083C01} (\bibinfo {year} {2022})}\BibitemShut {NoStop}%
\bibitem [{\citenamefont {Tiesinga}\ \emph {et~al.}(2021)\citenamefont
  {Tiesinga}, \citenamefont {Mohr}, \citenamefont {Newell},\ and\ \citenamefont
  {Taylor}}]{Tiesinga:2021myr}%
  \BibitemOpen
  \bibfield  {author} {\bibinfo {author} {\bibfnamefont {E.}~\bibnamefont
  {Tiesinga}}, \bibinfo {author} {\bibfnamefont {P.~J.}\ \bibnamefont {Mohr}},
  \bibinfo {author} {\bibfnamefont {D.~B.}\ \bibnamefont {Newell}}, \ and\
  \bibinfo {author} {\bibfnamefont {B.~N.}\ \bibnamefont {Taylor}},\ }\href
  {\doibase 10.1103/RevModPhys.93.025010} {\bibfield  {journal} {\bibinfo
  {journal} {Rev. Mod. Phys.}\ }\textbf {\bibinfo {volume} {93}},\ \bibinfo
  {pages} {025010} (\bibinfo {year} {2021})}\BibitemShut {NoStop}%
\bibitem [{\citenamefont {Zhang}\ and\ \citenamefont
  {Huang}(2009)}]{Zhang:2009re}%
  \BibitemOpen
  \bibfield  {author} {\bibinfo {author} {\bibfnamefont {J.-R.}\ \bibnamefont
  {Zhang}}\ and\ \bibinfo {author} {\bibfnamefont {M.-Q.}\ \bibnamefont
  {Huang}},\ }\href {\doibase 10.1016/j.physletb.2009.02.056} {\bibfield
  {journal} {\bibinfo  {journal} {Phys. Lett. B}\ }\textbf {\bibinfo {volume}
  {674}},\ \bibinfo {pages} {28} (\bibinfo {year} {2009})},\ \Eprint
  {http://arxiv.org/abs/0902.3297} {arXiv:0902.3297 [hep-ph]} \BibitemShut
  {NoStop}%
\bibitem [{\citenamefont {Aliev}\ \emph {et~al.}(2014)\citenamefont {Aliev},
  \citenamefont {Azizi},\ and\ \citenamefont {Savc\i{}}}]{Aliev:2014lxa}%
  \BibitemOpen
  \bibfield  {author} {\bibinfo {author} {\bibfnamefont {T.~M.}\ \bibnamefont
  {Aliev}}, \bibinfo {author} {\bibfnamefont {K.}~\bibnamefont {Azizi}}, \ and\
  \bibinfo {author} {\bibfnamefont {M.}~\bibnamefont {Savc\i{}}},\ }\href
  {\doibase 10.1088/0954-3899/41/6/065003} {\bibfield  {journal} {\bibinfo
  {journal} {J. Phys. G}\ }\textbf {\bibinfo {volume} {41}},\ \bibinfo {pages}
  {065003} (\bibinfo {year} {2014})},\ \Eprint {http://arxiv.org/abs/1404.2091}
  {arXiv:1404.2091 [hep-ph]} \BibitemShut {NoStop}%
\bibitem [{\citenamefont {Qin}\ \emph {et~al.}(2019)\citenamefont {Qin},
  \citenamefont {Roberts},\ and\ \citenamefont {Schmidt}}]{Qin:2019hgk}%
  \BibitemOpen
  \bibfield  {author} {\bibinfo {author} {\bibfnamefont {S.-x.}\ \bibnamefont
  {Qin}}, \bibinfo {author} {\bibfnamefont {C.~D.}\ \bibnamefont {Roberts}}, \
  and\ \bibinfo {author} {\bibfnamefont {S.~M.}\ \bibnamefont {Schmidt}},\
  }\href {\doibase 10.1007/s00601-019-1488-x} {\bibfield  {journal} {\bibinfo
  {journal} {Few Body Syst.}\ }\textbf {\bibinfo {volume} {60}},\ \bibinfo
  {pages} {26} (\bibinfo {year} {2019})},\ \Eprint
  {http://arxiv.org/abs/1902.00026} {arXiv:1902.00026 [nucl-th]} \BibitemShut
  {NoStop}%
\bibitem [{\citenamefont {Faessler}\ \emph {et~al.}(2006)\citenamefont
  {Faessler}, \citenamefont {Gutsche}, \citenamefont {Ivanov}, \citenamefont
  {Korner}, \citenamefont {Lyubovitskij}, \citenamefont {Nicmorus},\ and\
  \citenamefont {Pumsa-ard}}]{Faessler:2006ft}%
  \BibitemOpen
  \bibfield  {author} {\bibinfo {author} {\bibfnamefont {A.}~\bibnamefont
  {Faessler}}, \bibinfo {author} {\bibfnamefont {T.}~\bibnamefont {Gutsche}},
  \bibinfo {author} {\bibfnamefont {M.~A.}\ \bibnamefont {Ivanov}}, \bibinfo
  {author} {\bibfnamefont {J.~G.}\ \bibnamefont {Korner}}, \bibinfo {author}
  {\bibfnamefont {V.~E.}\ \bibnamefont {Lyubovitskij}}, \bibinfo {author}
  {\bibfnamefont {D.}~\bibnamefont {Nicmorus}}, \ and\ \bibinfo {author}
  {\bibfnamefont {K.}~\bibnamefont {Pumsa-ard}},\ }\href {\doibase
  10.1103/PhysRevD.73.094013} {\bibfield  {journal} {\bibinfo  {journal} {Phys.
  Rev. D}\ }\textbf {\bibinfo {volume} {73}},\ \bibinfo {pages} {094013}
  (\bibinfo {year} {2006})},\ \Eprint {http://arxiv.org/abs/hep-ph/0602193}
  {arXiv:hep-ph/0602193} \BibitemShut {NoStop}%
\bibitem [{\citenamefont {Bernotas}\ and\ \citenamefont
  {Simonis}(2009)}]{Bernotas:2008bu}%
  \BibitemOpen
  \bibfield  {author} {\bibinfo {author} {\bibfnamefont {A.}~\bibnamefont
  {Bernotas}}\ and\ \bibinfo {author} {\bibfnamefont {V.}~\bibnamefont
  {Simonis}},\ }\href {\doibase 10.3952/lithjphys.49110} {\bibfield  {journal}
  {\bibinfo  {journal} {Lith. J. Phys.}\ }\textbf {\bibinfo {volume} {49}},\
  \bibinfo {pages} {19} (\bibinfo {year} {2009})},\ \Eprint
  {http://arxiv.org/abs/0808.1220} {arXiv:0808.1220 [hep-ph]} \BibitemShut
  {NoStop}%
\bibitem [{\citenamefont {Brown}\ \emph {et~al.}(2014)\citenamefont {Brown},
  \citenamefont {Detmold}, \citenamefont {Meinel},\ and\ \citenamefont
  {Orginos}}]{Brown:2014ena}%
  \BibitemOpen
  \bibfield  {author} {\bibinfo {author} {\bibfnamefont {Z.~S.}\ \bibnamefont
  {Brown}}, \bibinfo {author} {\bibfnamefont {W.}~\bibnamefont {Detmold}},
  \bibinfo {author} {\bibfnamefont {S.}~\bibnamefont {Meinel}}, \ and\ \bibinfo
  {author} {\bibfnamefont {K.}~\bibnamefont {Orginos}},\ }\href {\doibase
  10.1103/PhysRevD.90.094507} {\bibfield  {journal} {\bibinfo  {journal} {Phys.
  Rev. D}\ }\textbf {\bibinfo {volume} {90}},\ \bibinfo {pages} {094507}
  (\bibinfo {year} {2014})},\ \Eprint {http://arxiv.org/abs/1409.0497}
  {arXiv:1409.0497 [hep-lat]} \BibitemShut {NoStop}%
\bibitem [{\citenamefont {Yang}\ \emph {et~al.}(2020)\citenamefont {Yang},
  \citenamefont {Ping}, \citenamefont {Ortega},\ and\ \citenamefont
  {Segovia}}]{Yang:2019lsg}%
  \BibitemOpen
  \bibfield  {author} {\bibinfo {author} {\bibfnamefont {G.}~\bibnamefont
  {Yang}}, \bibinfo {author} {\bibfnamefont {J.}~\bibnamefont {Ping}}, \bibinfo
  {author} {\bibfnamefont {P.~G.}\ \bibnamefont {Ortega}}, \ and\ \bibinfo
  {author} {\bibfnamefont {J.}~\bibnamefont {Segovia}},\ }\href {\doibase
  10.1088/1674-1137/44/2/023102} {\bibfield  {journal} {\bibinfo  {journal}
  {Chin. Phys. C}\ }\textbf {\bibinfo {volume} {44}},\ \bibinfo {pages}
  {023102} (\bibinfo {year} {2020})},\ \Eprint
  {http://arxiv.org/abs/1904.10166} {arXiv:1904.10166 [hep-ph]} \BibitemShut
  {NoStop}%
\bibitem [{\citenamefont {Flynn}\ \emph {et~al.}(2012)\citenamefont {Flynn},
  \citenamefont {Hernandez},\ and\ \citenamefont {Nieves}}]{Flynn:2011gf}%
  \BibitemOpen
  \bibfield  {author} {\bibinfo {author} {\bibfnamefont {J.~M.}\ \bibnamefont
  {Flynn}}, \bibinfo {author} {\bibfnamefont {E.}~\bibnamefont {Hernandez}}, \
  and\ \bibinfo {author} {\bibfnamefont {J.}~\bibnamefont {Nieves}},\ }\href
  {\doibase 10.1103/PhysRevD.85.014012} {\bibfield  {journal} {\bibinfo
  {journal} {Phys. Rev. D}\ }\textbf {\bibinfo {volume} {85}},\ \bibinfo
  {pages} {014012} (\bibinfo {year} {2012})},\ \Eprint
  {http://arxiv.org/abs/1110.2962} {arXiv:1110.2962 [hep-ph]} \BibitemShut
  {NoStop}%
\bibitem [{\citenamefont {Martynenko}(2008)}]{Martynenko:2007je}%
  \BibitemOpen
  \bibfield  {author} {\bibinfo {author} {\bibfnamefont {A.~P.}\ \bibnamefont
  {Martynenko}},\ }\href {\doibase 10.1016/j.physletb.2008.04.030} {\bibfield
  {journal} {\bibinfo  {journal} {Phys. Lett. B}\ }\textbf {\bibinfo {volume}
  {663}},\ \bibinfo {pages} {317} (\bibinfo {year} {2008})},\ \Eprint
  {http://arxiv.org/abs/0708.2033} {arXiv:0708.2033 [hep-ph]} \BibitemShut
  {NoStop}%
\bibitem [{\citenamefont {Wang}(2021{\natexlab{a}})}]{Wang:2020avt}%
  \BibitemOpen
  \bibfield  {author} {\bibinfo {author} {\bibfnamefont {Z.-G.}\ \bibnamefont
  {Wang}},\ }\href {\doibase 10.1007/s43673-021-00006-3} {\bibfield  {journal}
  {\bibinfo  {journal} {AAPPS Bull.}\ }\textbf {\bibinfo {volume} {31}},\
  \bibinfo {pages} {5} (\bibinfo {year} {2021}{\natexlab{a}})},\ \Eprint
  {http://arxiv.org/abs/2010.08939} {arXiv:2010.08939 [hep-ph]} \BibitemShut
  {NoStop}%
\bibitem [{\citenamefont {Wei}\ \emph {et~al.}(2015)\citenamefont {Wei},
  \citenamefont {Chen},\ and\ \citenamefont {Guo}}]{Wei:2015gsa}%
  \BibitemOpen
  \bibfield  {author} {\bibinfo {author} {\bibfnamefont {K.-W.}\ \bibnamefont
  {Wei}}, \bibinfo {author} {\bibfnamefont {B.}~\bibnamefont {Chen}}, \ and\
  \bibinfo {author} {\bibfnamefont {X.-H.}\ \bibnamefont {Guo}},\ }\href
  {\doibase 10.1103/PhysRevD.92.076008} {\bibfield  {journal} {\bibinfo
  {journal} {Phys. Rev. D}\ }\textbf {\bibinfo {volume} {92}},\ \bibinfo
  {pages} {076008} (\bibinfo {year} {2015})},\ \Eprint
  {http://arxiv.org/abs/1503.05184} {arXiv:1503.05184 [hep-ph]} \BibitemShut
  {NoStop}%
\bibitem [{\citenamefont {Wei}\ \emph {et~al.}(2017)\citenamefont {Wei},
  \citenamefont {Chen}, \citenamefont {Liu}, \citenamefont {Wang},\ and\
  \citenamefont {Guo}}]{Wei:2016jyk}%
  \BibitemOpen
  \bibfield  {author} {\bibinfo {author} {\bibfnamefont {K.-W.}\ \bibnamefont
  {Wei}}, \bibinfo {author} {\bibfnamefont {B.}~\bibnamefont {Chen}}, \bibinfo
  {author} {\bibfnamefont {N.}~\bibnamefont {Liu}}, \bibinfo {author}
  {\bibfnamefont {Q.-Q.}\ \bibnamefont {Wang}}, \ and\ \bibinfo {author}
  {\bibfnamefont {X.-H.}\ \bibnamefont {Guo}},\ }\href {\doibase
  10.1103/PhysRevD.95.116005} {\bibfield  {journal} {\bibinfo  {journal} {Phys.
  Rev. D}\ }\textbf {\bibinfo {volume} {95}},\ \bibinfo {pages} {116005}
  (\bibinfo {year} {2017})},\ \Eprint {http://arxiv.org/abs/1609.02512}
  {arXiv:1609.02512 [hep-ph]} \BibitemShut {NoStop}%
\bibitem [{\citenamefont {Patel}\ \emph {et~al.}(2009)\citenamefont {Patel},
  \citenamefont {Majethiya},\ and\ \citenamefont {Vinodkumar}}]{Patel:2008mv}%
  \BibitemOpen
  \bibfield  {author} {\bibinfo {author} {\bibfnamefont {B.}~\bibnamefont
  {Patel}}, \bibinfo {author} {\bibfnamefont {A.}~\bibnamefont {Majethiya}}, \
  and\ \bibinfo {author} {\bibfnamefont {P.~C.}\ \bibnamefont {Vinodkumar}},\
  }\href {\doibase 10.1007/s12043-009-0061-4} {\bibfield  {journal} {\bibinfo
  {journal} {Pramana}\ }\textbf {\bibinfo {volume} {72}},\ \bibinfo {pages}
  {679} (\bibinfo {year} {2009})},\ \Eprint {http://arxiv.org/abs/0808.2880}
  {arXiv:0808.2880 [hep-ph]} \BibitemShut {NoStop}%
\bibitem [{\citenamefont {Llanes-Estrada}\ \emph {et~al.}(2013)\citenamefont
  {Llanes-Estrada}, \citenamefont {Pavlova},\ and\ \citenamefont
  {Williams}}]{Llanes-Estrada:2013rwa}%
  \BibitemOpen
  \bibfield  {author} {\bibinfo {author} {\bibfnamefont {F.~J.}\ \bibnamefont
  {Llanes-Estrada}}, \bibinfo {author} {\bibfnamefont {O.~I.}\ \bibnamefont
  {Pavlova}}, \ and\ \bibinfo {author} {\bibfnamefont {R.}~\bibnamefont
  {Williams}},\ }\href {\doibase 10.5506/APhysPolBSupp.6.821} {\bibfield
  {journal} {\bibinfo  {journal} {Acta Phys. Polon. Supp.}\ }\textbf {\bibinfo
  {volume} {6}},\ \bibinfo {pages} {821} (\bibinfo {year} {2013})},\ \Eprint
  {http://arxiv.org/abs/1304.3636} {arXiv:1304.3636 [nucl-th]} \BibitemShut
  {NoStop}%
\bibitem [{\citenamefont {Guti\'errez-Guerrero}\ \emph
  {et~al.}(2019)\citenamefont {Guti\'errez-Guerrero}, \citenamefont {Bashir},
  \citenamefont {Bedolla},\ and\ \citenamefont
  {Santopinto}}]{Gutierrez-Guerrero:2019uwa}%
  \BibitemOpen
  \bibfield  {author} {\bibinfo {author} {\bibfnamefont {L.~X.}\ \bibnamefont
  {Guti\'errez-Guerrero}}, \bibinfo {author} {\bibfnamefont {A.}~\bibnamefont
  {Bashir}}, \bibinfo {author} {\bibfnamefont {M.~A.}\ \bibnamefont {Bedolla}},
  \ and\ \bibinfo {author} {\bibfnamefont {E.}~\bibnamefont {Santopinto}},\
  }\href {\doibase 10.1103/PhysRevD.100.114032} {\bibfield  {journal} {\bibinfo
   {journal} {Phys. Rev. D}\ }\textbf {\bibinfo {volume} {100}},\ \bibinfo
  {pages} {114032} (\bibinfo {year} {2019})},\ \Eprint
  {http://arxiv.org/abs/1911.09213} {arXiv:1911.09213 [nucl-th]} \BibitemShut
  {NoStop}%
\bibitem [{\citenamefont {Roberts}\ and\ \citenamefont
  {Pervin}(2008)}]{Roberts:2007ni}%
  \BibitemOpen
  \bibfield  {author} {\bibinfo {author} {\bibfnamefont {W.}~\bibnamefont
  {Roberts}}\ and\ \bibinfo {author} {\bibfnamefont {M.}~\bibnamefont
  {Pervin}},\ }\href {\doibase 10.1142/S0217751X08041219} {\bibfield  {journal}
  {\bibinfo  {journal} {Int. J. Mod. Phys. A}\ }\textbf {\bibinfo {volume}
  {23}},\ \bibinfo {pages} {2817} (\bibinfo {year} {2008})},\ \Eprint
  {http://arxiv.org/abs/0711.2492} {arXiv:0711.2492 [nucl-th]} \BibitemShut
  {NoStop}%
\bibitem [{\citenamefont {Wang}(2012)}]{Wang:2011ae}%
  \BibitemOpen
  \bibfield  {author} {\bibinfo {author} {\bibfnamefont {Z.-G.}\ \bibnamefont
  {Wang}},\ }\href {\doibase 10.1088/0253-6102/58/5/17} {\bibfield  {journal}
  {\bibinfo  {journal} {Commun. Theor. Phys.}\ }\textbf {\bibinfo {volume}
  {58}},\ \bibinfo {pages} {723} (\bibinfo {year} {2012})},\ \Eprint
  {http://arxiv.org/abs/1112.2274} {arXiv:1112.2274 [hep-ph]} \BibitemShut
  {NoStop}%
\bibitem [{\citenamefont {Yin}\ \emph {et~al.}(2019)\citenamefont {Yin},
  \citenamefont {Chen}, \citenamefont {Krein}, \citenamefont {Roberts},
  \citenamefont {Segovia},\ and\ \citenamefont {Xu}}]{Yin:2019bxe}%
  \BibitemOpen
  \bibfield  {author} {\bibinfo {author} {\bibfnamefont {P.-L.}\ \bibnamefont
  {Yin}}, \bibinfo {author} {\bibfnamefont {C.}~\bibnamefont {Chen}}, \bibinfo
  {author} {\bibfnamefont {G.~a.}\ \bibnamefont {Krein}}, \bibinfo {author}
  {\bibfnamefont {C.~D.}\ \bibnamefont {Roberts}}, \bibinfo {author}
  {\bibfnamefont {J.}~\bibnamefont {Segovia}}, \ and\ \bibinfo {author}
  {\bibfnamefont {S.-S.}\ \bibnamefont {Xu}},\ }\href {\doibase
  10.1103/PhysRevD.100.034008} {\bibfield  {journal} {\bibinfo  {journal}
  {Phys. Rev. D}\ }\textbf {\bibinfo {volume} {100}},\ \bibinfo {pages}
  {034008} (\bibinfo {year} {2019})},\ \Eprint
  {http://arxiv.org/abs/1903.00160} {arXiv:1903.00160 [nucl-th]} \BibitemShut
  {NoStop}%
\bibitem [{\citenamefont {Yan}\ \emph {et~al.}(2023)\citenamefont {Yan},
  \citenamefont {Zhang},\ and\ \citenamefont {Jia}}]{Yan:2023lvm}%
  \BibitemOpen
  \bibfield  {author} {\bibinfo {author} {\bibfnamefont {T.-Q.}\ \bibnamefont
  {Yan}}, \bibinfo {author} {\bibfnamefont {W.-X.}\ \bibnamefont {Zhang}}, \
  and\ \bibinfo {author} {\bibfnamefont {D.}~\bibnamefont {Jia}},\ }\href@noop
  {} {\  (\bibinfo {year} {2023})},\ \Eprint {http://arxiv.org/abs/2304.01684}
  {arXiv:2304.01684 [hep-ph]} \BibitemShut {NoStop}%
\bibitem [{\citenamefont {An}\ \emph {et~al.}(2021{\natexlab{b}})\citenamefont
  {An}, \citenamefont {Chen}, \citenamefont {Liu},\ and\ \citenamefont
  {Liu}}]{An:2020jix}%
  \BibitemOpen
  \bibfield  {author} {\bibinfo {author} {\bibfnamefont {H.-T.}\ \bibnamefont
  {An}}, \bibinfo {author} {\bibfnamefont {K.}~\bibnamefont {Chen}}, \bibinfo
  {author} {\bibfnamefont {Z.-W.}\ \bibnamefont {Liu}}, \ and\ \bibinfo
  {author} {\bibfnamefont {X.}~\bibnamefont {Liu}},\ }\href {\doibase
  10.1103/PhysRevD.103.074006} {\bibfield  {journal} {\bibinfo  {journal}
  {Phys. Rev. D}\ }\textbf {\bibinfo {volume} {103}},\ \bibinfo {pages}
  {074006} (\bibinfo {year} {2021}{\natexlab{b}})},\ \Eprint
  {http://arxiv.org/abs/2012.12459} {arXiv:2012.12459 [hep-ph]} \BibitemShut
  {NoStop}%
\bibitem [{\citenamefont {An}\ \emph {et~al.}(2022)\citenamefont {An},
  \citenamefont {Luo}, \citenamefont {Liu},\ and\ \citenamefont
  {Liu}}]{An:2022fvs}%
  \BibitemOpen
  \bibfield  {author} {\bibinfo {author} {\bibfnamefont {H.-T.}\ \bibnamefont
  {An}}, \bibinfo {author} {\bibfnamefont {S.-Q.}\ \bibnamefont {Luo}},
  \bibinfo {author} {\bibfnamefont {Z.-W.}\ \bibnamefont {Liu}}, \ and\
  \bibinfo {author} {\bibfnamefont {X.}~\bibnamefont {Liu}},\ }\href {\doibase
  10.1103/PhysRevD.105.074032} {\bibfield  {journal} {\bibinfo  {journal}
  {Phys. Rev. D}\ }\textbf {\bibinfo {volume} {105}},\ \bibinfo {pages}
  {074032} (\bibinfo {year} {2022})},\ \Eprint
  {http://arxiv.org/abs/2203.03448} {arXiv:2203.03448 [hep-ph]} \BibitemShut
  {NoStop}%
\bibitem [{\citenamefont {Yang}\ \emph {et~al.}(2022)\citenamefont {Yang},
  \citenamefont {Ping},\ and\ \citenamefont {Segovia}}]{Yang:2022bfu}%
  \BibitemOpen
  \bibfield  {author} {\bibinfo {author} {\bibfnamefont {G.}~\bibnamefont
  {Yang}}, \bibinfo {author} {\bibfnamefont {J.}~\bibnamefont {Ping}}, \ and\
  \bibinfo {author} {\bibfnamefont {J.}~\bibnamefont {Segovia}},\ }\href
  {\doibase 10.1103/PhysRevD.106.014005} {\bibfield  {journal} {\bibinfo
  {journal} {Phys. Rev. D}\ }\textbf {\bibinfo {volume} {106}},\ \bibinfo
  {pages} {014005} (\bibinfo {year} {2022})},\ \Eprint
  {http://arxiv.org/abs/2205.11548} {arXiv:2205.11548 [hep-ph]} \BibitemShut
  {NoStop}%
\bibitem [{\citenamefont {Yan}\ \emph {et~al.}(2022{\natexlab{b}})\citenamefont
  {Yan}, \citenamefont {Wu}, \citenamefont {Hu}, \citenamefont {Huang},\ and\
  \citenamefont {Ping}}]{Yan:2021glh}%
  \BibitemOpen
  \bibfield  {author} {\bibinfo {author} {\bibfnamefont {Y.}~\bibnamefont
  {Yan}}, \bibinfo {author} {\bibfnamefont {Y.}~\bibnamefont {Wu}}, \bibinfo
  {author} {\bibfnamefont {X.}~\bibnamefont {Hu}}, \bibinfo {author}
  {\bibfnamefont {H.}~\bibnamefont {Huang}}, \ and\ \bibinfo {author}
  {\bibfnamefont {J.}~\bibnamefont {Ping}},\ }\href {\doibase
  10.1103/PhysRevD.105.014027} {\bibfield  {journal} {\bibinfo  {journal}
  {Phys. Rev. D}\ }\textbf {\bibinfo {volume} {105}},\ \bibinfo {pages}
  {014027} (\bibinfo {year} {2022}{\natexlab{b}})},\ \Eprint
  {http://arxiv.org/abs/2110.10853} {arXiv:2110.10853 [hep-ph]} \BibitemShut
  {NoStop}%
\bibitem [{\citenamefont {Wang}(2021{\natexlab{b}})}]{Wang:2021xao}%
  \BibitemOpen
  \bibfield  {author} {\bibinfo {author} {\bibfnamefont {Z.-G.}\ \bibnamefont
  {Wang}},\ }\href {\doibase 10.1016/j.nuclphysb.2021.115579} {\bibfield
  {journal} {\bibinfo  {journal} {Nucl. Phys. B}\ }\textbf {\bibinfo {volume}
  {973}},\ \bibinfo {pages} {115579} (\bibinfo {year} {2021}{\natexlab{b}})},\
  \Eprint {http://arxiv.org/abs/2104.12090} {arXiv:2104.12090 [hep-ph]}
  \BibitemShut {NoStop}%
\bibitem [{\citenamefont {Zhang}(2021)}]{Zhang:2020vpz}%
  \BibitemOpen
  \bibfield  {author} {\bibinfo {author} {\bibfnamefont {J.-R.}\ \bibnamefont
  {Zhang}},\ }\href {\doibase 10.1103/PhysRevD.103.074016} {\bibfield
  {journal} {\bibinfo  {journal} {Phys. Rev. D}\ }\textbf {\bibinfo {volume}
  {103}},\ \bibinfo {pages} {074016} (\bibinfo {year} {2021})},\ \Eprint
  {http://arxiv.org/abs/2011.04594} {arXiv:2011.04594 [hep-ph]} \BibitemShut
  {NoStop}%
\end{thebibliography}
\end{document}